\PassOptionsToPackage{breaklinks=true}{hyperref}

\documentclass[acmsmall, nonacm]{acmart}

\makeatletter
\newif\if@taisap@known \@taisap@knowntrue
\let\@saved@ClassError\ClassError
\long\def\ClassError#1#2#3{\global\@taisap@knownfalse}
\acmJournal{TAISAP}
\let\ClassError\@saved@ClassError
\if@taisap@known\else
  \acmJournal{TOPS}%
  \gdef\@journalName{ACM Transactions on AI Security and Privacy}%
  \gdef\@journalNameShort{ACM Trans. AI Secur. Priv.}%
\fi
\makeatother

\acmYear{2026}
\acmVolume{1}      %
\acmNumber{1}      %
\acmArticle{1}     %
\acmMonth{1}       %

\setcopyright{none}              %

\usepackage{caption}
\usepackage{amsmath}
\usepackage{amsfonts}

\usepackage{amssymb}

\usepackage{graphicx}
\usepackage{xcolor}

\usepackage{booktabs}
\usepackage{multirow}
\usepackage{tabularx}

\usepackage{enumitem}
\usepackage{textcomp}

\usepackage{soul}
\setul{0.35ex}{0.10ex}   %
\newif\ifshowedits
\showeditsfalse   %

\usepackage{comment}
\ifshowedits\includecomment{cutblock}\else\excludecomment{cutblock}\fi
\soulregister{\emph}{1}
\soulregister{\textit}{1}
\soulregister{\textbf}{1}
\soulregister{\texttt}{1}
\soulregister{\textsc}{1}
\soulregister{\mbox}{1}
\soulregister{\url}{1}

\newcommand{\floatnote}[2][j]{%
  \par\vspace{2pt}%
  \begingroup
    \def\fn@centre{c}\def\fn@given{#1}%
    \begin{minipage}{\linewidth}%
      \setlength{\parindent}{0pt}\small
      \ifx\fn@given\fn@centre\centering\else\raggedright\fi
      #2%
    \end{minipage}%
  \endgroup
  \par\vspace{1pt}}

\usepackage{fontawesome5}

\usepackage{algorithm}
\usepackage{algorithmic}

\usepackage{url}

\usepackage{tikz}
\usetikzlibrary{positioning, arrows.meta, shapes.multipart, shapes.geometric, calc, fit, matrix, backgrounds}

\definecolor{cDefaultHeaderBg}{HTML}{E8EFF6} %

\definecolor{cDefaultHeaderBg1}{HTML}{B7D3E8} %
\definecolor{cDefaultHeaderBg2}{HTML}{E8EFF6} %
\definecolor{cDefaultHeaderFg}{HTML}{092030} %

\definecolor{cDefaultHeaderFg1}{HTML}{12365B} %
\definecolor{cDefaultHeaderFg2}{HTML}{092030} %
\definecolor{cDefaultHeaderRule}{HTML}{8DA5BC} %
\definecolor{cDefaultHeaderRule1}{HTML}{1F4E79} %
\definecolor{cDefaultHeaderRule2}{HTML}{8DA5BC} %
\definecolor{cDefaultSubHeaderBg}{HTML}{F4F7FB} %
\definecolor{cDefaultSubHeaderBg1}{HTML}{E8F1F8} %

\definecolor{cDefaultSubHeaderBg2}{HTML}{F4F7FB} %

\definecolor{cMarkGood}{HTML}{15496E}     %
\definecolor{cMarkGoodMid}{HTML}{2E7BAA}  %
\definecolor{cMarkGoodBg}{HTML}{DEEBF4}   %
\definecolor{cMarkFail}{HTML}{9E5A5A}     %
\definecolor{cMarkFailRing}{HTML}{CFA9A9} %
\definecolor{cMarkAmber}{HTML}{B45309}

\definecolor{cLKink}{HTML}{0B2E4F}
\definecolor{cLKslight}{HTML}{E9F0F7}
\definecolor{cLKfair}{HTML}{BCD5E9}
\definecolor{cLKmoderate}{HTML}{7FA8CE}
\definecolor{cLKsubstantial}{HTML}{3D6F9F}
\definecolor{cLKalmost}{HTML}{1B4368}
\tikzset{
  lkSlight/.style={fill=cLKslight, text=cLKink},
  lkFair/.style={fill=cLKfair, text=cLKink},
  lkModerate/.style={fill=cLKmoderate, text=cLKink},
  lkSubstantial/.style={fill=cLKsubstantial, text=white},
  lkAlmost/.style={fill=cLKalmost, text=white},
}
\newcommand{\lkchip}[1]{%
  \tikz[baseline=-0.30ex]{%
    \fill[#1] (0,0) rectangle (0.30,0.155);
    \draw[cLKink, line width=0.15pt] (0,0) rectangle (0.30,0.155);}}

\newcommand{\lkbandlegend}{%
  \begin{tabular}{@{}lll@{}}
    \textbf{Bands, Landis and Koch~\cite{landis1977observer}:} & \lkchip{lkSlight}\,slight 0.00 to 0.20 & \lkchip{lkFair}\,fair 0.21 to 0.40 \\
    \lkchip{lkModerate}\,moderate 0.41 to 0.60 & \lkchip{lkSubstantial}\,substantial 0.61 to 0.80 & \lkchip{lkAlmost}\,almost perfect 0.81 to 1.00
  \end{tabular}%
}

\usepackage{colortbl}
\colorlet{tblHeaderBg}{cDefaultHeaderBg}   %
\colorlet{tblHeaderFg}{cDefaultHeaderFg}    %
\colorlet{tblHeaderRule}{cDefaultHeaderRule}%

\newcommand{\thd}[1]{{\bfseries\color{tblHeaderFg}#1}}

\colorlet{tblSubHeaderBg}{cDefaultSubHeaderBg} %

\newcommand{\wtTickPath}[2]{%
  \draw[#1, line width=#2, line cap=round, line join=round]
    (-0.318em,0.019em) -- (-0.109em,-0.161em) -- (0.304em,0.282em);
}
\newcommand{\wtCrossPath}[2]{%
  \draw[#1, line width=#2, line cap=round]
    (-0.250em,-0.205em) -- (0.250em,0.295em)
    (-0.250em,0.295em) -- (0.250em,-0.205em);
}
\newcommand{\wtMarkBox}{\useasboundingbox (-0.60em,-0.56em) rectangle (0.60em,0.64em);}

\newcommand{\tMark}{\mbox{\begin{tikzpicture}[baseline=-0.34ex]\wtMarkBox
  \fill[cMarkGood] (0,0.045em) circle (0.545em);
  \wtTickPath{white}{0.150em}\end{tikzpicture}}}
\newcommand{\dMark}{\mbox{\begin{tikzpicture}[baseline=-0.34ex]\wtMarkBox
  \fill[cMarkGoodBg] (0,0.045em) circle (0.545em);
  \draw[cMarkGoodMid, line width=0.088em] (0,0.045em) circle (0.501em);
  \wtTickPath{cMarkGoodMid}{0.150em}\end{tikzpicture}}}
\newcommand{\xMark}{\mbox{\begin{tikzpicture}[baseline=-0.34ex]\wtMarkBox
  \draw[cMarkFailRing, line width=0.078em] (0,0.045em) circle (0.506em);
  \wtCrossPath{cMarkFail}{0.132em}\end{tikzpicture}}}

\tikzset{
  tblmat/.style={matrix of nodes,
    nodes={inner xsep=4pt, inner ysep=3pt, anchor=north west,
           font=\normalsize, align=left},
    row sep=0.3pt, column sep=0pt, nodes in empty cells,
    append after command={
      \pgfextra{
        \begin{scope}[on background layer]
          \fill[tblHeaderBg] (\tikzlastnode.north west) rectangle (\tikzlastnode.east|-\tikzlastnode-2-1.north);
        \end{scope}
      }
    }
  },
  tblhdr/.style={text=tblHeaderFg, minimum height=14pt, font=\bfseries\normalsize},
  Rtop/.style={line width=0.04pt, draw=tblHeaderRule},
  Rmid/.style={line width=.03pt, draw=tblHeaderRule},
  Rbot/.style={line width=0.04pt, draw=tblHeaderRule},
}

\newsavebox{\tbltmpbox}

\makeatletter
\@ifclassloaded{IEEEtran}{%
\long\def\@makecaption#1#2{%
\ifx\@captype\@IEEEtablestring%
\footnotesize\bgroup\par\centering\@IEEEtabletopskipstrut{\normalfont\footnotesize #1}\\{\normalfont\footnotesize #2}\par\addvspace{0.5\baselineskip}\egroup%
\@IEEEtablecaptionsepspace
\else
\@IEEEfigurecaptionsepspace
\setbox\@tempboxa\hbox{\normalfont\footnotesize {#1.}\nobreakspace\nobreakspace #2}%
\ifdim \wd\@tempboxa >\hsize%
\setbox\@tempboxa\hbox{\normalfont\footnotesize {#1.}\nobreakspace\nobreakspace}%
\parbox[t]{\hsize}{\normalfont\footnotesize\noindent\unhbox\@tempboxa#2}%
\else%
\ifCLASSOPTIONconference \hbox to\hsize{\normalfont\footnotesize\hfil\box\@tempboxa\hfil}%
\else \hbox to\hsize{\normalfont\footnotesize\box\@tempboxa\hfil}%
\fi\fi\fi}%
}{}%
\makeatother

\edef\figScaleScript{1}%
\newlength{\liveW}
\newsavebox{\figclampbox}
\newcommand{\figclamp}[2]{%
  \sbox\figclampbox{#2}%
  \ifdim\wd\figclampbox>\linewidth
    \GenericWarning{}{LaTeX Warning: TRUESIZE OVERFLOW (figure): content is
      \the\dimexpr\wd\figclampbox-\linewidth\relax\space wider than the line}%
  \fi
  \makebox[\ifdim\wd\figclampbox>\linewidth\wd\figclampbox\else\linewidth\fi][c]{\usebox\figclampbox}}
\usepackage{etoolbox}
\newcommand{\figShrink}{1}   %
\newsavebox{\figshrinkbox}
\ifdim\figShrink pt=1pt\relax\else
  \BeforeBeginEnvironment{tikzpicture}{\begin{lrbox}{\figshrinkbox}}
  \AfterEndEnvironment{tikzpicture}{\end{lrbox}%
    \scalebox{\figShrink}{\usebox{\figshrinkbox}}}
\fi

\newcommand{\figboxscript}[1]{\figclamp{\figScaleScript}{#1}}

\newcommand{\tblfitbox}{%
  \ifdim\wd\tbltmpbox>\linewidth
    \GenericWarning{}{LaTeX Warning: TRUESIZE OVERFLOW (table): content is
      \the\dimexpr\wd\tbltmpbox-\linewidth\relax\space wider than the line}%
  \fi
  \usebox{\tbltmpbox}}

\newcommand{\bfprop}{\ensuremath{\curvearrowright}}

\providecommand{\appendices}{\appendix}

\usepackage{placeins}
\usepackage{longtable}
\newenvironment{inlinefloat}{\begin{figure}[!htb]\centering}{\end{figure}}
\newenvironment{inlinebox}{\par\vspace{\intextsep}\noindent\begin{minipage}{\liveW}\centering}{\end{minipage}\par\vspace{\intextsep}}
\renewcommand{\footnotesize}{\small}
\renewcommand{\bibliofont}{\small}
\extrafloats{64}
\renewcommand{\topfraction}{0.92}
\renewcommand{\bottomfraction}{0.7}
\renewcommand{\dbltopfraction}{0.92}
\renewcommand{\textfraction}{0.07}
\renewcommand{\floatpagefraction}{0.9}
\renewcommand{\dblfloatpagefraction}{0.9}
\newcommand{\appendixfloatexception}{%
  \setcounter{topnumber}{3}%
  \setcounter{bottomnumber}{2}%
  \setcounter{totalnumber}{4}%
  \setcounter{dbltopnumber}{3}%
  \renewcommand{\topfraction}{0.92}%
  \renewcommand{\bottomfraction}{0.75}%
  \renewcommand{\dbltopfraction}{0.92}%
  \renewcommand{\textfraction}{0.08}%
  \renewcommand{\floatpagefraction}{0.7}%
  \renewcommand{\dblfloatpagefraction}{0.7}%
}

\newcommand{\gemmaModel}{Gemma~4 31B}
\newcommand{\gptossModel}{GPT-OSS 120B}
\newcommand{\qwenModel}{Qwen3 14B}
\newcommand{\gemmaId}{\texttt{gemma\_4\_31b}}
\newcommand{\gptossId}{\texttt{gpt\_oss\_120b}}
\newcommand{\qwenId}{\texttt{qwen3\_14b}}

\soulregister{\gemmaModel}{0}
\soulregister{\gptossModel}{0}
\soulregister{\qwenModel}{0}
\soulregister{\gemmaId}{0}
\soulregister{\gptossId}{0}
\soulregister{\qwenId}{0}

\def\wtCorpusPooled{660}               %
\def\wtCorpusIncluded{20}              %
\def\wtDbSpringerLink{414}             %
\def\wtDbSemanticScholar{171}          %
\def\wtDbGoogleScholar{41}             %
\def\wtDbArxiv{32}                     %
\def\wtDbIEEEXplore{22}                %
\def\wtDbACMDL{11}                     %

\def\wtIrrCVEs{13}                   %
\def\wtIrrCells{52}                  %
\def\wtIrrAnnotators{2}              %

\def\wtKappaCause{0.876}             %
\def\wtKappaStyleCause{lkAlmost}     %
\def\wtAgreePctCause{92.3}           %
\def\wtKappaOperation{0.900}         %
\def\wtKappaStyleOperation{lkAlmost}  %
\def\wtAgreePctOperation{92.3}       %
\def\wtKappaConsequence{0.626}       %
\def\wtKappaStyleConsequence{lkSubstantial}  %
\def\wtAgreePctConsequence{69.2}     %
\def\wtKappaAttribute{0.235}         %
\def\wtKappaStyleAttribute{lkFair}   %
\def\wtAgreePctAttribute{46.2}       %

\def\wtKappaPooled{0.732}            %
\def\wtKappaStylePooled{lkSubstantial}  %
\def\wtAgreePctPooled{75.0}          %

\def\wtIrrNINP{9}                    %
\def\wtIrrNMEM{2}                    %
\def\wtIrrNDAT{2}                    %

\def\wtVerifiedN{20}            %
\def\wtPoolSlots{16}            %

\def\wtCaseStudies{4}           %

\def\wtFrameParsed{363{,}394}     %
\def\wtFrameMinWords{20}          %
\def\wtFrameRmEone{17{,}642}      %
\def\wtFrameRemEone{345{,}752}    %
\def\wtFrameRmEtwo{27{,}792}      %
\def\wtFrameRemEtwo{317{,}960}    %
\def\wtFrameRmEthree{0}           %
\def\wtFrameRemEthree{317{,}960}  %
\def\wtFrameRmEfour{21}           %
\def\wtFrameRemEfour{317{,}939}   %

\def\wtStratumNSone{55{,}798}     %
\def\wtStratumAllocSone{175}      %
\def\wtStratumNStwo{40{,}803}     %
\def\wtStratumAllocStwo{128}      %
\def\wtStratumNSthree{17{,}013}   %
\def\wtStratumAllocSthree{54}     %
\def\wtStratumNSfoura{36{,}245}   %
\def\wtStratumAllocSfoura{114}    %
\def\wtStratumNSfourb{168{,}080}  %
\def\wtStratumAllocSfourb{529}    %
\def\wtStrataTotal{317{,}939}     %
\def\wtDrawN{1{,}000}             %
\def\wtDrawSeed{20260706}         %

\def\wtWindowCVEs{85}
\def\wtWindowRounds{3}
\def\wtArmCount{2}
\def\wtRoundGridFracs{{0}/0.000000, {1/3}/0.333333, {2/3}/0.666667, {1}/1.000000}
\def\wtRoundGridTeX{$\{0,\tfrac{1}{3},\tfrac{2}{3},1\}$}

\def\wtDmeanGemmaOwnCap{0.890}
\def\wtSmeanGemmaOwnCap{0.968}

\def\wtDmeanGptOssOwnCap{0.094}
\def\wtSmeanGptOssOwnCap{0.577}

\def\wtOwnNGemmaOwnCap{352}
\def\wtOwnAxesGemmaOwnCap{1/0.979, 2/0.994, 3/0.909, 4/0.934, 5/0.951, 6/0.979, 7/0.955, 8/0.989}
\def\wtOwnVerNGemmaOwnCap{19}

\def\wtOwnVerShareGemmaOwnCap{0.789}
\def\wtOwnVerLoGemmaOwnCap{0.567}
\def\wtOwnVerHiGemmaOwnCap{0.915}
\def\wtOwnHighNGemmaOwnCap{145}
\def\wtOwnHighDGemmaOwnCap{0.938}

\def\wtOwnHighFullyGemmaOwnCap{132}
\def\wtOwnHighShareGemmaOwnCap{0.910}
\def\wtOwnHighLoGemmaOwnCap{0.853}
\def\wtOwnHighHiGemmaOwnCap{0.947}
\def\wtOwnMedNGemmaOwnCap{185}
\def\wtOwnMedDGemmaOwnCap{0.793}

\def\wtOwnMedFullyGemmaOwnCap{132}
\def\wtOwnMedShareGemmaOwnCap{0.714}
\def\wtOwnMedLoGemmaOwnCap{0.645}
\def\wtOwnMedHiGemmaOwnCap{0.774}
\def\wtOwnDrawnNGemmaOwnCap{330}

\def\wtOwnDrawnShareGemmaOwnCap{0.800}
\def\wtOwnDrawnLoGemmaOwnCap{0.754}
\def\wtOwnDrawnHiGemmaOwnCap{0.840}
\def\wtOwnWholeShareGemmaOwnCap{0.801}
\def\wtOwnWholeKGemmaOwnCap{282}
\def\wtOwnSpineShareGemmaOwnCap{0.886}
\def\wtOwnSpineKGemmaOwnCap{312}
\def\wtGoldPtsGemmaOwnCap{1.000/0.000, 1.000/0.000, 1.000/0.000, 1.000/0.000, 0.333/0.000, 1.000/0.000, 1.000/0.000, 1.000/0.000, 1.000/1.000, 0.333/0.000, 0.000/0.000, 1.000/0.000, 1.000/0.000, 1.000/0.000}
\def\wtGoldNGemmaOwnCap{14}
\def\wtOwnNGptOssOwnCap{94}
\def\wtOwnAxesGptOssOwnCap{1/0.514, 2/0.652, 3/0.351, 4/0.376, 5/0.518, 6/0.734, 7/0.798, 8/0.599}
\def\wtOwnVerNGptOssOwnCap{15}

\def\wtOwnVerShareGptOssOwnCap{0.067}
\def\wtOwnVerLoGptOssOwnCap{0.012}
\def\wtOwnVerHiGptOssOwnCap{0.298}
\def\wtOwnHighNGptOssOwnCap{28}

\def\wtOwnHighFullyGptOssOwnCap{3}
\def\wtOwnHighShareGptOssOwnCap{0.107}
\def\wtOwnHighLoGptOssOwnCap{0.037}
\def\wtOwnHighHiGptOssOwnCap{0.272}
\def\wtOwnMedNGptOssOwnCap{7}

\def\wtOwnMedFullyGptOssOwnCap{0}
\def\wtOwnMedShareGptOssOwnCap{0.000}
\def\wtOwnMedLoGptOssOwnCap{0.000}
\def\wtOwnMedHiGptOssOwnCap{0.354}
\def\wtOwnDrawnNGptOssOwnCap{35}

\def\wtOwnDrawnShareGptOssOwnCap{0.086}
\def\wtOwnDrawnLoGptOssOwnCap{0.030}
\def\wtOwnDrawnHiGptOssOwnCap{0.224}
\def\wtOwnWholeShareGptOssOwnCap{0.043}
\def\wtOwnWholeKGptOssOwnCap{4}
\def\wtOwnSpineShareGptOssOwnCap{0.202}
\def\wtOwnSpineKGptOssOwnCap{19}
\def\wtGoldPtsGptOssOwnCap{0.000/0.000, 0.000/0.000, 0.000/0.000, 0.000/0.000, 0.000/0.000, 0.000/0.000, 0.333/0.000, 0.000/0.000, 0.000/0.667, 0.000/0.000, 0.000/0.000, 0.000/0.000}
\def\wtGoldNGptOssOwnCap{12}
\def\wtGoldRowLabels{%
  1/{CVE-2013-4934 (2)}/{MAD$\to$MMN},
  2/{CVE-2015-5221 (2)}/{MMN$\to$MUS},
  3/{CVE-2017-17833 (2)}/{MAD$\to$MMN},
  4/{CVE-2006-2362 (3)}/{DVR$\to$MAD$\to$MUS},
  5/{CVE-2007-1320 (3)}/{DVR$\to$MAD$\to$MUS},
  6/{CVE-2008-4539 (3)}/{DVR$\to$MAD$\to$MUS},
  7/{CVE-2014-0160 (3)}/{DVR$\to$MAD$\to$MUS},
  8/{CVE-2018-14557 (3)}/{DCL$\to$MAD$\to$MUS},
  9/{CVE-2019-14814 (3)}/{DVR$\to$MAD$\to$MUS},
  10/{CVE-2007-6429 (4)}/{DVR$\to$TCM$\to$MMN$\to$MUS},
  11/{CVE-2013-4930 (4)}/{DVR$\to$TCM$\to$TCV$\to$MMN},
  12/{CVE-2015-0235 (4)}/{TCM$\to$MMN$\to$MAD$\to$MUS},
  13/{CVE-2021-21834 (5)}/{DVR$\to$TCM$\to$MMN$\to$MAD$\to$MUS},
  14/{CVE-2022-34835 (7)}/{DCL$\to$TCV$\to$TCM$\to$TCV$\to$TCV$\to$MAD$\to$MUS}}
\def\wtGoldRowCount{14}
\def\wtGoldSeps{3,9,12,13}
\def\wtGoldMarksGemmaOwnCap{1/0.000, 2/1.000, 3/0.000, 4/0.000, 5/0.000, 6/0.000, 7/0.000, 8/0.000, 9/0.000, 10/0.000, 11/0.000, 12/0.000, 13/0.000, 14/0.000}
\def\wtGoldMarksGptOssOwnCap{1/0.000, 2/0.667, 4/0.000, 5/0.000, 6/0.000, 7/0.000, 8/0.000, 9/0.000, 10/0.000, 11/0.000, 12/0.000, 14/0.000}
\def\wtSpineRowsGemmaOwnCap{1/{DVL,DVL}/81/cBspRampDark/white, 2/{DVL}/74/cBspRampDark/white, 3/{MAD,DVL}/66/cBspRampDark/white, 4/{DVR,DVL}/23/cBspRampDark/white}
\def\wtSpineOtherGemmaOwnCap{28/108}
\def\wtSpinePanelNGemmaOwnCap{352}
\def\wtSpineRowsGptOssOwnCap{1/{DVL}/16/cBspRampDark/white, 2/{DVL,MUS}/14/cBspRampDark/white, 3/{NRS,DVL}/8/cBspRampMid/cBspText, 4/{TCM,DVL}/6/cBspRampMid/cBspText}
\def\wtSpineOtherGptOssOwnCap{32/50}
\def\wtSpinePanelNGptOssOwnCap{94}

\providecommand{\wtKappaStyleCause}{lkSlight}
\providecommand{\wtKappaStyleOperation}{lkSlight}
\providecommand{\wtKappaStyleConsequence}{lkSlight}
\providecommand{\wtKappaStyleAttribute}{lkSlight}
\providecommand{\wtKappaStylePooled}{lkSlight}
\providecommand{\wtVerifiedN}{\wtMISSING}
\providecommand{\wtCorpusPooled}{\wtMISSING}
\providecommand{\wtCorpusIncluded}{\wtMISSING}
\providecommand{\wtIrrCVEs}{\wtMISSING}
\providecommand{\wtIrrCells}{\wtMISSING}
\providecommand{\wtKappaPooled}{\wtMISSING}
\soulregister{\wtVerifiedN}{0}
\soulregister{\wtCorpusPooled}{0}
\soulregister{\wtCorpusIncluded}{0}
\soulregister{\wtIrrCVEs}{0}
\soulregister{\wtIrrCells}{0}
\soulregister{\wtKappaCause}{0}
\soulregister{\wtKappaOperation}{0}
\soulregister{\wtKappaConsequence}{0}
\soulregister{\wtKappaAttribute}{0}
\soulregister{\wtKappaPooled}{0}
\soulregister{\wtAgreePctPooled}{0}
\soulregister{\wtPoolSlots}{0}
\soulregister{\wtCaseStudies}{0}

\title[]{Evaluating the {NIST} Bugs Framework Against CWE as a Successor for Automated Vulnerability Classification}

\author{Md Nazmul Hoque}
\email{mhoque4@ua.edu}
\affiliation{%
  \institution{The University of Alabama}
  \city{Tuscaloosa}
  \state{Alabama}
  \country{USA}}
\author{Shaswata Mitra}
\email{smitra3@ua.edu}
\affiliation{%
  \institution{The University of Alabama}
  \city{Tuscaloosa}
  \state{Alabama}
  \country{USA}
}
\author{Subash Neupane}
\email{sneupane4@ua.edu}
\affiliation{%
  \institution{The University of Alabama}
  \city{Tuscaloosa}
  \state{Alabama}
  \country{USA}
}
\author{Sudip Mittal}
\email{sudip.mittal@ua.edu}
\affiliation{%
  \institution{The University of Alabama}
  \city{Tuscaloosa}
  \state{Alabama}
  \country{USA}
}
\author{Shahram Rahimi}
\email{shahram.rahimi@ua.edu}
\affiliation{
  \institution{The University of Alabama}
  \city{Tuscaloosa}
  \state{Alabama}
  \country{USA}
}

\renewcommand{\shortauthors}{Hoque et al.}

\begin{document}

\begin{abstract}

Vulnerability classification based on root cause weaknesses is essential for numerous cybersecurity activities, where the Common Weakness Enumeration (CWE) serves as a public repository of such flaws. However, its overlapping entries create a non-orthogonal structure. The result is the same vulnerability being mapped to multiple weaknesses, complicating Root Cause Analysis (RCA) and triage. To address this, NIST Special Publication 800-231 introduces the Bugs Framework (BF), which organizes vulnerabilities into $\langle\text{cause},\,\text{operation},\,\text{consequence}\rangle$ triples and links such triples into a causal chain, so that a vulnerability carries its root cause and its sink together instead of a single terminal label. To date, however, BF has been specified but not evaluated regarding its performance against the challenges to automated classification. Additionally, the evidence required for adoption has not been investigated empirically. In this paper, we evaluate BF as a classification target and a complement to CWE using a systematically screened corpus of automated Common Vulnerabilities and Exposures (CVEs) linked to CWE research. We assess the reproducibility of CVE-to-BF classification through two evaluations. The first is qualitative: an anonymized inter-rater study in which \wtIrrAnnotators\ subject-matter experts (SMEs) independently mapped \wtIrrCVEs\ CVEs onto the four BF axes. Annotators showed strong agreement on the cause and operation axes, while the attribute axis indicated fair agreement. We also tested our automated framework\footnote{\label{github}Experiment and Framework Code: \url{github.com/shaswata09/cve2bf}} across two large language model (LLM) deployments under different budgets for reproducibility analysis. Despite limitations, such as evidence availability and the absence of retrievable fix commits for closed-source software, our findings support the claim that BF is a more structured and automation-friendly framework than CWE. Our exploration reveals specific gaps in BF, including under-specified guidance on attributes.

\end{abstract}



\maketitle

\FloatBarrier
\section{Introduction}\label{sec:introduction}
Vulnerability classification is the semantic layer on which much of modern security operations depends. When this layer captures only the surface form of a weakness, and not the causal chain that produces it, remediation can miss the real defect at serious cost. The 2017 Equifax breach shows the operational impact of this distinction. Attackers exploited a known Apache Struts flaw (CVE-2017-5638)~\cite{nvd_cve20175638} that had been publicly disclosed and flagged by US-CERT months earlier, but the vulnerable component was never patched. The National Vulnerability Database labels that flaw CWE-755 (Improper Handling of Exceptional Conditions)~\cite{nvd_nist}, a generic weakness class that names the failure mode, not the root cause an engineer must locate and fix. Intruders operated undetected for 76 days and exposed the personal records of roughly 148 million consumers~\cite{equifax_house2018,gao2018equifax}. A published technical analysis attributes the breach to the gap between the disclosed weakness and the specific vulnerable library inside a legacy application~\cite{luszcz2018struts}. Equifax later settled with United States regulators for at least \$575 million~\cite{ftc2019equifax}. The quality of vulnerability classification is therefore an operational concern, not only a theoretical one. 
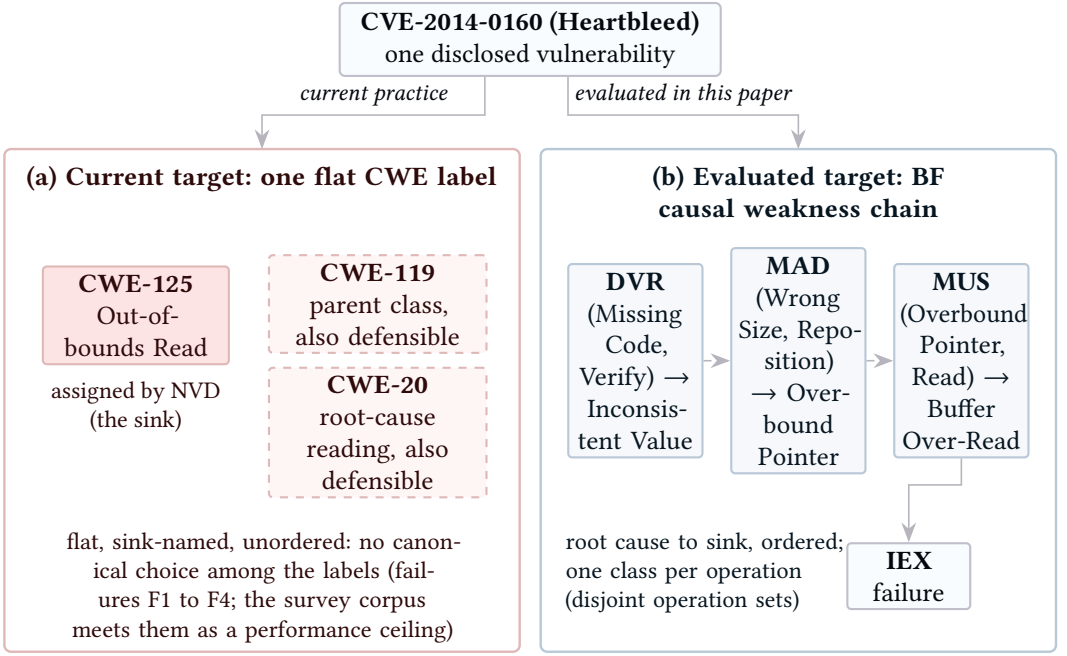
\definecolor{cTldrFailBorder}{HTML}{DFB8B8} %
\definecolor{cTldrFailFillD}{HTML}{FDE4E4} %
\definecolor{cTldrFailFillL}{HTML}{FEF5F5} %
\definecolor{cTldrFailText}{HTML}{2E0A0A} %
\definecolor{cTldrBfBorder}{HTML}{B9C8D7} %
\definecolor{cTldrBfFillL}{HTML}{F6F8FB} %
\definecolor{cTldrBfFillD}{HTML}{E8EFF6} %
\definecolor{cTldrBfText}{HTML}{061520} %
\definecolor{cTldrSlateBorder}{HTML}{BFC0CC} %
\definecolor{cTldrSlateFill}{HTML}{F8FCFE} %
\definecolor{cTldrSlateText}{HTML}{05080F} %
\definecolor{cTldrArrLabelSlateText}{HTML}{05080F} %

\definecolor{cTldrArrLineSlateLine}{HTML}{B7B4BD} %
\begin{inlinefloat}
\begin{tikzpicture}[font=\normalsize,
  cve/.style={draw=cTldrSlateBorder, fill=cTldrSlateFill, text=cTldrSlateText,
              rounded corners=2pt, line width=0.8pt, align=center,
              inner xsep=8pt, inner ysep=3pt},
  panel/.style={draw, rounded corners=2.5pt, line width=0.9pt},
  ptitle/.style={align=center, font=\normalsize\bfseries, anchor=north},
  cwebox/.style={draw=cTldrFailBorder, fill=cTldrFailFillD, text=cTldrFailText,
                 rounded corners=1.5pt, line width=0.8pt, align=center,
                 text width=0.16\liveW, inner ysep=3pt},
  cwealt/.style={cwebox, fill=cTldrFailFillL, dashed, text width=0.19\liveW},
  wbox/.style={draw=cTldrBfBorder, fill=cTldrBfFillL, text=cTldrBfText,
               rounded corners=1.5pt, line width=0.8pt, align=center,
               text width=0.115\liveW, inner xsep=2.5pt, inner ysep=3pt},
  fchip/.style={draw=cTldrSlateBorder, fill=cTldrSlateFill, text=cTldrSlateText,
                rounded corners=1.5pt, line width=0.8pt, align=center,
                text width=0.10\liveW, inner ysep=3pt},
  note/.style={align=center, font=\small},
  arr/.style={-{Stealth[scale=1.3]}, semithick, cTldrArrLineSlateLine},
  alabel/.style={font=\small, text=cTldrArrLabelSlateText, align=center, fill=white, inner sep=1.2pt}]

\node[cve] (cve) at ({0.5\liveW}, 0)
  {\textbf{CVE-2014-0160 (Heartbleed)}\\one disclosed vulnerability};

\draw[panel, draw=cTldrFailBorder] (0,-1.45) rectangle ({0.49\liveW},-8.10);
\node[ptitle, text=cTldrFailText, text width=0.46\liveW] at ({0.245\liveW},-1.55)
  {(a) Current target: one flat CWE label};
\node[cwebox] (cwe125) at ({0.125\liveW},-3.65)
  {\textbf{CWE-125}\\Out-of-bounds Read};
\node[note, text=cTldrFailText, anchor=north] at ({0.125\liveW},-4.40) {assigned by NVD\\(the sink)};
\node[cwealt] (cwe119) at ({0.355\liveW},-3.50)
  {\textbf{CWE-119}\\parent class, also defensible};
\node[cwealt] (cwe20)  at ({0.355\liveW},-5.20)
  {\textbf{CWE-20}\\root-cause reading, also defensible};
\node[note, text=cTldrFailText, text width=0.45\liveW, anchor=north] at ({0.245\liveW},-6.40)
  {flat, sink-named, unordered: no canonical choice among the labels
   (failures F1 to F4; the survey corpus meets them as a performance ceiling)};

\draw[panel, draw=cTldrBfBorder] ({0.51\liveW},-1.45) rectangle ({\liveW},-8.10);
\node[ptitle, text=cTldrBfText, text width=0.46\liveW] at ({0.755\liveW},-1.55)
  {(b) Evaluated target: BF causal weakness chain};
\node[wbox] (w1) at ({0.60\liveW},-4.25)
  {\textbf{DVR}\\(Missing Code, Verify) $\rightarrow$ Inconsistent Value};
\node[wbox] (w2) at ({0.755\liveW},-4.25)
  {\textbf{MAD}\\(Wrong Size, Reposition) $\rightarrow$ Overbound Pointer};
\node[wbox] (w3) at ({0.91\liveW},-4.25)
  {\textbf{MUS}\\(Overbound Pointer, Read) $\rightarrow$ Buffer Over-Read};
\draw[arr, dashed] (w1) -- (w2);
\draw[arr, dashed] (w2) -- (w3);
\node[fchip] (iex) at ({0.86\liveW},-7.10) {\textbf{IEX}\\failure};
\draw[arr] (w3.south) -- ++(0,-0.35) -| (iex.north);
\node[note, text=cTldrBfText, text width=0.27\liveW, align=left, anchor=north west] at ({0.525\liveW},-6.40)
  {root cause to sink, ordered; one class per operation (disjoint operation sets)};

\draw[arr] ([xshift=-0.6cm]cve.south) |- node[pos=0.75, above, alabel, yshift=2pt] {\textit{current practice}} ({0.245\liveW},-1.0) -- ({0.245\liveW},-1.45);
\draw[arr] ([xshift=0.5cm]cve.south) |- node[pos=0.75, above, alabel, yshift=2pt] {\textit{evaluated in this paper}} ({0.755\liveW},-1.0) -- ({0.755\liveW},-1.45);
\node[anchor=north, text=cTldrBfText, inner sep=3pt, font=\small,
      text width=0.9\liveW, align=center] at ({0.5\liveW}, -8.20)
  {\textit{The CWE label admits defensible alternatives. The full BF chain is in Figure~\ref{fig:heartbleed_chain}.}};
\end{tikzpicture}%
\caption{The same CVE mapped to both targets: a flat sink-level CWE label (left) against the ordered BF weakness chain (right).}
\label{fig:tldr_contrast}
\Description{Two side-by-side panels beneath a header box naming CVE-2014-0160 (Heartbleed). The left panel, tinted red, holds three CWE boxes (CWE-125 assigned by NVD as the sink, CWE-119 the parent class, CWE-20 the root-cause reading) with no ordering between them. The right panel, tinted blue, holds four boxes in a left-to-right chain (DVR, MAD, MUS, and the IEX failure) connected by arrows from root cause to sink. Arrows from the header label the left panel as current practice and the right panel as the target evaluated in this paper.}
\end{inlinefloat}

CVE records~\cite{cve_mitre} identify such disclosed vulnerabilities, but downstream systems require additional metadata: severity, affected platforms, exploitability, cause of the weakness, likely consequences, and defensive relevance. The volume of disclosures continues to grow at approximately 263\% between 2020 and 2025~\cite{nist2026nvdupdate}, with more than 40,000 CVEs published in 2024 and over 48,000 in 2025. Each disclosure must be enriched with a severity score (CVSS)~\cite{mell2007cvss,first2019cvss31}, a weakness label (CWE~\cite{cwe_mitre}), and context that maps the disclosure into adversary models such as MITRE ATT\&CK~\cite{mitre_attack} and D3FEND~\cite{mitre_d3fend}. In practice, CWE has become the principal target of weakness for this enrichment because of its breadth and community adoption. 

The same breadth, however, exposes a central problem for trustworthy automation. CWE entries overlap, sit at different abstraction levels, and often describe the final manifestation of a defect instead of the defect's root cause~\cite{bojanova2024bf}, so the same CVE admits several defensible labels and independent annotators or classifiers legitimately diverge. The operational cost of this divergence is directly observable: tool-assisted CVE classification becomes less definitive, cross-report correlation weakens, and remediation priorities are obscured (Figure~\ref{fig:tldr_contrast} previews the pattern, alongside the alternative target this paper evaluates). This issue is addressed in NIST Special Publication 800-231~\cite{bojanova2024bf} by introducing the Bugs Framework (BF), which represents a weakness as a structured relation among a cause, an operation, and a consequence. It organizes weakness classes around disjoint operation sets: the specification requires that no two BF classes share an operation. However, beyond theoretical declarations, applied research on the BF specification remains 

\definecolor{cRqBoxHeaderBg}{HTML}{B7D3E8} %
\definecolor{cRqBoxHeaderFg}{HTML}{061520} %
\definecolor{cRqBoxHeaderRule}{HTML}{B9C8D7} %
\definecolor{cRqBoxSubHeaderBg}{HTML}{E8F1F8} %
\begin{inlinebox}

\colorlet{tblHeaderBg}{cRqBoxHeaderBg}
\colorlet{tblHeaderFg}{cRqBoxHeaderFg}
\colorlet{tblHeaderRule}{cRqBoxHeaderRule}
\colorlet{tblSubHeaderBg}{cRqBoxSubHeaderBg}
\begin{tikzpicture}
\matrix (m) [tblmat,
  column 1/.style={nodes={text width=\dimexpr\liveW-8pt\relax, align=left, font=\normalsize}},
]{
  |[tblhdr, align=center, font=\bfseries\normalsize]| Research Questions (RQ) \\
  \textbf{RQ-1:} What are the structural limitations of CWE that automated CVE-to-CWE classification repeatedly encounters, and how can those limitations be addressed in a successor framework and tested against? \\
  \textbf{RQ-2:} Does the BF specification, by design, address each of the operationally defined CWE limitations, and can the correspondence be traced directly to the specification language and to published BF chains? \\
  \textbf{RQ-3:} To what degree is CVE-to-BF classification reproducible in practice, both across independent SMEs and repeated runs in an automated pipeline under a constant configuration? \\
};
\draw[Rtop] (m.north west) rectangle (m.south east);
\draw[Rmid] (m.west|-m-2-1.north) -- (m.east|-m-2-1.north);
\end{tikzpicture}
\end{inlinebox}

scarce compared against the two decades of CVE-to-CWE work. Therefore, a literature-grounded evaluation of the NIST Bugs Framework is both timely and necessary before the community invests in CVE-to-BF automation. We therefore ask the following three research questions (RQs).

To answer these questions, we conducted the research as a sequence of methodological steps, reported in Section~\ref{sec:methodology}: a systematic literature review with a grounded characterization schema for RQ1, a specification-level correspondence analysis for RQ2, and a two-study empirical evaluation for RQ3. Those steps produced the contribution (C$_x$) of this work, a literature-grounded case study that evaluates NIST SP 800-231 as a classification target, in five parts:

\begin{enumerate}[leftmargin=1.5em,itemsep=2pt,topsep=3pt,label=C\arabic*.,ref=C\arabic*]
  \item \textbf{Corpus and schema.} We assemble and characterize a systematically screened corpus of automated CVE-to-CWE research using the Preferred Reporting Items for Systematic Reviews and Meta-Analyses (PRISMA) procedure across six databases, and report the corpus according to a defensible five-axis characterization schema.

  \item \textbf{Failure analysis.} We read the corpus against four operationally defined CWE failures, namely a non-orthogonal hierarchy, an intractable target space, sink-only labeling, and the absence of a causal chain, and show that the surveyed systems repeatedly accommodate those failures rather than overcome them.

  \item \textbf{Structural correspondence.} We demonstrate how BF addresses each failure by construction and support the claim with verbatim specification language.

  \item \textbf{Empirical evaluation.} We report two complementary studies: an anonymous inter-rater reliability study that assesses reproducibility per BF axis, and an automated determinism study that executes a reproducible, evidence-based CVE-to-BF pipeline and measures chain-level determinism across repeated runs of \wtArmCount\ deployments and budget combinations under baseline configurations.

  \item \textbf{Case studies.} We present case studies, one per BF class type, that tie documented CWE assignments and reassignments to the structural failures they expose and to the BF properties that address them.
\end{enumerate}

In the remainder of the manuscript, Section~\mbox{\ref{sec:methodology}} provides the survey protocol, the characterization schema, and the design of the evaluations; Section~\mbox{\ref{sec:preliminaries_and_key_concepts}} defines the classification properties at issue and the four structural failures, worked on a single CVE; and Section~\mbox{\ref{sec:cti_pipeline}} situates vulnerability classification within the Cyber Threat Intelligence (CTI) pipeline. The research questions are answered in turn: Section~\mbox{\ref{sec:related_work}} synthesizes the corpus (RQ1), Section~\mbox{\ref{sec:bf_addresses}} traces each documented CWE failure to the corresponding BF specification language (RQ2), Section~\mbox{\ref{sec:worked_examples}} presents one case study per BF class type, and Section~\mbox{\ref{sec:evaluation}} reports the inter-rater and determinism studies (RQ3). Section~\mbox{\ref{sec:scope}} states the boundary conditions of every claim, and Section~\mbox{\ref{sec:conclusion}} concludes; the appendices carry the acronym table, the threats to validity, and research protocol.
\FloatBarrier
\section{Methodology}\label{sec:methodology}
This section reports how the research was conducted, one methodological step per research question. To answer RQ1, we collected a corpus of CVE-to-CWE research under a registered systematic procedure, characterized each included paper with a schema derived from a citable taxonomy-development method, and let the recurring limitation patterns emerge from that characterization, not from any prior position. To answer RQ2, we traced each limitation to the verbatim language of NIST SP 800-231 and to published BF chains (Section~\ref{sec:bf_addresses}). The correspondence between the limitations the corpus surfaced and the properties the BF specification formalizes is an empirical finding of this procedure, not its premise. To answer RQ3, we designed two studies: an anonymous inter-rater reliability study and an automated determinism study, both conducted over a curated, evidence-based pipeline, whose apparatus and metrics are specified along with their results (Section~\ref{sec:evaluation}).

The corpus procedure follows the PRISMA reporting structure~\cite{page2021prisma}, and the review protocol follows the guidelines of Kitchenham et al.~\cite{kitchenham2004}. The characterization schema is constructed using the systematic-mapping classification practice of Petersen et al.~\cite{petersen2015guidelines} and the taxonomy-development method of Nickerson et al.~\cite{nickerson2013taxonomy}, so that every axis of the schema is grounded in a named, citable methodology rather than ad hoc judgment.

\subsection{Corpus Collection: A PRISMA Procedure}
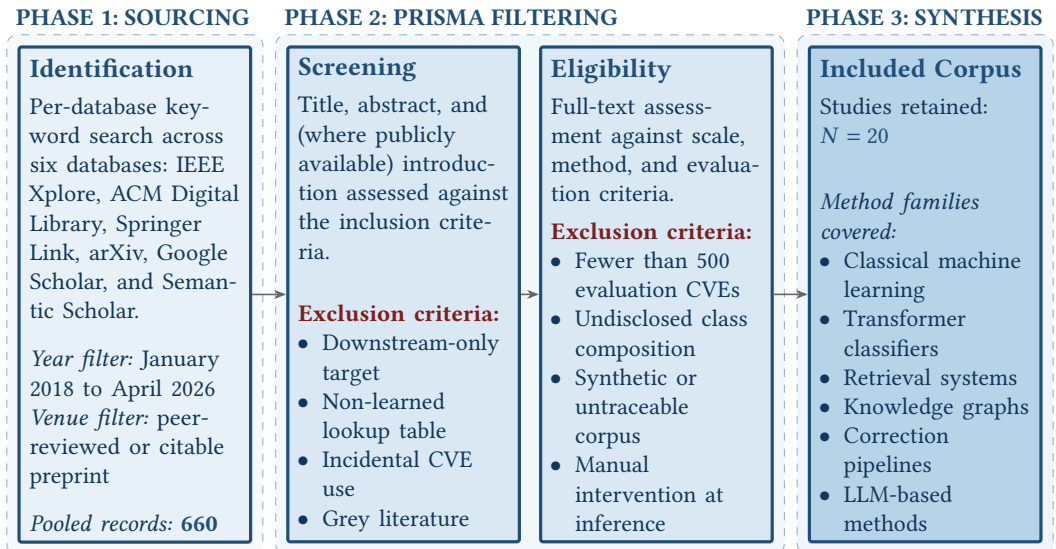
\definecolor{cPrismaCorpusBorder}{HTML}{1F4E79} %
\definecolor{cPrismaCorpusFillD}{HTML}{B7D3E8} %
\definecolor{cPrismaCorpusFillL}{HTML}{E8F1F8} %
\definecolor{cPrismaCorpusText}{HTML}{12365B} %
\definecolor{cPrismaFailText}{HTML}{7F1D1D} %
\begin{inlinefloat}

\pgfmathsetlengthmacro{\bw}{0.20\liveW+8pt}
\ifdefined\prismaH\else\newlength{\prismaH}\newsavebox{\prismaBox}\fi
\newcommand{\prismaRagged}{\raggedright\rightskip0pt plus2em \spaceskip.3333em \xspaceskip.5em\relax}
\newcommand{\prismaBodyId}{%
  \textbf{\textcolor{cPrismaCorpusBorder}{Identification}}\\[3pt]
  \small
  Per-database keyword search across six databases:
  IEEE Xplore, ACM Digital Library, Springer Link,
  arXiv, Google Scholar, and Semantic Scholar.\par
  \vspace{3pt plus 1fill}
  \textit{Year filter:} January 2018 to April 2026\\
  \textit{Venue filter:} peer-reviewed or citable preprint\par
  \vspace{3pt plus 1fill}
  \textit{Pooled records:} \textcolor{cPrismaCorpusBorder}{\textbf{\wtCorpusPooled}}\par}
\newcommand{\prismaBodyScr}{%
  \textbf{\textcolor{cPrismaCorpusBorder}{Screening}}\\[3pt]
  \small
  Title, abstract, and (where publicly available) introduction assessed against
  the inclusion criteria.\par
  \vspace{3pt plus 1fill}
  \textcolor{cPrismaFailText}{\small\textbf{Exclusion criteria:}}
  \begin{itemize}[leftmargin=1em, noitemsep, topsep=0pt]
    \item Downstream-only target
    \item Non-learned lookup table
    \item Incidental CVE use
    \item Grey literature
  \end{itemize}}
\newcommand{\prismaBodyElig}{%
  \textbf{\textcolor{cPrismaCorpusBorder}{Eligibility}}\\[3pt]
  \small
  Full-text assessment against scale, method,
  and evaluation criteria.\par
  \vspace{3pt plus 1fill}
  \textcolor{cPrismaFailText}{\small\textbf{Exclusion criteria:}}
  \begin{itemize}[leftmargin=1em, noitemsep, topsep=0pt]
    \item Fewer than 500 evaluation CVEs
    \item Undisclosed class composition
    \item Synthetic or untraceable corpus
    \item Manual intervention at inference
  \end{itemize}}
\newcommand{\prismaBodyInc}{%
  \textbf{\textcolor{cPrismaCorpusBorder}{Included Corpus}}\\[3pt]
  \small
  Studies retained: \textcolor{cPrismaCorpusBorder}{\textbf{\mbox{$N=\wtCorpusIncluded$}}}\par
  \vspace{3pt plus 1fill}
  \textit{Method families covered:}
  \begin{itemize}[leftmargin=1em, noitemsep, topsep=0pt]
    \item Classical machine learning
    \item Transformer classifiers
    \item Retrieval systems
    \item Knowledge graphs
    \item Correction pipelines
    \item LLM-based methods
  \end{itemize}}
\newcommand{\prismaMeasure}[1]{%
  \sbox{\prismaBox}{\parbox[t]{0.20\liveW}{\normalsize\prismaRagged #1}}%
  \ifdim\dimexpr\ht\prismaBox+\dp\prismaBox\relax>\prismaH
    \setlength{\prismaH}{\dimexpr\ht\prismaBox+\dp\prismaBox\relax}\fi}
\setlength{\prismaH}{0pt}
\prismaMeasure{\prismaBodyId}\prismaMeasure{\prismaBodyScr}%
\prismaMeasure{\prismaBodyElig}\prismaMeasure{\prismaBodyInc}%
\newcommand{\prismaBody}[1]{\parbox[t][\prismaH][s]{\linewidth}{\prismaRagged #1}}
\begin{tikzpicture}[
  font=\normalsize,
  basebox/.style={
    draw=cPrismaCorpusBorder, very thick, rounded corners=1pt,
    align=left, inner sep=4pt, text width=0.20\liveW, text=cPrismaCorpusText,
  },
  idbox/.style={basebox, fill=cPrismaCorpusFillL},
  filterbox/.style={basebox, fill=cPrismaCorpusFillL!70!cPrismaCorpusFillD},
  includebox/.style={basebox, fill=cPrismaCorpusFillD},
  arrow/.style={-{Stealth[scale=0.8]}, semithick, black!60},
]

\node[idbox, anchor=north west] (id) at (0,0) {\prismaBody{\prismaBodyId}};
\node[filterbox, anchor=north west] (scr) at ({\bw+14pt},0) {\prismaBody{\prismaBodyScr}};
\node[filterbox, anchor=north west] (elig) at ({2*\bw+22pt},0) {\prismaBody{\prismaBodyElig}};
\node[includebox, anchor=north west] (inc) at ({3*\bw+36pt},0) {\prismaBody{\prismaBodyInc}};

\draw[arrow] (id) -- (scr);
\draw[arrow] (scr) -- (elig);
\draw[arrow] (elig) -- (inc);

\begin{pgfonlayer}{background}
  \node[draw=cPrismaCorpusBorder!45, fill=cPrismaCorpusFillL!40, rounded corners,
        fit=(id), inner sep=4pt, dashed] (g1) {};
  \node[above right, font=\bfseries\small,
        text=cPrismaCorpusBorder] at (g1.north west) {PHASE 1: SOURCING};

  \node[draw=cPrismaCorpusBorder!45, fill=cPrismaCorpusFillL!55, rounded corners,
        fit=(scr)(elig), inner sep=4pt, dashed] (g2) {};
  \node[above right, font=\bfseries\small,
        text=cPrismaCorpusBorder] at (g2.north west) {PHASE 2: PRISMA FILTERING};

  \node[draw=cPrismaCorpusBorder!45, fill=cPrismaCorpusFillL!70, rounded corners,
        fit=(inc), inner sep=4pt, dashed] (g3) {};
  \node[above right, font=\bfseries\small,
        text=cPrismaCorpusBorder] at (g3.north west) {PHASE 3: SYNTHESIS};
\end{pgfonlayer}

\end{tikzpicture}%
\floatnote{\itshape Sourcing across six databases, filtering through Screening and Eligibility, and synthesis on the retained corpus grouped by method family. Solid arrows denote flow; dashed backgrounds group the phases.}
\caption{PRISMA-style corpus construction for the \wtCorpusIncluded-paper survey.}
\label{fig:prisma}
\Description{Four rounded boxes in a left-to-right flow joined by solid arrows: Identification (keyword search across six databases, pooled record count), Screening (title, abstract and introduction against the inclusion criteria, with a red exclusion list), Eligibility (full-text assessment, with a red exclusion list), and Included Corpus (studies retained, with the method families covered). Dashed background bands group the boxes into three phases labeled sourcing, PRISMA filtering, and synthesis.}
\end{inlinefloat}

The full search protocol and the anchor list are reproduced in Appendix~\ref{app:search}. The search used keyword queries across six databases (IEEE Xplore, the ACM Digital Library, SpringerLink, arXiv, Google Scholar, and Semantic Scholar). The master tracking sheet was created by merging six database exports and deduplicating records based on persistent identifiers and normalized titles. To be included, a work needed to use a CVE artifact, description text, NVD metadata, or scanner output as input and generate a CWE label as a primary output or an independently evaluated stage. Predictions had to be produced using learned, retrieval-based, or generative methods without manual intervention, and had to utilize at least 500 distinct CVE records from a named public registry with disclosed class composition. The publication must have been peer-reviewed or a citable preprint with a reimplementable method. Exclusions included downstream-only targets, non-learned lookup tables, incidental use of CVEs, untraceable or synthetic corpora, undisclosed evaluation details, grey literature, and non-substantive duplication. Two criteria allowed the inclusion of a corpus with fewer than 500 records if it occupied a unique design-space cell and maintained a downstream pipeline, provided the CWE stage was evaluated independently. After screening, \wtCorpusIncluded\ papers (Table~\ref{tab:methodology_overview}) met all inclusion criteria and form the final corpus.

\subsection{Manual Data Extraction}

The authors performed abstract screening and full-text data extraction using a pre-defined form, following Kitchenham's guidelines~\cite{kitchenham2004}. They applied inclusion, exclusion, and safety-valve criteria to each title, abstract, and available introduction, documenting the decisions along with the criteria met or failed. Records labeled as borderline or unclear were set aside for full-text eligibility assessment and discussed among the authors before making a decision. For eligible papers, they recorded properties based on five schema axes and retained relevant excerpts for traceability. To minimize transcription errors, a second author verified a sample of the records against the original papers, resolving any discrepancies through joint review.

\subsection{Characterization Schema}\label{subsec:char_schema}

Each included paper is characterized along five axes that together form
the Characterization Schema used throughout the manuscript: input
artifact, target form, method family, evaluation setting, and stated
limitation. The schema was derived with the method of Nickerson et
al.~\cite{nickerson2013taxonomy}: the meta-characteristic is
``properties of a CVE-to-CWE system that determine how its evidence bears
on the suitability of the classification target,'' and the five axes were
obtained through the empirical-to-conceptual iteration that the method
prescribes.

The schema is designed to support synthesis, not only inventory.
When many systems narrow the target space, the pattern is interpreted as
evidence of target-space intractability. When systems report parent-child
or sibling errors, the pattern is interpreted as evidence of
non-orthogonality (the terms are defined operationally in
Section~\ref{sec:preliminaries_and_key_concepts}). When repair systems identify invalid or generic labels,
the pattern is interpreted as evidence that CWE assignments often fail to
encode root cause. The schema is reported in
the integrated Tables~\ref{tab:methodology_overview} and~\ref{tab:performance_reported}, which give both the
methodological dimensions and the headline performance with its test-set
conditions, and the same schema positions the \wtCorpusIncluded\ papers in the
visual taxonomy of Figure~\ref{fig:schema_overview}. Performance is reported
beside those conditions because a number is interpretable only against its
label-space size, class composition and evaluation metric, and the table is the
evidentiary basis for the synthesis in Section~\ref{sec:related_work} and for
the failure analysis in Section~\ref{sec:bf_addresses}.
\FloatBarrier
\section{Preliminaries}\label{sec:preliminaries_and_key_concepts}

    This section defines each of the terms the comparison relies on operationally. Two terms describe properties a classification scheme may or may not have; four name the structural failures of CWE; the last composes them.
    
    \subsection{Operational Definitions}\label{subsec:definitions}
  \paragraph{Orthogonality of a classification scheme}
        A taxonomy is \emph{orthogonal} when its categories partition the space of classified objects without semantic overlap: for any object $O$ there exists exactly one category $C$ such that $O$ belongs to $C$. That assignment is invariant under reasonable refinements of analysis~\cite{nickerson2013taxonomy, sangupamba2015taxonomy}. BF realizes this property by associating each class with a disjoint operation set, so that for any two classes $c_i, c_j$ with $i\neq j$ one has $\text{ops}(c_i) \cap \text{ops}(c_j) = \emptyset$~\cite{bojanova2024bf}.

\definecolor{cFailNotRoseFillD}{HTML}{FAF3F3} %
\definecolor{cFailNotRoseFillL}{HTML}{FCF8F8} %
\definecolor{cFailNotRoseText}{HTML}{2E0A0A} %
\definecolor{cFailNotRoseBorder}{HTML}{E9D6D6} %

\begin{inlinefloat}
\colorlet{tblHeaderBg}{cFailNotRoseFillD}
\colorlet{tblHeaderFg}{cFailNotRoseText}
\colorlet{tblHeaderRule}{cFailNotRoseBorder}
\colorlet{tblSubHeaderBg}{cFailNotRoseFillL}

\begin{lrbox}{\tbltmpbox}%
\begin{tikzpicture}
\matrix (m) [tblmat,
  column 1/.style={nodes={text width=0.045\liveW, align=left}},
  column 2/.style={nodes={text width=0.17\liveW, align=left}},
  column 3/.style={nodes={text width=\dimexpr0.785\liveW-24pt\relax, align=left}},
]{
  |[tblhdr]| & |[tblhdr]| & |[tblhdr]| \\
  |[tblhdr]| ID & |[tblhdr]| Failure & |[tblhdr]| Definition and documented example \\
  F1 & Non-orthogonal hierarchy &
  Labels at different abstraction levels co-apply to one defect, so no
  single label is canonical. \emph{Example:} near-identical descriptions
  receive the class-level CWE-74 (CVE-2019-5404) and its base-level
  descendant CWE-94 (CVE-2018-0461)~\cite{pan2023bigru}. \\
  F2 & Intractable target space &
  A large, long-tailed label set that supervised classifiers cannot learn
  across, forcing restriction to a subset. \emph{Example:} about 70\% of
  CWE classes have fewer than 100 mapped CVEs and only 10\% have more
  than 500~\cite{das2021v2wbert}. \\
  F3 & Sink-only labeling &
  The assigned label names the final manifestation (sink), not the
  upstream root cause. \emph{Example:} Heartbleed (CVE-2014-0160) is
  recorded as CWE-125, the out-of-bounds read at the sink, while the
  missing length verification goes unrecorded~\cite{bojanova2024bf,owasp_heartbleed}. \\
  F4 & Absence of a causal chain &
  Weaknesses are recorded as an unordered label set with no per-instance
  causal order. \emph{Example:} the CVE-2018-5907 record first carried
  the set \{CWE-20, CWE-119, CWE-190\}, then was remapped to CWE-190
  alone~\cite{nvd_cve20185907,bojanova2024bf}. \\
};

\node[tblhdr, anchor=center, yshift=1pt] at ($(m-1-1.center)!0.52!(m-1-3.center)$) {CWE Structural Failures};

\draw[Rtop, line width=0.02pt] (m.north west) -- (m.north east);
\draw[Rmid, line width=0.02pt] (m.west|-m-2-1.north) -- (m.east|-m-2-1.north); %
\draw[Rmid, line width=0.02pt] (m.west|-m-3-1.north) -- (m.east|-m-3-1.north); %
\draw[Rmid, line width=0.02pt] (m.west|-m-4-1.north) -- (m.east|-m-4-1.north);
\draw[Rmid, line width=0.02pt] (m.west|-m-5-1.north) -- (m.east|-m-5-1.north);
\draw[Rmid, line width=0.02pt] (m.west|-m-6-1.north) -- (m.east|-m-6-1.north);
\draw[Rbot, line width=0.03pt] (m.south west) -- (m.south east);

\begin{scope}[on background layer]
  \fill[tblSubHeaderBg] (m.west|-m-2-1.north) rectangle (m.east|-m-3-1.north);
\end{scope}
\end{tikzpicture}%
\end{lrbox}
\usebox{\tbltmpbox}
\captionof{table}{The four structural CWE failures: notation, definition, and one
documented example of each.}
\label{tab:failure_notation}
\end{inlinefloat}
    \paragraph{Determinism of a mapping procedure}
        A mapping procedure is \emph{deterministic} when repeated independent classification of the same object by qualified experts converges on the same category above an agreed agreement threshold~\cite{cohen1960kappa, landis1977observer}. Determinism is a property of the procedure and its target jointly: an orthogonal target removes the principal source of inter-annotator divergence. Table~\ref{tab:failure_notation} fixes the notation for the four structural CWE failures, F1 through F4, and pairs each with one documented example. The following paragraphs provide the operational definition and test for each failure.
    
    \paragraph{Non-orthogonal hierarchy (CWE failure F1)}
        This failure denotes a hierarchical taxonomy in which labels at different abstraction levels can simultaneously apply to the same defect. The operational test is that the same CVE plausibly maps to two or more CWE entries that are ancestors and descendants, or siblings, in the CWE tree. Pan et al.~\cite{pan2023bigru} document the operational test directly on the two-CVE pair recorded in Table~\ref{tab:failure_notation}.
    
    \paragraph{Intractable target space (CWE failure F2)}
        This failure is structural in origin but automation-facing in effect: it denotes a label space whose cardinality is large and whose label frequency is skewed enough that supervised classifiers cannot reliably learn the long tail, thereby forcing systems to restrict themselves to a subset and trade coverage for accuracy. What makes it specific to CWE is the source of the skew: CWE encodes specificity by enumeration and assigns each weakness its own flat entry, so the catalog grows beyond 900 interdependent entries~\cite{cwe_mitre}. Each rare type receives its own sparsely populated label. In operational terms, the test is that surveyed systems restrict to a fraction of the available labels and that accuracy degrades sharply on out-of-restriction labels. V2W-BERT quantifies the skew as recorded in Table~\ref{tab:failure_notation}~\cite{das2021v2wbert}, and the corpus-wide pattern in Table~\ref{tab:methodology_overview} shows the majority of the \wtCorpusIncluded\ systems declining the full CWE label space and operating instead on CWE-1003, a frequent-class subset, or a custom-curated view.

    \paragraph{Sink-only labeling (CWE failure F3)}
        This failure denotes a labeling convention in which the assigned category names the defect's final manifestation, the sink, rather than the upstream cause, the source. The operational test for F3 follows directly. Two CVEs that share a sink but differ in their upstream causes receive the same sink-level CWE label, yet they would require different mitigations under that label. Within the corpus, FixV2W finds this loss to be widespread: more than half of all CVEs carry invalid or insufficiently detailed CWE mappings~\cite{simsek2025fixing,10.1145/3815425}.

    \paragraph{Absence of a per-instance causal chain (CWE failure F4)}
        This failure denotes a taxonomy that records a defect's weaknesses as an unordered identifier set, with no per-instance mechanism that fixes the causal order in which one weakness creates the conditions for the next. Consider Heartbleed (CVE-2014-0160), where a missing bounds check permits an out-of-bounds read, yet the NVD records the defect under a single buffer-error weakness~\cite{owasp_heartbleed}. The documented CVE-2018-5907 case in Table~\ref{tab:failure_notation}, whose narrated chain runs from improper input validation through integer overflow to a memory-buffer error, shows this two-step loss in the record itself: the record first carried the three weaknesses as an unordered set, with no causal order among them, and was then remapped to the single label CWE-190~\cite{nvd_cve20185907,bojanova2024bf}.
    
\definecolor{cScreenTreeCorpusBorder}{HTML}{1F4E79}
\definecolor{cScreenTreeCorpusFillL}{HTML}{E8F1F8}
\definecolor{cScreenTreeCorpusText}{HTML}{12365B}
\definecolor{cScreenTreeFailBorder}{HTML}{991B1B}
\definecolor{cScreenTreeFailFillL}{HTML}{FEE2E2}
\definecolor{cScreenTreeFailText}{HTML}{7F1D1D}
\definecolor{cScreenTreeGoodBorder}{HTML}{0284C7}
\definecolor{cScreenTreeGoodFillD}{HTML}{B9E2F5}
\definecolor{cScreenTreeGoodText}{HTML}{075985}
\definecolor{cScreenTreeValveBorder}{HTML}{B4D5D5}
\definecolor{cScreenTreeValveColor}{HTML}{F9FEFF}
\definecolor{cScreenTreeValveText}{HTML}{061C1A}
\definecolor{cScreenTreeIconGray}{HTML}{64748B} 

\begin{inlinefloat}
\begin{tikzpicture}[
  font=\normalsize,
  test/.style={draw=cScreenTreeCorpusBorder, line width=0.9pt, rounded corners=4pt, fill=cScreenTreeCorpusFillL,
               align=center, text width=3.6cm, inner sep=4pt, text=cScreenTreeCorpusText},
  keep/.style={draw=cScreenTreeGoodBorder, line width=1.0pt, rounded corners=4pt, fill=cScreenTreeGoodFillD,
               align=center, text width=2.9cm, inner sep=4pt, text=cScreenTreeGoodText, font=\bfseries\normalsize},
  excl/.style={draw=cScreenTreeFailBorder, line width=0.9pt, rounded corners=4pt, fill=cScreenTreeFailFillL,
               align=center, text width=2.8cm, inner sep=4pt, text=cScreenTreeFailText},
  valve/.style={draw=cScreenTreeValveBorder, double, double distance=1pt, line width=1.2pt, dashed, rounded corners=4pt, fill=cScreenTreeValveColor,
                align=center, text width=2.9cm, inner sep=4pt, text=cScreenTreeValveText},
  yes/.style={semithick, -{Stealth[length=2.4mm,scale=1.3]}, line width=0.9pt, draw=cScreenTreeCorpusBorder, rounded corners=3pt},
  no/.style={-{Stealth[length=2.4mm]}, line width=0.5pt, draw=cScreenTreeFailBorder, rounded corners=3pt},
  vv/.style={-{Stealth[length=2.4mm]}, line width=1.2pt, dashed, draw=cScreenTreeValveBorder, rounded corners=3pt},
  ylab/.style={midway, fill=white, inner xsep=2pt, inner ysep=2pt,
               font=\normalsize\bfseries, text=cScreenTreeCorpusText},
  nlab/.style={midway, fill=white, inner xsep=2pt, inner ysep=2pt,
               font=\normalsize \bfseries, text=cScreenTreeFailText}
]

\node[test] (t0) {\textcolor{cScreenTreeIconGray}{\faIcon{question-circle}}~~Pooled, deduplicated candidate records ($N=\wtCorpusPooled$)\\[1pt]
  Does the work take a \textbf{CVE artifact} as input and produce a CWE label
  as output or as an evaluated intermediate stage?};
\node[test, below=9mm of t0] (t1)
  {\textcolor{cScreenTreeIconGray}{\faIcon{question-circle}}~~Is the prediction produced by a learned, retrieval-based, or
  generative method \textbf{without manual intervention} at inference?};
\node[test, below=9mm of t1] (t2)
  {\textcolor{cScreenTreeIconGray}{\faIcon{question-circle}}~~Does the evaluation employ at least 500 distinct CVE records
  from a named \textbf{public registry} with disclosed class composition?};
\node[test, below=9mm of t2] (t3)
  {\textcolor{cScreenTreeIconGray}{\faIcon{question-circle}}~~Is the venue \textbf{peer-reviewed}, or is the work a \textbf{citable preprint}
  with a reimplementable method?};
\node[keep, below=9mm of t3] (inc)
  {\faIcon{check-circle}~~Included corpus\\ $N=\wtCorpusIncluded$ papers};

\node[excl, right=26mm of t0] (e0)
  {\faIcon{times-circle}~~Excluded: downstream-only target, non-learned lookup table, or
  incidental CVE use};
\node[excl, right=26mm of t1] (e1)
  {\faIcon{times-circle}~~Excluded: manual intervention at inference};
\node[excl, right=26mm of t2] (e2)
  {\faIcon{times-circle}~~Excluded: sub-500 corpus, undisclosed composition, or synthetic or
  untraceable corpus};
\node[excl, right=26mm of t3] (e3)
  {\faIcon{times-circle}~~Excluded: grey literature or non-substantive duplication};

\node[valve, right=8mm of e2] (v1)
  {\faIcon{shield-alt}~~Safety valve A: sub-500 corpus retained when it occupies a design-space
  cell no other work occupies};
\node[valve, right=8mm of e0] (v2)
  {\faIcon{shield-alt}~~Safety valve B: downstream pipeline retained when its CWE stage is
  evaluated in its own right};

\draw[yes] (t0) -- node[ylab] {Yes} (t1);
\draw[yes] (t1) -- node[ylab] {Yes} (t2);
\draw[yes] (t2) -- node[ylab] {Yes} (t3);
\draw[yes] (t3) -- node[ylab] {Yes} (inc);

\draw[no] (t0) -- node[nlab] {No} (e0);
\draw[no] (t1) -- node[nlab] {No} (e1);
\draw[no] (t2) -- node[nlab] {No} (e2);
\draw[no] (t3) -- node[nlab] {No} (e3);

\draw[vv] (e2.east) -- (v1.west);
\coordinate (aY) at ($(t2.south)!0.14!(t3.north)$);
\coordinate (aL) at ($(t3.west)+(-4mm,0)$);
\draw[vv] (v1.south) -- (v1.south|-aY) -- (aL|-aY) -- (aL) -- (t3.west);

\draw[vv] (e0.east) -- (v2.west);
\coordinate (bY) at ($(t0.south)!0.14!(t1.north)$);
\coordinate (bL) at ($(t1.west)+(-4mm,0)$);
\draw[vv] (v2.south) -- (v2.south|-bY) -- (bL|-bY) -- (bL) -- (t1.west);

\end{tikzpicture}%
\floatnote{\itshape Each internal node is one inclusion test; a Yes edge descends toward inclusion and a No edge exits to an exclusion category. The two dashed safety valves are the documented overrides (criteria of Section~\ref{sec:methodology}).}
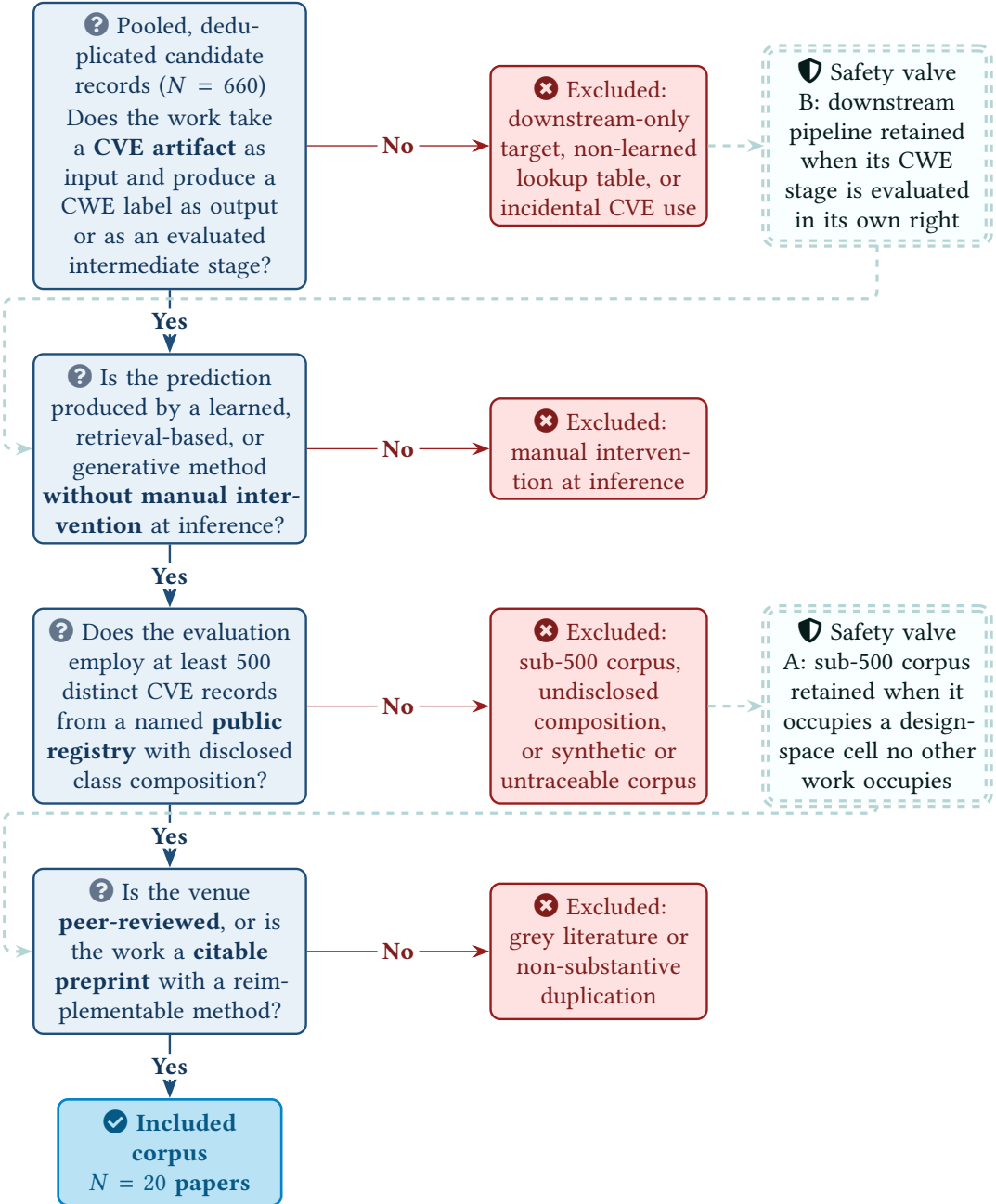
\caption{Decision-tree rendering of the screening and eligibility procedure.}
\label{fig:screening_tree}
\Description{A vertical column of four blue decision nodes, each a yes-or-no question: CVE artifact in and CWE label out; learned or generative prediction without manual intervention; at least 500 CVE records from a public registry; peer-reviewed or citable preprint. A Yes edge descends to the next node and a red No edge branches right to a red exclusion box. Two dashed grey boxes at the far right mark the documented safety-valve overrides on the first and third nodes, and the column ends in a green Included Corpus box.}
\end{inlinefloat}

    \paragraph{Structural inconsistency of the CVE-CWE classification system}
        We define the \emph{structural inconsistency} of the CVE-CWE classification system, as enriched and published in the NVD, as the simultaneous presence of failures F1 through F4 in the target taxonomy and the mapping procedure. However, we restrict the claim to the operational classification: CWE supplies mechanisms for both root-cause mapping and named weakness chains (CWE-709). However, a per-CVE NVD record encodes neither, so the four failures persist at the point of use.

    \subsection{Demonstration of Multiple Defensible CWE Mappings}\label{subsec:cwe_ambiguity_demo}
        CVE-2021-3156, the Sudo heap-based buffer overflow analyzed in Section~\ref{ex:sudo}, admits several defensible CWE mappings at once: original record characterizes the defect as an off-by-one error (CWE-193)~\cite{nvd_cve20213156,cwe193}, current primary assignment is the heap-based buffer overflow CWE-122~\cite{nvd_cve20213156,cwe122}, and class-level CWE-119 together with base-level CWE-787 and CWE-120 plausibly apply to the same defect at other hierarchy levels (see Figure~\ref{fig:tldr_contrast} for end to end ambiguity preview).

    \subsection{Overview of the Bugs Framework (BF)}\label{subsec:bf_in_one_page}
\definecolor{cTaxonomyBfBorder}{HTML}{B9C8D7}
\definecolor{cTaxonomyBfFillD}{HTML}{B7D3E8}
\definecolor{cTaxonomyCorpusBorder}{HTML}{B9C8D7}
\definecolor{cTaxonomyCorpusFillD}{HTML}{B7D3E8}
\definecolor{cTaxonomyCorpusFillL}{HTML}{E8F1F8}
\definecolor{cTaxonomyCorpusText}{HTML}{12365B}
\definecolor{cTaxonomyFailBorder}{HTML}{DFB8B8}
\definecolor{cTaxonomyFailFillD}{HTML}{FCA5A5}
\definecolor{cTaxonomyFailFillL}{HTML}{FEE2E2}
\definecolor{cTaxonomyFailText}{HTML}{7F1D1D}


\begin{inlinefloat}
\begin{tikzpicture}[
  font=\normalsize,
  typeCategoryW/.style={draw=cTaxonomyCorpusBorder, fill=cTaxonomyCorpusFillD,  text=cTaxonomyCorpusText, line width=0.8pt,
    rounded corners=2pt, font=\bfseries\normalsize, align=center,
    text width=0.22\liveW, inner ysep=1.25mm},
  typeCategoryF/.style={typeCategoryW, draw=cTaxonomyFailBorder, fill=cTaxonomyFailFillD, text=cTaxonomyFailText},
  classTypeW/.style={draw=cTaxonomyCorpusBorder, fill=cTaxonomyCorpusFillL,  text=cTaxonomyCorpusText,  line width=0.8pt,
    rounded corners=2pt, font=\bfseries\normalsize, align=center,
    text width=0.19\liveW, inner ysep=1.0mm},
  classTypeF/.style={classTypeW, draw=cTaxonomyFailBorder, fill=cTaxonomyFailFillL, text=cTaxonomyFailText},
  classW/.style={draw=cTaxonomyCorpusBorder, fill=white, text=cTaxonomyCorpusText, line width=0.7pt,
    rounded corners=2pt, font=\small, align=center,
    text width=0.16\liveW, inner xsep=2pt, inner ysep=1.0mm},
  classF/.style={classW, draw=cTaxonomyFailBorder, text=cTaxonomyFailText},
  root/.style={draw=cTaxonomyBfBorder, fill=cTaxonomyBfFillD, line width=1pt,
    rounded corners=2pt, font=\bfseries\normalsize,
    align=center, inner xsep=0.4cm, inner ysep=1.5mm, text=black},
  line/.style={draw=black!55, semithick},
  arr/.style={draw=black!55, semithick, -{Stealth[scale=1.3]}},
  level/.style={font=\itshape\small, text=black!75}
]
\node[root] (BF) at ({0.52\liveW}, 0) {BUGS FRAMEWORK (BF)};

\node[typeCategoryW] (WEAK) at ({0.4025\liveW}, -1.35) {BF Weakness Type};
\node[typeCategoryF] (FAIL) at ({0.881\liveW}, -1.35) {BF Failure Type};
\draw[line] (BF.south) -- ++(0,-.3) coordinate (hubRoot);
\draw[arr] (hubRoot) -| (WEAK.north);
\draw[arr] (hubRoot) -| (FAIL.north);

\node[classTypeW] (INP) at ({0.164\liveW}, -2.85) {BF\_INP\\I/O Check};
\node[classTypeW] (MEM) at ({0.4025\liveW}, -2.85) {BF\_MEM\\Memory};
\node[classTypeW] (DAT) at ({0.642\liveW}, -2.85) {BF\_DAT\\Data Type};
\node[classTypeF] (FLR) at ({0.881\liveW}, -2.85) {BF\_FLR\\Failure};
\draw[line] (WEAK.south) -- ++(0,-0.30) coordinate (hubWeak);
\draw[arr] (hubWeak) -| (INP.north);
\draw[arr] (hubWeak) -- (MEM.north);
\draw[arr] (hubWeak) -| (DAT.north);
\draw[arr] (FAIL.south) -- (FLR.north);

\newcommand{\leafcol}[3]{
  \coordinate (sp) at ([xshift=0.02\liveW]#1.south west);
  \foreach \abbr/\name [count=\i from 1, remember=\abbr as \prev] in {#3}{
    \ifnum\i=1
      \node[#2, anchor=north west] (\abbr) at ([xshift=0.045\liveW, yshift=-0.35cm]#1.south west) {\textbf{\abbr}\\\name};
    \else
      \node[#2, anchor=north west] (\abbr) at ([yshift=-4pt]\prev.south west) {\textbf{\abbr}\\\name};
    \fi
    \draw[arr] (sp|-\abbr.west) -- (\abbr.west);
    \coordinate (spend) at (sp|-\abbr.west);
  }
  \draw[line] (sp) -- (spend);}
\leafcol{INP}{classW}{DVL/Data Validation, DVR/Data Verification}
\leafcol{MEM}{classW}{MAD/Memory Addressing, MMN/Memory Management, MUS/Memory Use}
\leafcol{DAT}{classW}{DCL/Declaration, NRS/Name Resolution, TCV/Type Conversion, TCM/Type Computation}
\leafcol{FLR}{classF}{IEX/Information Exposure, ACE/Arbitrary Code Execution, DOS/Denial of Service, TPR/Data Tampering}

\node[level, rotate=0, anchor=south] at ({0.32\liveW}, -1.1) {Type categories};
\node[level, rotate=0, anchor=south] at ({0.10\liveW}, -2.4) {Class types};
\node[level, rotate=90, anchor=south] at ({0.08\liveW}, -4.0) {Classes};
\end{tikzpicture}
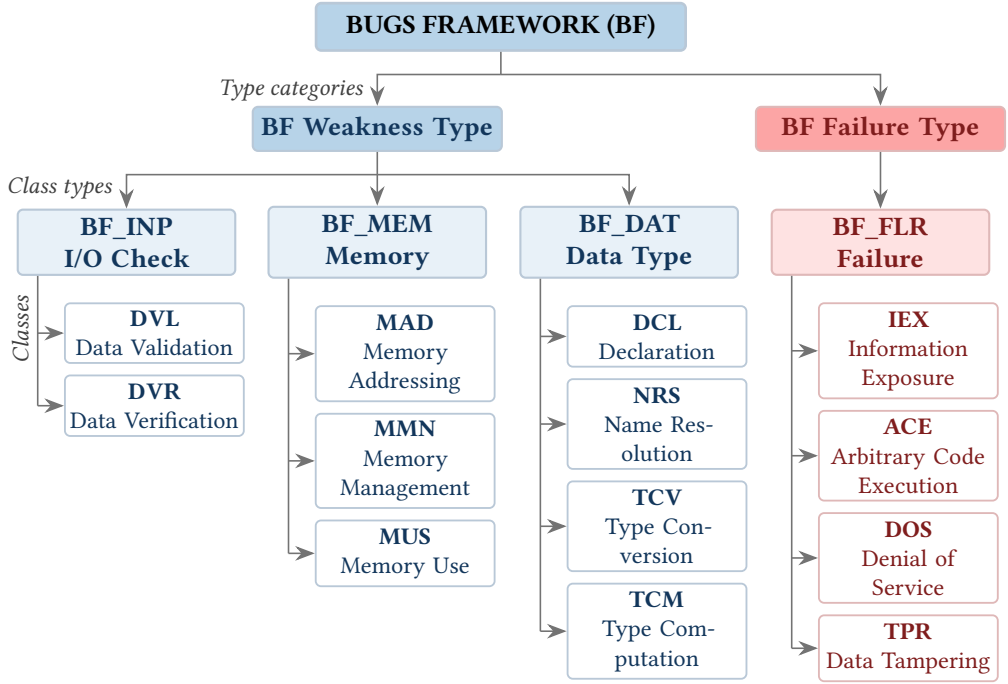
\caption{BF taxonomy structure up to the class level. The figure separates the
BF weakness branch from the BF failure branch, with each class type expanded
into its constituent classes.}
\label{fig:bf_taxonomy}
\Description{A tree with the Bugs Framework at the top splitting into two type-category boxes: BF Weakness Type (blue) on the left and BF Failure Type (red) on the right. Under the weakness type sit three class-type boxes, Input/Output Check, Memory, and Data Type, each expanding downward into its class abbreviations (DVL and DVR; MAD, MMN and MUS; DCL, NRS, TCV and TCM). The failure type expands into IEX, ACE, DOS and TPR. Blue tinting marks weakness classes and red tinting marks failure classes.}
\end{inlinefloat}
        NIST SP 800-231 defines BF as ``a classification of security bugs and related faults that features a formal language for the unambiguous specification of software and hardware security weaknesses and vulnerabilities''~\cite{bojanova2024bf}. The specification states four properties: \emph{structured}, \emph{orthogonal}, \emph{multidimensional}, and \emph{context-free}. The comparison developed in this paper rests on the last two in particular. ``Structured means that a weakness is expressed as a $\langle$cause, operation$\rangle$$\rightarrow$consequence triple with a precise causal relation''~\cite{bojanova2024bf}. ``Orthogonal means that the intersection of the sets of operations of any two BF classes is the empty set''~\cite{bojanova2024bf}. ``Multidimensional means that weaknesses are organized not only by their operations but also by their causes, consequences, and operation and operand attributes''~\cite{bojanova2024bf}. ``Context-free means an operation cannot have different meanings depending on the language or domain''~\cite{bojanova2024bf}. A weakness is an instance of a BF class consisting of one cause, one operation, one consequence, and operation and operand attributes~\cite{bojanova2024bf}. A vulnerability is ``a chain of weaknesses linked by causality via a consequence$\curvearrowright$cause propagation that eventually enables a security failure''~\cite{bojanova2024bf}. The classes used in this manuscript follow the taxonomy of NIST SP 800-231~\cite{bojanova2024bf}: the Input/Output Check (INP) classes Data Validation (DVL) and Data Verification (DVR); the Memory (MEM) classes Memory Addressing (MAD), Memory Management (MMN), and Memory Use (MUS); the Data Type (DAT) classes Type Conversion (TCV) and Type Computation (TCM); and the Failure (FLR) classes Information Exposure (IEX) and Arbitrary Code Execution (ACE), a subset of the full SP 800-231 taxonomy selected for the vulnerability classes this study analyzes. Figure~\mbox{\ref{fig:bf_taxonomy}} renders the full class taxonomy specified in NIST SP 800-231 in detail.

    \subsection{The BF Research Program as a Design Response}\label{subsec:bf_programme}
    
        NIST SP 800-231~\cite{bojanova2024bf} also defines BF as a classification of security bugs and related faults with multi-dimensional weakness and failure taxonomies and a formal language. The antecedent BF paper~\cite{bojanova2016bf} establishes the $\langle$cause, operation$\rangle$$\rightarrow$consequence triple as the unit of weakness specification. Galhardo et al.~\cite{galhardo2020dap} measurements sit alongside the BF program as a metric-design contribution.

\FloatBarrier
\section{CTI Frameworks and Defensive Operations Pipeline}\label{sec:cti_pipeline}
The frameworks introduced in Section~\ref{sec:preliminaries_and_key_concepts} do not operate as isolated artifacts; rather, they compose a directed \emph{CTI}~\cite{mitra2024localintel} pipeline in which the output of one stage becomes the input of the next. Within this pipeline, a disclosed vulnerability is entered as a CVE record and assigned a CWE weakness label and a CVSS severity score through NVD enrichment. It then gains an optional Exploit Prediction Scoring System (EPSS) probability of in-the-wild exploitation~\cite{jacobs2021epss}, maps through CAPEC and ATT\&CK to adversary behavior, and finally links to a defensive countermeasure in MITRE D3FEND~\cite{mitre_d3fend, kaloroumakis2021d3fend}. As depicted in Figure~\ref{fig:cti_pipeline}, this end-to-end flow runs from CVE disclosure on the left to a D3FEND countermeasure on the right. However, each stage is downstream of, and bounded by, the fidelity of the one preceding it; NIST itself documents that CWE and CVE deficiencies propagate into the NVD and can yield imprecise or wrong weakness assignments~\cite{bojanova2024bf}, and the structural failures of the CVE-to-CWE step therefore constrain every artifact that the later stages can recover. This bottleneck is the central concern of the present analysis.
\definecolor{cCtiPipeCorpusBorder}{HTML}{1F4E79} %
\definecolor{cCtiPipeCorpusFillL}{HTML}{E8F1F8} %
\definecolor{cCtiPipeCorpusText}{HTML}{12365B} %
\definecolor{cCtiPipeFailBorder}{HTML}{991B1B} %
\definecolor{cCtiPipeFailFillL}{HTML}{FEE2E2} %
\definecolor{cCtiPipeFailText}{HTML}{7F1D1D} %
\definecolor{cCtiPipeGoodBorder}{HTML}{0284C7} %
\definecolor{cCtiPipeGoodFillL}{HTML}{EAF6FD} %
\definecolor{cCtiPipeGoodText}{HTML}{075985} %
\definecolor{cCtiPipeArrRuleGrayLine}{HTML}{4D4D4D} %
\definecolor{cCtiPipeArrLabelGrayText}{HTML}{171717} %
\definecolor{cCtiPipeArrLabelSlateText}{HTML}{05080F} %
\definecolor{cCtiPipeArrRuleSlateLine}{HTML}{05080F} %

\begin{inlinefloat}
\begin{tikzpicture}[font=\normalsize,
  stage/.style={draw, thick, rounded corners, align=center,
                text width=0.14\liveW, inner xsep=1.5pt, minimum height=0.8cm},
  corpus/.style={stage, draw=cCtiPipeCorpusBorder, fill=cCtiPipeCorpusFillL, text=cCtiPipeCorpusText},
  fail/.style={stage, draw=cCtiPipeFailBorder, fill=cCtiPipeFailFillL, text=cCtiPipeFailText},
  good/.style={stage, draw=cCtiPipeGoodBorder, fill=cCtiPipeGoodFillL, text=cCtiPipeGoodText},
  arrow/.style={-{Stealth[scale=1.0]}, semithick, draw=cCtiPipeArrRuleGrayLine}]

\node[corpus] (cve) at ({0.08\liveW},0) {\textbf{CVE}\\{\small disclosure}};
\node[fail]   (cwe) at ({0.248\liveW},0) {\textbf{CWE}\\{\small weakness label}};
\node[corpus] (cvss) at ({0.416\liveW},0) {\textbf{CVSS / EPSS}\\{\small severity, exploit}};
\node[corpus] (capec) at ({0.584\liveW},0) {\textbf{CAPEC}\\{\small attack pattern}};
\node[corpus] (attack) at ({0.752\liveW},0) {\textbf{ATT\&CK}\\{\small tactic, technique}};
\node[good]   (d3fend) at ({0.92\liveW},0) {\textbf{D3FEND}\\{\small counter\-measure}};

\draw[arrow] (cve) -- (cwe);
\draw[arrow] (cwe) -- (cvss);
\draw[arrow] (cvss) -- (capec);
\draw[arrow] (capec) -- (attack);
\draw[arrow] (attack) -- (d3fend);

\coordinate (invR) at ([yshift=-0.5cm]d3fend.south);
\coordinate (invL) at (cwe.south |- invR);
\draw[arrow, dashed, draw=cCtiPipeGoodBorder]
      (d3fend.south) -- (invR) --
      node[below, font=\small\itshape, text=cCtiPipeArrLabelGrayText]
      {inverse mapping: defense to posture}
      (invL) -- (cwe.south);
\end{tikzpicture}%
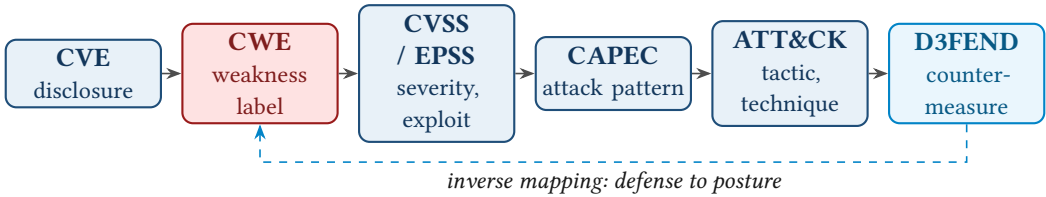
\caption{End-to-end CTI pipeline, from a CVE through to a D3FEND countermeasure.}
\label{fig:cti_pipeline}
\Description{Six boxes in a single horizontal row joined by right-pointing arrows: CVE disclosure, CWE weakness label, CVSS and EPSS severity and exploit, CAPEC attack pattern, ATT\&CK tactic and technique, and D3FEND countermeasure. The CWE box is highlighted in red as the stage this paper studies. A dashed return arrow runs beneath the row from D3FEND back to CWE, labeled as the inverse mapping from defense to pattern.}
    
\end{inlinefloat}

\subsection{Operational Workflow of the CTI Pipeline}
A typical enterprise vulnerability-management workflow applies these stages in sequence. After a CVE is disclosed, NVD analysts, or increasingly an automated classifier of the kind surveyed in Section~\ref{sec:related_work}, assign a CWE label and a CVSS vector. The Stakeholder-Specific Vulnerability Categorization (SSVC) methodology~\cite{spring2019prioritizing} then supplements this severity signal with decision trees over stakeholder decision points to yield a remediation priority~\cite{patel2026agentra, mitra2025falcon}. At the same time, EPSS contributes a complementary probabilistic estimate of exploitation. Bridging weakness to adversary, Common Attack Pattern Enumeration and Classification (CAPEC)~\cite{capec_mitre} maps each weakness to the attacker methodologies that exploit it, and ATT\&CK~\cite{mitre_attack, strom2020attack} supplies the operational tactics, techniques, and procedures~\cite{tamanna2026adversaries} against which detections are authored. However, the severity signal supplied at this stage is a coarse first-order filter rather than a causal account of the defect: CVSS Base scores correlate poorly with observed exploitation~\cite{allodi2014comparing}, which is the gap that EPSS is designed to narrow.

The ATT\&CK situates a vulnerability within a broader family of adversary-modeling. Among these, the \emph{Cyber Kill Chain}~\cite{hutchins2011kill} and the \emph{Diamond Model}~\cite{caltagirone2013diamond} structure incident reasoning. Neither yields a machine-readable weakness label, and ATT\&CK itself sits at a coarser granularity than CWE. The design documentation states that ATT\&CK does not enumerate attack vectors against software and instead defers to CAPEC and CWE, so a single technique, such as Exploit Public-Facing Application (EPFA), subsumes vulnerabilities arising from multiple distinct underlying weaknesses.

\subsection{Bidirectional Mapping between ATT\&CK and D3FEND}

D3FEND is MITRE's defensive analog to ATT\&CK. Where ATT\&CK enumerates the techniques an adversary employs, D3FEND enumerates the digital artifacts and countermeasures that defenders deploy to disrupt those techniques~\cite{mitre_d3fend, kaloroumakis2021d3fend}. The two knowledge graphs are explicitly cross-linked. By construction, this bidirectional mapping can produce a defensive recommendation only as specific as the upstream CWE and CAPEC labels permit. For instance, a vulnerability tagged only with the broad CWE-787 (Out-of-bounds Write) yields correspondingly coarse D3FEND countermeasures, even where the root cause would call for precise control over a single length field. The same bound applies on the failure side, where mitigations aimed at a denial-of-service failure, such as adaptive proof-of-work admission control~\cite{chakraborty2022ai_adaptive_pow}, are selected against the failure class rather than against the weakness chain that produced it. Consequently, the pipeline inherits the ambiguity introduced at the classification stage, a propagation that NIST documents for the CWE-to-NVD step itself~\cite{bojanova2024bf}, and each subsequent mapping can only preserve, never restore, the causal precision lost there. This is the same sink-only and non-orthogonal behavior (see Section~\ref{sec:preliminaries_and_key_concepts}) that the Heartbleed study (see Section~\ref{sec:worked_examples}) specifies. Replacing the CWE label with a BF causal chain removes this bound at its source.

\FloatBarrier
\section{CVE-to-CWE Automation Corpus}\label{sec:related_work}

    During 2020 to 2025, the automated mapping of CVE records to CWE labels has advanced from shallow lexical classifiers toward fine-tuned transformer encoders, retrieval over learned embeddings, knowledge-graph inference, post-prediction label repair, and large language model (LLM) in-context inference. To organize this body of work on common axes, we apply the Characterization Schema of Section~\ref{subsec:char_schema} to the \wtCorpusIncluded-paper corpus summarized in the integrated Tables~\ref{tab:methodology_overview} and~\ref{tab:performance_reported}. Amid this heterogeneity, a single pattern recurs: the reported performance ceiling for CVE-to-CWE classification is set less by model capacity than by the structure of the target taxonomy, a thesis that this section develops by reading the corpus against the four structural failures defined in Section~\ref{subsec:definitions}.

    \subsection{Methods and Reported Performance}\label{subsec:methods_perf}
\definecolor{cPerfCeilBorder}{HTML}{0284C7} %
\definecolor{cPerfCeilFillD}{HTML}{B9E2F5} %
\definecolor{cPerfCeilFillL}{HTML}{EAF6FD} %
\definecolor{cPerfCeilText}{HTML}{075985} %
\begin{inlinefloat}
\begin{tikzpicture}[font=\normalsize,
  axis/.style={draw=cPerfCeilText, line width=0.8pt, -{Stealth[scale=1.3]}},
  curve/.style={draw=cPerfCeilBorder, line width=1.3pt, line cap=round},
  ceil/.style={draw=cPerfCeilBorder, line width=0.9pt, dash pattern=on 3pt off 2pt},
  lead/.style={draw=cPerfCeilBorder, line width=0.5pt},
  ttl/.style={text=cPerfCeilText, align=center},
  note/.style={text=cPerfCeilText, align=center, font=\normalsize}]

\pgfmathsetlengthmacro{\W}{0.82\liveW}\def\H{5.2}\def\Cy{3.85}
\pgfmathsetlengthmacro{\Wh}{0.5*\W}

\fill[cPerfCeilFillL] (0,\Cy) rectangle (\W,\H);
\node[note, text=cPerfCeilBorder, anchor=center] at (\W/2,\Cy + \H/2 - \Cy/2)
  {unreachable: capped by CWE label-space ambiguity};

\draw[axis] (0,0) -- ({\W+10pt},0);
\draw[axis] (0,0) -- (0,\H+0.30);

\draw[ceil] (0,\Cy) -- (\W,\Cy);
\node[fill=white, inner xsep=3pt, inner ysep=1pt, text=cPerfCeilText,
      font=\normalsize\bfseries] at (\Wh,\Cy) {performance ceiling};

\draw[curve] ({0.04*\W},3.66)
  .. controls ({0.18*\W},3.74) and ({0.29*\W},3.58) .. ({0.41*\W},2.90)
  .. controls ({0.53*\W},2.25) and ({0.67*\W},1.65) .. ({0.94*\W},1.10);

\fill[cPerfCeilFillD, draw=cPerfCeilBorder, line width=0.8pt] ({0.14*\W},3.70) circle (2.4pt);
\draw[lead] ({0.14*\W},3.70) -- ({0.19*\W},3.05);
\node[note, anchor=north] at ({0.20*\W},3.00)
  {restrict or repair\\the target labels};

\node[note, anchor=center] at ({0.72*\W},3.05)
  {long tail, sparse classes,\\parent-child and sibling overlap};

\node[ttl, anchor=north] at (\Wh,-0.42)
  {CWE label space covered (long tail admitted $\rightarrow$)};
\node[ttl, rotate=90, anchor=south] at (-0.42,\H/2) {Reported accuracy};
\end{tikzpicture}%
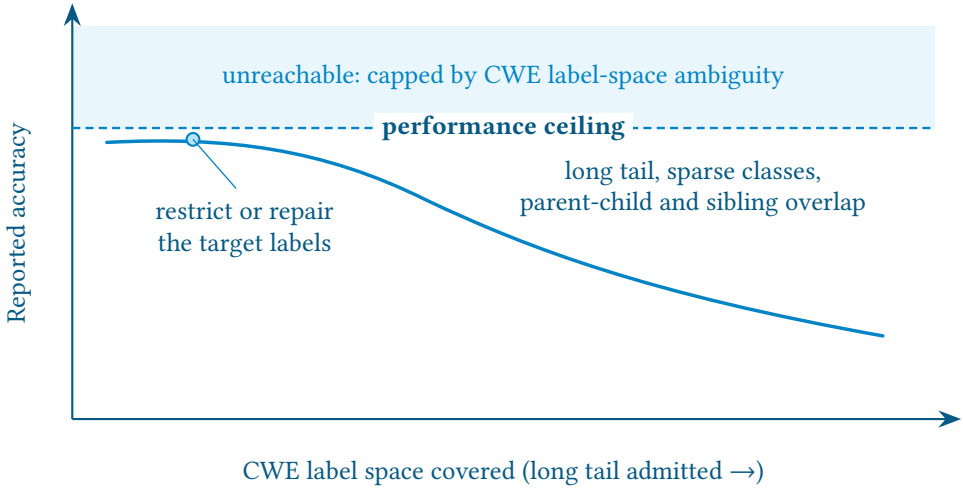
\caption{Conceptual performance ceiling created by CWE label-space ambiguity.}
\label{fig:perf_ceiling}
\Description{A schematic line chart with reported accuracy on the vertical axis and CWE label space covered on the horizontal axis. A dashed horizontal line labeled performance ceiling sits below a shaded band marked unreachable. A single curve starts just under the ceiling at the left, where a dot marks the restrict-or-repair strategies, and falls steadily to the right as the long tail, sparse classes, and hierarchy overlap are admitted. No numeric values are plotted.}
\end{inlinefloat}
        The strongest headline numbers in the corpus are obtained under conditions that narrow the task. The LLM-based and LLM-hybrid entries in the corpus, Text2Weak~\cite{66a872db52274544800b98018fdfa15d}, Key Term Extraction~\cite{10.1007/978-981-95-4434-9_22}, LLM-VulnClass~\cite{exp_llms_vulnclass}, and LLM-Triage~\cite{exp_triage_llm}, show that modern LLMs help most at the term-extraction and label-ranking stages and help least at single-label discrimination across the full hierarchy. Across Table~\ref{tab:performance_reported}, the reported figures share a common structure: accuracy is high where the label space is small or skewed toward common classes, and degrades wherever the long tail is admitted~\cite{10.1145/3815425,li2025out}. That is the empirical signature that motivates the failure analysis below. Figure~\ref{fig:perf_ceiling} renders this signature as a conceptual performance ceiling: reported accuracy rises as the label space narrows and falls as the long tail is admitted, regardless of model family.

\definecolor{cPerfMergedSteelBorder}{HTML}{ADBCD3} %
\definecolor{cPerfMergedSteelFillD}{HTML}{EDF2F5} %
\definecolor{cPerfMergedSteelFillL}{HTML}{EDF2FA} %
\definecolor{cPerfMergedSteelText}{HTML}{0D162D} %
\providecommand{\nr}{\textnormal{\textit{n.r.}}}%
\begin{inlinefloat}
\renewcommand{\thetable}{\arabic{table}a}%
\colorlet{tblHeaderBg}{cPerfMergedSteelFillD}
\colorlet{tblHeaderFg}{cPerfMergedSteelText}
\colorlet{tblHeaderRule}{cPerfMergedSteelBorder}
\colorlet{tblSubHeaderBg}{cPerfMergedSteelFillL}
\begin{lrbox}{\tbltmpbox}%
\begin{tikzpicture}
\matrix (m) [tblmat,
  nodes={inner xsep=3pt, inner ysep=2.4pt},
  column 1/.style={nodes={text width=0.36\liveW, align=left}},
  column 2/.style={nodes={text width=0.09\liveW, align=center}},
  column 3/.style={nodes={text width=0.09\liveW, align=center}},
  column 4/.style={nodes={text width=0.22\liveW, align=left}},
  column 5/.style={nodes={text width=0.10\liveW, align=center}},
]{
  |[tblhdr]| Study & |[tblhdr]| Mod. & |[tblhdr]| Trn. & |[tblhdr]| Dataset & |[tblhdr]| Labels \\
  Text2Weak~\cite{66a872db52274544800b98018fdfa15d}, 2024 & TR & OTH & NVD & 57 \\
  Pure Self-Attention~\cite{9724608}, 2021 & TR & FT & NVD & 11 \\
  Key Term Extraction~\cite{10.1007/978-981-95-4434-9_22}, 2026 & TR & HYB & NVD & 57 \\
  VulnBERTa~\cite{exp_vulnberta}, 2024 & TR & FT & NVD & 160 \\
  V2W-BERT~\cite{das2021v2wbert}, 2021 & TR & FT & NVD\,+\,MITRE & 124 \\
  Temporal LR~\cite{exp_temporal_cwe}, 2024 & ML & SUP & CVE (NVD-derived) & \nr \\
  FixV2W~\cite{simsek2025fixing}, 2025 & KG & OTH & NVD & 130 \\
  CVEDrill~\cite{aghaei2023}, 2023 & SB & FT & NVD & 100 \\
  ThreatZoom~\cite{aghaei2020threatzoom}, 2020 & DL & SUP & Mixed & 364 \\
  VulnScopper~\cite{exp_vulnscopper}, 2024 & HY & FT & Mixed & 393 \\
  VulnBERTa-XAI~\cite{exp_vulnberta_xai}, 2026 & TR & FT & NVD & 160 \\
  Semantic Sim.~\cite{exp_semantic_sim}, 2024 & TR & FT & NVD & 130 \\
  Multi-Taxonomy~\cite{exp_multitax}, 2026 & TR & FT & Mixed & 680 \\
  RoBERTa-125M~\cite{exp_roberta125m}, 2026 & TR & FT & Mixed & 205 \\
  LLM-VulnClass~\cite{exp_llms_vulnclass}, 2025 & HY & OTH & NVD & \nr \\
  VWC-BERT~\cite{exp_vwcbert}, 2022 & TR & FT & NVD & 124 \\
  NVD-Correct~\cite{exp_poster_nvd}, 2024 & OTH & OTH & NVD\,+\,MITRE & 130 \\
  CVE2CWE~\cite{exp_cve2cwe}, 2024 & ML & OTH & NVD\,+\,MITRE & 25 \\
  CWE-CVE-CPE KG~\cite{exp_cwe_cve_cpe_kg}, 2024 & KG & OTH & NVD & 924 \\
  LLM-Triage~\cite{exp_triage_llm}, 2025 & TR & FT & NVD & \nr \\
};
\draw[Rtop] (m.north west) -- (m.north east);
\draw[Rmid] (m.west|-m-2-1.north) -- (m.east|-m-2-1.north);
\draw[Rbot] (m.south west) -- (m.south east);
\begin{scope}[on background layer]
  \foreach \r [evaluate=\r as \rr using int(\r+1)] in {3,5,7,9,11,13,15,17,19}{
    \fill[cPerfMergedSteelFillL] (m.west|-m-\r-1.north) rectangle (m.east|-m-\rr-1.north);
  }
  \fill[cPerfMergedSteelFillL] (m.west|-m-21-1.north) rectangle (m.south east);
\end{scope}
\begin{scope}[lchip/.style={rounded corners=2pt, inner xsep=3.5pt,
    inner ysep=1.5pt, font=\small\bfseries,
    fill=cPerfMergedSteelFillD, text=cPerfMergedSteelText}]
\node[anchor=north west, font=\small, text=cPerfMergedSteelText,
      inner xsep=4pt, inner ysep=3pt, text width=\dimexpr\liveW-8pt\relax, align=left]
  at (m.south west)
  {\renewcommand{\arraystretch}{1.25}%
   \begin{tabular}{@{}l@{\hspace{8pt}}l@{\hspace{14pt}}l@{\hspace{14pt}}l@{}}
   \textbf{Mod.}\ (model family): & \tikz[baseline=(c.base)]\node(c)[lchip]{TR}; Transformer/LLM & \tikz[baseline=(c.base)]\node(c)[lchip]{ML}; Classical ML & \tikz[baseline=(c.base)]\node(c)[lchip]{KG}; Knowledge Graph\\
    & \tikz[baseline=(c.base)]\node(c)[lchip]{DL}; Deep Learning & \tikz[baseline=(c.base)]\node(c)[lchip]{HY}; Hybrid & \tikz[baseline=(c.base)]\node(c)[lchip]{SB}; SecureBERT\\
   \textbf{Trn.}\ (training): & \tikz[baseline=(c.base)]\node(c)[lchip]{FT}; Fine-tuning & \tikz[baseline=(c.base)]\node(c)[lchip]{SUP}; Supervised & \tikz[baseline=(c.base)]\node(c)[lchip]{HYB}; Hybrid\\
    & \tikz[baseline=(c.base)]\node(c)[lchip]{OTH}; Other & & \\
   \end{tabular}};
\end{scope}
\end{tikzpicture}%
\end{lrbox}
\usebox{\tbltmpbox}
\floatnote[c]{\textit{Column codes are expanded in the legend below the table; ``\nr'' marks a value not reported in the source. The evaluation details of the same twenty studies are in Table~\ref{tab:performance_reported}.}}
\captionof{table}{Part (a) of the corpus table: methodological approach of the 20-paper CVE-to-CWE automation corpus, with model family, training regime, dataset, and label-space size.}
\label{tab:methodology_overview}
\end{inlinefloat}

\begingroup\small
\addtocounter{table}{-1}\renewcommand{\thetable}{\arabic{table}b}%
\setlength{\LTleft}{0pt}\setlength{\LTright}{0pt}\setlength{\tabcolsep}{3pt}
\renewcommand{\arraystretch}{1.12}
\begin{longtable}{@{}>{\raggedright\arraybackslash}p{0.21\liveW}>{\raggedright\arraybackslash}p{0.19\liveW}>{\raggedright\arraybackslash}p{0.24\liveW}>{\raggedright\arraybackslash}p{0.26\liveW}@{}}
\arrayrulecolor{cPerfMergedSteelBorder}\specialrule{0.4pt}{0pt}{0pt}
\rowcolor{cPerfMergedSteelFillD}\textbf{\textcolor{cPerfMergedSteelText}{Study}} & \textbf{\textcolor{cPerfMergedSteelText}{Headline result}} & \textbf{\textcolor{cPerfMergedSteelText}{Test-set characteristics}} & \textbf{\textcolor{cPerfMergedSteelText}{BF-versus-CWE relevance}} \\
\specialrule{0.4pt}{0pt}{2pt}
\endfirsthead
\multicolumn{4}{@{}l}{\small\itshape Table~\thetable{} (continued)}\\[2pt]
\arrayrulecolor{cPerfMergedSteelBorder}\specialrule{0.4pt}{0pt}{0pt}
\rowcolor{cPerfMergedSteelFillD}\textbf{\textcolor{cPerfMergedSteelText}{Study}} & \textbf{\textcolor{cPerfMergedSteelText}{Headline result}} & \textbf{\textcolor{cPerfMergedSteelText}{Test-set characteristics}} & \textbf{\textcolor{cPerfMergedSteelText}{BF-versus-CWE relevance}} \\
\specialrule{0.4pt}{0pt}{2pt}
\endhead
\multicolumn{4}{r@{}}{\small\itshape continued on the next page}\\
\endfoot
\specialrule{0.4pt}{2pt}{0pt}
\multicolumn{4}{@{}p{\liveW}@{}}{\small\itshape Headline results are reported beside their test-set conditions because a number is interpretable only against its label-space size, class composition, and metric; ``\nr'' marks a value not reported in the source. The methodological columns of the same studies are in Table~\ref{tab:methodology_overview}.}\\[2pt]
\caption{Part (b) of the corpus table: headline performance and BF-versus-CWE relevance of the same twenty studies.}
\label{tab:performance_reported}
\endlastfoot

Text2Weak~\cite{66a872db52274544800b98018fdfa15d}, 2024 & Macro-$F_1$: 22.33\% & Top@5 acc.\ 66.38\%; 57-class CWE-1003. & Parent-child errors expose hierarchy ambiguity. \\
\rowcolor{cPerfMergedSteelFillL}Pure Self-Attention~\cite{9724608}, 2021 & Accuracy: 90.35\% & Top-10 CWE only; weighted-$F_1$ 89.31\%. & High accuracy depends on drastic label-space narrowing. \\
Key Term Extraction~\cite{10.1007/978-981-95-4434-9_22}, 2026 & Weighted-$F_1$: 70.36\% & Macro-$F_1$ 59.66\%; 57-class; +8.88\% over full description. & Root-cause term extraction aligns with BF axis extraction. \\
\rowcolor{cPerfMergedSteelFillL}VulnBERTa~\cite{exp_vulnberta}, 2024 & Accuracy: 88.5\% & Tiered over 160 classes; MCC 0.721 at Tier 1. & Tiering exposes CWE class-imbalance pressure. \\
V2W-BERT~\cite{das2021v2wbert}, 2021 & Accuracy: up to 97\% & Relaxed prediction; 48-76\% on rare, 61\% zero-shot. & Non-disjoint hierarchy and rare classes are central barriers. \\
\rowcolor{cPerfMergedSteelFillL}Temporal LR~\cite{exp_temporal_cwe}, 2024 & Accuracy: 66\% & $F_1$ 0.64; skewed CWE distribution. & Temporal validation does not remove skew or ambiguity. \\
FixV2W~\cite{simsek2025fixing}, 2025 & Exact-match: 52\% & 93\% same-branch within top-10 (CWE-1003); 69\% top-10 on exploited CVEs. & Mapping repair reveals invalid, generic, and root-cause-poor labels. \\
\rowcolor{cPerfMergedSteelFillL}CVEDrill~\cite{aghaei2023}, 2023 & $F_1$: 84.91-96.08\% & Hierarchical hit-rate 90.77-97.51\%. & Strong engineering remains bound to sink-oriented CWE. \\
ThreatZoom~\cite{aghaei2020threatzoom}, 2020 & Accuracy: 92\% (NVD) & 75\% on MITRE fine-grain; full hierarchy. & Full-hierarchy ambition exposes multi-path weakness difficulty. \\
\rowcolor{cPerfMergedSteelFillL}VulnScopper~\cite{exp_vulnscopper}, 2024 & Hits@10: 71\% (NVD) & CVE-CWE linking; +11.7\% over LLMs. & Context helps; the target lacks causal-chain syntax. \\
VulnBERTa-XAI~\cite{exp_vulnberta_xai}, 2026 & Accuracy: 88-97.7\% & Per-tier; prediction-tree 90.1\%. & Explainability exposes over-generalization and ``Other'' pressure. \\
\rowcolor{cPerfMergedSteelFillL}Semantic Sim.~\cite{exp_semantic_sim}, 2024 & Accuracy & Similarity-based retrieval over CWE-1003 view. & Better similarity does not dissolve overlap in the target. \\
Multi-Taxonomy~\cite{exp_multitax}, 2026 & Accuracy & Cross-taxonomy classification; error propagation observed. & Error propagation worsens as taxonomic breadth expands. \\
\rowcolor{cPerfMergedSteelFillL}RoBERTa-125M~\cite{exp_roberta125m}, 2026 & Top-1 acc.: 87.4\% & Macro-$F_1$ 60.7\%; 205-class. & Top-1 accuracy coexists with weaker macro coverage. \\
LLM-VulnClass~\cite{exp_llms_vulnclass}, 2025 & Accuracy & Full-hierarchy LLM; ambiguity persists. & Generic LLM ability does not eliminate CWE ambiguity. \\
\rowcolor{cPerfMergedSteelFillL}VWC-BERT~\cite{exp_vwcbert}, 2022 & Accuracy & Cascade: CVE$\rightarrow$CWE$\rightarrow$CAPEC; upstream uncertainty propagates. & Upstream CWE uncertainty contaminates downstream mappings. \\
NVD-Correct~\cite{exp_poster_nvd}, 2024 & Top-K acc. & Repair-ranking; targets contestable initial NVD labels. & Label correction becomes necessary because mappings are unstable. \\
\rowcolor{cPerfMergedSteelFillL}CVE2CWE~\cite{exp_cve2cwe}, 2024 & Top-1 acc.: 69.9\% & Top-3 acc.\ 87.5\%; 25-class. & Accuracy degrades as class count rises. \\
CWE-CVE-CPE KG~\cite{exp_cwe_cve_cpe_kg}, 2024 & Accuracy & 924-class repository graph; sparsity limits inference. & Structured relations help but sparsity limits inference. \\
\rowcolor{cPerfMergedSteelFillL}LLM-Triage~\cite{exp_triage_llm}, 2025 & Accuracy & Full-hierarchy LLM; sibling/parent-child confusions persist. & Sibling and parent-child confusions persist in newer LLM workflows. \\
\end{longtable}
\endgroup

    \subsection{Four Adaptive Strategies for Thematic Synthesis}\label{subsec:thematic}
\definecolor{cSchemaBfBorder}{HTML}{1F4E79} %
\definecolor{cSchemaBfFillD}{HTML}{E1EBF4} %
\definecolor{cSchemaBfFillL}{HTML}{E8F1F8} %
\definecolor{cSchemaBfText}{HTML}{12365B} %
\definecolor{cSchemaCorpusBorder}{HTML}{1F4E79} %
\definecolor{cSchemaCorpusFillD}{HTML}{E1EBF4} %
\definecolor{cSchemaCorpusFillL}{HTML}{E8F1F8} %
\definecolor{cSchemaCorpusText}{HTML}{12365B} %
\begin{inlinefloat}
\begin{tikzpicture}[
  font=\normalsize,
  clusterTitle/.style={
    draw=cSchemaCorpusBorder, very thick, rounded corners=1.5pt,
    fill=cSchemaCorpusFillD, font=\bfseries\normalsize, align=center, text=cSchemaCorpusText,
    text width=0.20\liveW, inner sep=3pt, minimum height=1.05cm, anchor=north
  },
  paper/.style={
    draw=gray!60, thin, rounded corners=1pt,
    fill=white, font=\normalsize, align=center,
    text width=0.20\liveW, inner xsep=1.5pt, inner ysep=2.5pt
  },
  bfTitle/.style={
    draw=cSchemaBfBorder, very thick, rounded corners=1.5pt,
    fill=cSchemaBfFillD, font=\bfseries\normalsize, align=center, text=cSchemaBfText,
    text width=0.20\liveW, inner sep=5pt
  },
  arrow/.style={-{Stealth[scale=.9]}, thick, dashed, draw=gray!70}
]
\node[clusterTitle] (T1) at ({0.125\liveW}, 8.0) {\textit{Strategy 1}\\Restrict the label space};
\node[clusterTitle] (T2) at ({0.375\liveW}, 8.0) {\textit{Strategy 2}\\Change the prediction form};
\node[clusterTitle] (T3) at ({0.625\liveW}, 8.0) {\textit{Strategy 3}\\Repair labels};
\node[clusterTitle] (T4) at ({0.875\liveW}, 8.0) {\textit{Strategy 4}\\Push to LLMs};

\newcommand{\papercol}[3]{
  \foreach \name [count=\i from 0, remember=\i as \j] in {#3}{
    \ifnum\i=0 \node[paper, anchor=north] (#2\i) at ([yshift=-4pt]#1.south) {\name};
    \else \node[paper, anchor=north] (#2\i) at ([yshift=-3pt]#2\j.south) {\name};
    \fi}}
\papercol{T1}{s1}{Pure Self-Attention, CVE2CWE, VulnBERTa, VulnBERTa-XAI, RoBERTa-125M, Text2Weak, Temporal LR}
\papercol{T2}{s2}{V2W-BERT, CVEDrill (SecureBERT), ThreatZoom, Semantic Similarity, Multi-Taxonomy, VulnScopper, CWE-CVE-CPE KG}
\papercol{T3}{s3}{FixV2W, NVD-Correct, VWC-BERT (cascade)}
\papercol{T4}{s4}{Key Term Extraction, LLM-VulnClass, LLM-Triage}

\node[draw=cSchemaCorpusBorder, dashed, thick, rounded corners=2pt,
      fit=(T1) (s10) (s16), inner sep=4pt] (C1) {};
\node[draw=cSchemaCorpusBorder, dashed, thick, rounded corners=2pt,
      fit=(T2) (s20) (s26), inner sep=4pt] (C2) {};
\node[draw=cSchemaCorpusBorder, dashed, thick, rounded corners=2pt,
      fit=(T3) (s30) (s32), inner sep=4pt] (C3) {};
\node[draw=cSchemaCorpusBorder, dashed, thick, rounded corners=2pt,
      fit=(T4) (s40) (s42), inner sep=4pt] (C4) {};

\node[fit=(C1) (C2) (C3) (C4), inner sep=0pt, draw=none] (allPanels) {};

\coordinate (BFcenter) at ({0.5\liveW}, 0);
\node[bfTitle, anchor=north, yshift=-16pt] (BF) at (BFcenter |- allPanels.south) {NIST\\Bugs Framework};

\node[font=\normalsize, align=center, text width=0.60\liveW, anchor=north] (BFdesc)
  at ([yshift=-4pt]BF.south)
 {\textit {Orthogonal classes; small structured class set; explicit cause-operation axes; cause-to-consequence chain}};

\draw[arrow] ($(C1.south west)!0.45!(C1.south east)$) |- ($(BF.west)-(0,0.12)$);
\draw[arrow] ($(C2.south west)!0.45!(C2.south east)$) |- ($(BF.west)+(0,0.12)$);
\draw[arrow] ($(C3.south west)!0.55!(C3.south east)$) |- ($(BF.east)+(0,0.12)$);
\draw[arrow] ($(C4.south west)!0.55!(C4.south east)$) |- ($(BF.east)-(0,0.12)$);
\end{tikzpicture}%
\floatnote{\itshape Each cluster groups papers by the strategy with which they cope with CWE's structural failures (F1 to F4); BF responds to the same failures by design rather than by adaptation. Placement reflects the Characterization Schema axes; a paper that also contributes to a secondary strategy appears at its primary cluster only. Strategies of Section~\ref{subsec:thematic}.}
\caption{Schema overview of the 20-paper CVE-to-CWE automation corpus, clustered by the four adaptive strategies.}
\label{fig:schema_overview}
\Description{Four dashed rounded panels in a row, one per adaptive strategy (restrict the label space, change the prediction form, repair labels, push to LLMs), each holding a vertical stack of small boxes naming the corpus systems assigned to that strategy. Dashed lines from all four panels converge on a single box at the bottom labeled NIST Bugs Framework, annotated with its three design properties.}
\end{inlinefloat}

        Reading the corpus against the operational criteria of Section~\ref{subsec:definitions} yields four strategies by which systems accommodate a target taxonomy that resists clean single-label prediction. Each strategy demonstrates one or more of the four CWE failures from a different angle, and the resulting acknowledgment pattern is recorded in Table~\ref{tab:performance_reported}'s ``BF-versus-CWE relevance'' column.

        \paragraph{Strategy 1: Restrict the label space.} A first group of systems addresses the intractability of CWEs by narrowing the target. Pure Self-Attention~\cite{9724608} reduces to the top ten CWE identifiers. CVE2CWE~\cite{exp_cve2cwe} reduces to a 25-class view. VulnBERTa~\cite{exp_vulnberta} and its explainability extension VulnBERTa-XAI~\cite{exp_vulnberta_xai} operate over 160 frequent classes with tiering, RoBERTa-125M~\cite{exp_roberta125m} over 205 classes, and Key-Term Extraction~\cite{10.1007/978-981-95-4434-9_22} and Text2Weak~\cite{66a872db52274544800b98018fdfa15d} over a 57-class view drawn from CWE-1003~\cite{cwe1003}, the simplified-mapping view that the NVD itself applies when labeling CVEs~\cite{nvd_nist}. The Temporal LR baseline~\cite{exp_temporal_cwe} reinforces the intractability reading: even with temporal validation, the skewed label distribution sets a hard performance ceiling.

        \paragraph{Strategy 2: Change the prediction form.} A second group changes the shape of the prediction, not the size of the target. V2W-BERT~\cite{das2021v2wbert} predicts hierarchically; CVEDrill, built on SecureBERT~\cite{aghaei2023}, predicts top-$K$ over the CWE hierarchy with explicit hierarchical hit-rates; ThreatZoom~\cite{aghaei2020threatzoom} predicts across the full CWE hierarchy; Semantic Similarity~\cite{exp_semantic_sim} reframes the problem as embedding retrieval rather than direct classification; the Multi-Taxonomy transformer~\cite{exp_multitax} extends the prediction across taxonomies; and the CWE-CVE-CPE Knowledge Graph~\cite{exp_cwe_cve_cpe_kg} replaces flat classification with relation prediction over a structured graph. The errors of this group cluster around parent-child and sibling CWE relationships; consistent with this, VulnScopper~\cite{exp_vulnscopper} reports that $44.3\%$ of CVEs cataloged in two databases carry a different CWE, a divergence it attributes to the hierarchy permitting either a general or a specific weakness for the same defect.
        
        \paragraph{Strategy 3: Repair labels after prediction.} A third group treats CWE labels as objects requiring repair. FixV2W~\cite{simsek2025fixing} applies knowledge-graph embeddings to correct invalid (Prohibited) or insufficiently-specific (Discouraged) NVD mappings; for Prohibited mappings it recovers the exact CWE within the top-10 in $65\%$ of cases, with the correct label at the first rank in $52\%$, and a same-branch candidate in $93\%$; and NVD-Correct~\cite{exp_poster_nvd} applies a ranking-and-repair pipeline that targets the cases where the initial NVD assignment is contestable. This group is especially important for the BF argument because it demonstrates that the ecosystem already spends effort repairing the target representation after the fact: more than half ($55\%$) of NVD CVEs carry invalid or insufficiently detailed CWE mappings~\cite{simsek2025fixing}, a direct empirical signature of structural inconsistency in the target taxonomy.
        
        \paragraph{Strategy 4: Push to LLMs at the boundary.} A fourth group reaches for LLMs at the boundary where the prior strategies stop helping. Key Term Extraction~\cite{10.1007/978-981-95-4434-9_22} uses an LLM to extract root-cause terms before mapping into CWE, which aligns naturally with BF's axis-by-axis extraction model. LLM-VulnClass~\cite{exp_llms_vulnclass} and LLM-Triage~\cite{exp_triage_llm} apply LLMs more directly to the classification task. The counterintuitive headline of LLM-VulnClass~\cite{exp_llms_vulnclass}, namely that a simple TF-IDF baseline outperforms LLM embeddings on the same task, provides an empirical anchor for how classical baselines compare to LLMs at this label-space scale. VWC-BERT~\cite{exp_vwcbert} reinforces this conclusion from a cascade perspective: upstream CWE uncertainty propagates into the downstream CWE-to-CAPEC mapping, so an inherited target-space problem at the CVE-CWE stage contaminates everything that depends on it.

    \subsection{Historical Progression}\label{subsec:historical}
        Chronologically, the corpus traces a clear progression in which more capable systems do not dissolve the failures observed in the earlier ones. They redirect effort to mitigations (narrower targets, hierarchical ranking, label repair, LLM term extraction) that an orthogonal target with explicit cause and chain syntax would not require. Figure~\ref{fig:schema_overview} renders this progression as a thematic clustering across the five Characterization Schema axes.
        
\FloatBarrier
\section{Comparative Analysis of the Bugs Framework Against CWE's Structural Failures}\label{sec:bf_addresses}

    \subsection{Property-by-Failure Mapping}\label{subsec:bf_addresses_failures}
        This subsection shows, for each of the four operational structural failures of CWE (see Section~\ref{subsec:definitions}), the specific design property of the BF that addresses it. The mapping is grounded in the BF specification, where each property is named as the specification names it, and each is supported by verbatim text from NIST Special Publication 800-231 or from the Bojanova et al.\ class papers that document it. We demonstrate a one-to-one correspondence between these properties and failures F1 through F4, and the bottom panel of Figure~\ref{fig:cwe_failure_panels} records the resulting matrix. Figure~\ref{fig:bf_fix_panels} schematically renders the four resolutions, corresponding to the panel-for-panel failure schematics of Figure~\ref{fig:cwe_failure_panels}.


\definecolor{cFailPanBorder}{HTML}{CB8B8B}
\definecolor{cFailPanFillD}{HTML}{FDD2D2}
\definecolor{cFailPanFillL}{HTML}{FEEFEF}
\definecolor{cFailPanText}{HTML}{450F0F}
\definecolor{cFailPanGrayBorder}{HTML}{BCBCBC}
\definecolor{cFailPanGrayFill}{HTML}{E8E8E8}
\definecolor{cFailPanGrayText}{HTML}{232323}
\definecolor{cCorrSteelBorder}{HTML}{ADBCD3} %
\definecolor{cCorrSteelFillD}{HTML}{EDF2F5} %
\definecolor{cCorrSteelFillL}{HTML}{F3F6F6} %
\definecolor{cCorrSteelText}{HTML}{0D162D} %

\begin{inlinefloat}
\pgfmathsetlengthmacro{\PW}{0.49\liveW}\pgfmathsetlengthmacro{\PX}{0.51\liveW}
\begin{tikzpicture}[font=\normalsize,
  frame/.style={draw=cFailPanBorder, rounded corners=2.5pt, line width=0.9pt},
  badge/.style={draw=cFailPanBorder, fill=cFailPanFillD, text=cFailPanText,
                rounded corners=1.2pt, line width=0.6pt, font=\normalsize\bfseries,
                inner xsep=3.5pt, inner ysep=1.8pt},
  plab/.style={text=cFailPanText, font=\normalsize\bfseries, align=center},
  mini/.style={draw=cFailPanBorder, fill=cFailPanFillL, text=cFailPanText,
               rounded corners=1.2pt, line width=0.7pt, align=center,
               font=\normalsize, inner xsep=3pt, inner ysep=2pt, minimum height=0.55cm},
  minid/.style={mini, fill=cFailPanFillD},
  minig/.style={draw=cFailPanGrayBorder, fill=cFailPanGrayFill, text=cFailPanGrayText,
                rounded corners=1.2pt, line width=0.7pt, align=center, font=\normalsize,
                inner xsep=3pt, inner ysep=2pt, minimum height=0.55cm},
  note/.style={text=cFailPanGrayText, font=\small, align=center},
  arr/.style={-{Stealth[scale=1.2]}, semithick, draw=cFailPanBorder},
  tedge/.style={draw=cFailPanGrayBorder, semithick}]
\draw[frame] (0,0) rectangle (\PW,-3.9);
\node[badge, anchor=north west] at (0.12,-0.12) {F1};
\node[minig] (f1cve) at ({0.22*\PW},-1.65) {CVE};
\node[mini] (f1p) at ({0.70*\PW},-0.75) {CWE-119 (class)};
\node[mini] (f1c) at ({0.70*\PW},-2.30) {CWE-122 (base)};
\draw[tedge] (f1p.south) -- (f1c.north);
\node[note, rotate=90] at ({0.78*\PW},-1.55) {parent};
\draw[arr] (f1cve.east) -- (f1p.west);
\draw[arr] (f1cve.east) -- (f1c.west);
\node[note] at ({0.5*\PW},-2.95) {both labels defensible};
\node[plab] at ({0.5*\PW},-3.50) {labels co-apply across levels};
\draw[frame] (\PX,0) rectangle (\liveW,-3.9);
\node[badge, anchor=north west] at ({\PX+0.12cm},-0.12) {F2};
\draw[tedge] ({\PX+0.14*\PW},-2.75) -- ({\PX+0.90*\PW},-2.75);
\foreach \k/\h/\c in {0/1.85/cFailPanFillD,1/1.18/cFailPanFillD,2/0.78/cFailPanFillL,3/0.50/cFailPanFillL,4/0.34/cFailPanFillL,5/0.25/cFailPanFillL,6/0.18/cFailPanFillL,7/0.13/cFailPanFillL,8/0.09/cFailPanFillL}
  \fill[\c] ({\PX+(0.17+0.08*\k)*\PW},-2.75) rectangle ++({0.05*\PW},\h);
\node[note] at ({\PX+0.68*\PW},-1.40) {$>$900 entries,\\long tail};
\node[note] at ({\PX+0.52*\PW},-3.02) {CWE labels by mapped-CVE count};
\node[plab] at ({\PX+0.5*\PW},-3.50) {long tail defeats supervision};
\draw[frame] (0,-4.3) rectangle (\PW,-8.2);
\node[badge, anchor=north west] at (0.12,-4.42) {F3};
\node[minig, dashed] (f3a) at ({0.18*\PW},-5.85) {root\\cause};
\node[minig] (f3b) at ({0.5*\PW},-5.85) {$\cdots$};
\node[minid] (f3c) at ({0.82*\PW},-5.85) {sink};
\draw[arr, draw=cFailPanGrayBorder] (f3a.east) -- (f3b.west);
\draw[arr, draw=cFailPanGrayBorder] (f3b.east) -- (f3c.west);
\node[note] at ({0.18*\PW},-6.80) {not recorded};
\node[note, text=cFailPanText] at ({0.82*\PW},-6.80) {= the CWE label};
\node[plab] at ({0.5*\PW},-7.80) {label names the sink, not the cause};
\draw[frame] (\PX,-4.3) rectangle (\liveW,-8.2);
\node[badge, anchor=north west] at ({\PX+0.12cm},-4.42) {F4};
\node[mini] (f4b) at ({\PX+0.5*\PW},-5.85) {CWE-190};
\node[mini, left=3mm of f4b] (f4a) {CWE-20};
\node[mini, right=3mm of f4b] (f4c) {CWE-119};
\node[note, font=\large, text=cFailPanText, anchor=east] at ([xshift=1mm]f4a.west) {$\{$};
\node[note, font=\large, text=cFailPanText, anchor=west] at ([xshift=-1mm]f4c.east) {$\}$};
\node[note, text width={0.9*\PW}] at ({\PX+0.5*\PW},-6.85)
  {an unordered set; which weakness enables which is not recorded};
\node[plab] at ({\PX+0.5*\PW},-7.80) {causal order is lost};
\end{tikzpicture}

\vspace{4pt}
\colorlet{tblHeaderBg}{cCorrSteelFillD}
\colorlet{tblHeaderFg}{cCorrSteelText}
\colorlet{tblHeaderRule}{cCorrSteelBorder}
\colorlet{tblSubHeaderBg}{cCorrSteelFillL}
\begin{lrbox}{\tbltmpbox}%
\begin{tikzpicture}
\matrix (m) [tblmat,
  nodes={inner xsep=3pt},
  column 1/.style={nodes={text width=0.27\liveW, align=left}},
  column 2/.style={nodes={text width=0.15\liveW, align=center, anchor=north}},
  column 3/.style={nodes={text width=0.15\liveW, align=center, anchor=north}},
  column 4/.style={nodes={text width=0.15\liveW, align=center, anchor=north}},
  column 5/.style={nodes={text width=0.15\liveW, align=center, anchor=north}},
]{
  |[tblhdr]| CWE structural failure & |[tblhdr]| & |[tblhdr]| & |[tblhdr]| & |[tblhdr]| \\
  |[tblhdr]| & |[tblhdr, font=\bfseries\small]| Orthogonal classes & |[tblhdr, font=\bfseries\small]| Small, structured class set & |[tblhdr, font=\bfseries\small]| Cause and operation axes & |[tblhdr, font=\bfseries\small]| Cause-to-consequence chain \\
  F1: Non-orthogonal hierarchy & \tMark & \xMark & \xMark & \xMark \\
  F2: Intractable target space & \xMark & \tMark & \xMark & \xMark \\
  F3: Sink-only labeling & \xMark & \xMark & \tMark & \dMark \\
  F4: Absence of a causal chain & \xMark & \xMark & \dMark & \tMark \\
};
\path ($(m-1-2.center)!0.5!(m-1-5.center)$) coordinate (superCenterX);
\node[tblhdr, anchor=center] at (superCenterX |- m-1-1.center) {Bugs Framework (BF) Design Properties};
\draw[Rtop] (m.north west) -- (m.north east);
\draw[Rmid] (m.west|-m-2-1.north) -- (m.east|-m-2-1.north);
\draw[Rmid] (m.west|-m-3-1.north) -- (m.east|-m-3-1.north);
\draw[Rbot] (m.south west) -- (m.south east);
\begin{scope}[on background layer]
  \fill[tblSubHeaderBg] (m.west|-m-2-1.north) rectangle (m.east|-m-3-1.north);
\end{scope}
\node[anchor=north, font=\small, text=cCorrSteelText, inner xsep=0pt, inner ysep=5pt]
  at (m.south)
  {\textbf{Legend:}\quad \tMark~Primary correspondence\quad \dMark~Reinforcing correspondence\quad \xMark~Not applicable};
\end{tikzpicture}%
\end{lrbox}
\usebox{\tbltmpbox}
\caption{Top: minimal schematic of each structural CWE failure, keyed to
Table~\ref{tab:failure_notation}; Figure~\ref{fig:bf_fix_panels} maps onto
these panels with the BF property that addresses each failure. Bottom:
correspondence between the four structural CWE failures
(Section~\ref{subsec:definitions}, rows) and the BF design properties
that address them (columns); the legend below the table defines the three marks.}
\label{fig:cwe_failure_panels}
\label{tab:bf_failure_matrix}
\Description{Top: a two-by-two grid of red-bordered panels F1 to F4. F1 shows one CVE fanning out to two CWE labels at different levels; F2 shows a bar chart with a tall head and a long flat tail of sparsely used entries; F3 shows a chain of root cause, propagation, and sink in which only the sink box is filled and labeled; F4 shows three CWE boxes as an unordered set with no connecting arrows. Bottom: a matrix with the four failures as rows and orthogonal classes, small structured class set, and cause-to-consequence chain as columns, using a filled tick for primary correspondence, a ringed tick for reinforcing correspondence, and a cross for not applicable; the primary ticks fall on the diagonal.}

\end{inlinefloat}
        %
    
        \subsubsection{Orthogonal weakness classes address the non-orthogonal hierarchy (F1)}
        
            The non-orthogonal hierarchy is defined in Section~\ref{subsec:definitions}. BF addresses this failure through the orthogonality of its weakness classes, a property that the specification directly asserts for these classes~\cite{bojanova2024bf}. The partition is exact at the level of operations: ``Orthogonal means that the intersection of the sets of operations of any two BF classes is the empty set''~\cite{bojanova2024bf}. Because the operation sets are disjoint, a defect represented by one class cannot also belong to another class at a competing level of the same hierarchy. The orthogonal partition removes the freedom, central to the non-orthogonal hierarchy, to select among ancestor, descendant, or sibling labels for one defect. We note a wording difference for accuracy: the specification applies \emph{orthogonal} to the classes, realized through disjoint operation sets, and \emph{multidimensional} to the attribute axes that supply the within-class structure~\cite{bojanova2024bf}. Orthogonality removes overlap by design; it does not by itself guarantee that an automated procedure will select the single correct class for a given CVE, a practical reduction that a CVE-to-BF system would be intended to test.

        \subsubsection{Structured composition over a small class set addresses the intractable target space (F2)}
        
            The intractable target space (F2) is stated operationally in Section~\mbox{\ref{subsec:definitions}}. BF addresses this failure through structured composition over a small and complete set of weakness classes~\cite{bojanova2024bf,bojanova2021bfio}.\footnote{BF project site: \url{https://samate.nist.gov/BF/}.} BF does not enumerate each weakness as a separate entry; it expresses specificity by composition: a weakness is one value per axis drawn from a small per-phase class, and a complex type is described by combining axis values without adding a new entry. The composing base is deliberately small; a single execution phase is covered by a handful of classes, as when ``four language-independent, orthogonal classes that cover all possible kinds of memory-related software bugs and weaknesses'' span the memory phase~\cite{bojanova_mem_bugs}. The contrast is stated in the same literature, which observes that the exhaustive CWE list ``is prone to having gaps and overlaps in coverage''~\cite{bojanova_mem_bugs}, whereas the CWE catalog has grown beyond 900 entries~\cite{cwe_mitre}, the long tail of which Table~\ref{tab:methodology_overview} shows the surveyed systems decline. Because specificity is carried by axis combinations rather than by entries, the effective target along each axis is bounded by design, thereby addressing the cardinality and skew that drive 

\definecolor{sSky800}{HTML}{075985}
\definecolor{cArchAmberBorder}{HTML}{B45309}
\definecolor{sTailwindAmber}{HTML}{D97706}
\definecolor{cArchBfBorder}{HTML}{1F4E79}
\definecolor{cArchBfFillD}{HTML}{B7D3E8}
\definecolor{cArchBfFillL}{HTML}{E8F1F8}
\definecolor{cArchBfText}{HTML}{12365B}
\definecolor{cArchCorpusBorder}{HTML}{1F4E79}
\definecolor{cArchCorpusFillD}{HTML}{B7D3E8}
\definecolor{cArchCorpusFillL}{HTML}{E8F1F8}
\definecolor{cArchCorpusText}{HTML}{12365B}
\definecolor{cArchFailBorder}{HTML}{991B1B}
\definecolor{cArchFailFillL}{HTML}{FEE2E2}
\definecolor{cArchFailText}{HTML}{7F1D1D}
\definecolor{cArchGoodBorder}{HTML}{0284C7}
\definecolor{sSky900}{HTML}{0284C7}
\definecolor{cArchGoodFillD}{HTML}{B9E2F5}
\definecolor{cArchGoodFillL}{HTML}{EAF6FD}
\definecolor{cArchGoodText}{HTML}{075985}
\definecolor{cArchIndigoBorder}{HTML}{3730A3}
\definecolor{sDeepAbyss}{HTML}{171444}
\definecolor{cArchIndigoFillL}{HTML}{EEF2FF}
\definecolor{cArchIndigoText}{HTML}{312E81}
\definecolor{cArchSlateBorder}{HTML}{334155}
\definecolor{cArchSlateText}{HTML}{0F172A}
\definecolor{cArchTealBorder}{HTML}{B4D5D5}
\definecolor{sDarkMutedTeal}{HTML}{2E5959}
\definecolor{cArchVivOrangeBorder}{HTML}{FADFCC}
\definecolor{sDeeperPeach}{HTML}{ED8D4A}

\begin{inlinefloat}
\definecolor{bandbg}{HTML}{F2F2F2}
\newcommand{\archGlyph}[2]{\def\aiW{0.075em}\csname archG#2\endcsname{#1}}
\newcommand{\archIcon}[2]{\mbox{\begin{tikzpicture}[baseline=-0.32em, rounded corners=0pt]
  \useasboundingbox (-0.68em,-0.70em) rectangle (0.68em,0.54em);
  \fill[#1!14!white, rounded corners=0.15em] (-0.63em,-0.70em) rectangle (0.63em,0.54em);
  \draw[#1!55!white, line width=0.05em, rounded corners=0.15em]
       (-0.63em,-0.70em) rectangle (0.63em,0.54em);
  \begin{scope}[yshift=-0.08em]\archGlyph{#1}{#2}\end{scope}
\end{tikzpicture}}}
\newcommand{\archIconRev}[2]{\mbox{\begin{tikzpicture}[baseline=-0.32em, rounded corners=0pt]
  \useasboundingbox (-0.68em,-0.70em) rectangle (0.68em,0.54em);
  \draw[#1, line width=0.05em, rounded corners=0.15em]
       (-0.63em,-0.70em) rectangle (0.63em,0.54em);
  \begin{scope}[yshift=-0.08em]\archGlyph{#1}{#2}\end{scope}
\end{tikzpicture}}}
\newcommand{\archGhash}[1]{
  \draw[#1, line width=\aiW, line cap=round]
    (-0.14em,-0.34em) -- (-0.04em,0.34em) (0.08em,-0.34em) -- (0.18em,0.34em)
    (-0.30em,-0.13em) -- (0.30em,-0.13em) (-0.26em,0.15em) -- (0.34em,0.15em);}
\newcommand{\archGfile}[1]{
  \draw[#1, line width=\aiW, line join=round]
    (-0.22em,-0.36em) -- (-0.22em,0.36em) -- (0.08em,0.36em) -- (0.24em,0.20em)
    -- (0.24em,-0.36em) -- cycle;
  \draw[#1, line width=0.06em] (0.08em,0.36em) -- (0.08em,0.20em) -- (0.24em,0.20em);}
\newcommand{\archGcode}[1]{
  \draw[#1, line width=\aiW, line cap=round, line join=round]
    (-0.12em,0.26em) -- (-0.34em,0.00em) -- (-0.12em,-0.26em)
    (0.12em,0.26em) -- (0.34em,0.00em) -- (0.12em,-0.26em);}
\newcommand{\archGglobe}[1]{
  \draw[#1, line width=\aiW] (0,0) circle (0.33em);
  \draw[#1, line width=0.06em] (0,0) ellipse (0.15em and 0.33em);
  \draw[#1, line width=0.06em] (-0.32em,0em) -- (0.32em,0em);}
\newcommand{\archGfilter}[1]{
  \draw[#1, line width=\aiW, line join=round]
    (-0.33em,0.30em) -- (0.33em,0.30em) -- (0.08em,-0.04em) -- (0.08em,-0.33em)
    -- (-0.08em,-0.24em) -- (-0.08em,-0.04em) -- cycle;}
\newcommand{\archGbox}[1]{
  \draw[#1, line width=\aiW, line join=round]
    (-0.28em,0.05em) -- (-0.28em,-0.31em) -- (0.28em,-0.31em) -- (0.28em,0.05em) -- cycle;
  \draw[#1, line width=\aiW, line cap=round]
    (-0.28em,0.05em) -- (-0.40em,0.26em) (0.28em,0.05em) -- (0.40em,0.26em);}
\newcommand{\archGtarget}[1]{
  \draw[#1, line width=\aiW] (0,0) circle (0.26em);
  \fill[#1] (0,0) circle (0.075em);
  \draw[#1, line width=\aiW, line cap=round]
    (0,0.33em) -- (0,0.42em) (0,-0.33em) -- (0,-0.42em)
    (0.33em,0) -- (0.42em,0) (-0.33em,0) -- (-0.42em,0);}
\newcommand{\archGask}[1]{
  \draw[#1, line width=\aiW, line cap=round]
    (-0.16em,0.16em) arc[start angle=180, end angle=-45, radius=0.16em];
  \draw[#1, line width=\aiW, line cap=round] (0.047em,0.047em) -- (0.00em,-0.10em);
  \fill[#1] (0.00em,-0.30em) circle (0.07em);}
\newcommand{\archGbranch}[1]{
  \fill[#1] (-0.18em,0.26em) circle (0.085em);
  \fill[#1] (-0.18em,-0.26em) circle (0.085em);
  \fill[#1] (0.24em,0.00em) circle (0.085em);
  \draw[#1, line width=0.065em] (-0.18em,0.18em) -- (-0.18em,-0.18em);
  \draw[#1, line width=0.065em] (-0.18em,0.05em) .. controls (-0.18em,-0.03em)
    and (0.06em,0.00em) .. (0.16em,0.00em);}
\newcommand{\archGloop}[1]{
  \draw[#1, line width=\aiW, -{Stealth[scale=0.7]}]
    (0.10em,0.31em) arc[start angle=72, end angle=-330, radius=0.33em];}
\newcommand{\archGbug}[1]{
  \draw[#1, line width=\aiW] (0,-0.06em) ellipse (0.18em and 0.26em);
  \fill[#1] (0,0.26em) circle (0.09em);
  \draw[#1, line width=0.065em, line cap=round]
    (-0.18em,0.06em) -- (-0.34em,0.14em) (0.18em,0.06em) -- (0.34em,0.14em)
    (-0.17em,-0.20em) -- (-0.31em,-0.29em) (0.17em,-0.20em) -- (0.31em,-0.29em);}
\newcommand{\archGtable}[1]{
  \draw[#1, line width=\aiW] (-0.33em,-0.28em) rectangle (0.33em,0.30em);
  \draw[#1, line width=0.06em] (-0.33em,0.105em) -- (0.33em,0.105em)
    (-0.02em,0.105em) -- (-0.02em,-0.28em);}
\newcommand{\archGtree}[1]{
  \draw[#1, line width=\aiW] (-0.10em,0.14em) rectangle (0.10em,0.36em);
  \draw[#1, line width=\aiW] (-0.38em,-0.36em) rectangle (-0.18em,-0.14em);
  \draw[#1, line width=\aiW] (0.18em,-0.36em) rectangle (0.38em,-0.14em);
  \draw[#1, line width=0.06em] (0,0.14em) -- (0,-0.02em)
    (-0.28em,-0.14em) -- (-0.28em,-0.02em) -- (0.28em,-0.02em) -- (0.28em,-0.14em);}
\newcommand{\archGrobot}[1]{
  \draw[#1, line width=\aiW, rounded corners=0.06em]
    (-0.30em,-0.28em) rectangle (0.30em,0.18em);
  \fill[#1] (-0.125em,-0.02em) circle (0.06em);
  \fill[#1] (0.125em,-0.02em) circle (0.06em);
  \draw[#1, line width=0.06em] (0,0.18em) -- (0,0.30em);
  \fill[#1] (0,0.345em) circle (0.055em);}
\newcommand{\archGchain}[1]{
  \fill[#1] (-0.30em,0.16em) circle (0.10em);
  \fill[#1] (0.00em,-0.18em) circle (0.10em);
  \fill[#1] (0.30em,0.16em) circle (0.10em);
  \draw[#1, line width=0.065em] (-0.23em,0.09em) -- (-0.07em,-0.11em)
                               (0.07em,-0.11em) -- (0.23em,0.09em);}
\newcommand{\archGcheckdouble}[1]{
  \draw[#1, line width=\aiW, line cap=round, line join=round]
    (-0.36em,0.06em) -- (-0.20em,-0.14em) -- (0.10em,0.22em);
  \draw[#1, line width=\aiW, line cap=round, line join=round]
    (-0.05em,-0.14em) -- (0.04em,-0.25em) -- (0.36em,0.11em);}
\newcommand{\archGcheckcircle}[1]{
  \draw[#1, line width=\aiW] (0,0) circle (0.34em);
  \draw[#1, line width=\aiW, line cap=round, line join=round]
    (-0.16em,0.01em) -- (-0.05em,-0.12em) -- (0.18em,0.15em);}
\newcommand{\archGrosette}[1]{
  \foreach \a in {0,45,...,315}{\fill[#1] (\a:0.31em) circle (0.08em);}
  \fill[#1!22!white] (0,0) circle (0.25em);
  \draw[#1, line width=0.05em] (0,0) circle (0.25em);
  \draw[#1, line width=\aiW, line cap=round, line join=round]
    (-0.13em,0.01em) -- (-0.035em,-0.105em) -- (0.155em,0.125em);}

\begin{tikzpicture}[
  font=\normalsize,
  band/.style={draw=cArchSlateBorder, dashed, rounded corners=6pt, inner sep=7pt},
  bandlabel/.style={font=\bfseries\normalsize, text=cArchSlateText, fill=bandbg,
    draw=cArchSlateBorder, rounded corners=5pt, inner xsep=5pt, inner ysep=3pt,
    align=center},
  evid/.style={draw=cArchCorpusBorder, fill=cArchCorpusFillL, text=cArchCorpusText,
    rounded corners=3pt, align=center, text width=0.20\liveW, minimum height=0.74cm, line width=0.7pt},
  evidD/.style={evid, fill=cArchCorpusFillD, font=\bfseries\normalsize},
  eng/.style={draw=cArchBfBorder, fill=cArchBfFillL, text=cArchBfText,
    rounded corners=3pt, align=center, text width=0.24\liveW, minimum height=0.74cm, line width=0.7pt},
  engD/.style={eng, fill=cArchBfFillD, font=\bfseries\normalsize},
  store/.style={draw=cArchIndigoBorder, fill=cArchIndigoFillL, text=cArchIndigoText,
    rounded corners=3pt, align=center, text width=0.20\liveW, minimum height=0.72cm, line width=0.7pt},
  ver/.style={draw=cArchGoodBorder, fill=cArchGoodFillL, text=cArchGoodText,
    rounded corners=3pt, align=center, text width=0.20\liveW,
    minimum height=0.68cm, line width=0.7pt},
  verD/.style={ver, fill=cArchGoodFillD, font=\bfseries\normalsize},
  stopn/.style={draw=cArchFailBorder, fill=cArchFailFillL, text=cArchFailText,
    rounded corners=3pt, align=center, text width=0.24\liveW, minimum height=0.74cm, line width=0.7pt},
  arr/.style={draw=cArchSlateBorder, semithick, -{Stealth[scale=1.3]}},
  arrR/.style={draw=cArchFailBorder, semithick, -{Stealth[scale=1.3]}},
  arrG/.style={draw=cArchGoodBorder, semithick, -{Stealth[scale=1.3]}},
  lbl/.style={font=\small, text=cArchSlateText, fill=white, inner sep=1.5pt}
]

\def\gapBandFirstNode{4mm}      %
\def\gapDblLineBandFirstNode{6.5mm}      %
\def\gapLastNodeBandBorder{2mm} %
\def\gapInterNodeNoArr{2mm}     %
\def\gapInterNodeArr{4mm}       %
\def\gapInterNodeArrLbl{4mm}    %
\def\gapInterBand{27mm}          %

\node[evidD] (cve)    at ({0.14\liveW}, 0) {\archIcon{cArchIndigoBorder}{hash}~CVE identifier $v$};
\node[evid]  (nvd)    at ([yshift=-\gapInterNodeNoArr]cve.south) [anchor=north] {\archIcon{sTailwindAmber}{file}~Advisory text\\$E_{\mathrm{dsc}}$};
\node[evid]  (code)   at ([yshift=-\gapInterNodeNoArr]nvd.south) [anchor=north] {\archIcon{cArchFailBorder}{code}~Fix commit,\\buggy code $E_{\mathrm{cod}}$};
\node[evid]  (adv)    at ([yshift=-\gapInterNodeNoArr]code.south) [anchor=north] {\archIcon{cArchGoodBorder}{globe}~Advisory pages\\$E_{\mathrm{adv}}$};
\node[evid]  (strip)  at ([yshift=-\gapInterNodeNoArr]adv.south) [anchor=north] {\archIcon{cArchCorpusText}{filter}~Contamination strip\\(clean policy)};
\node[evidD] (bundle) at ([yshift=-\gapInterNodeNoArr]strip.south) [anchor=north] {\archIcon{cArchFailBorder}{box}~Evidence bundle $E$};

\coordinate (E_top) at ([yshift=\gapBandFirstNode]cve.north);
\coordinate (E_bot) at ([yshift=-\gapLastNodeBandBorder]bundle.south);
\begin{scope}[on background layer]
\node[band, fit=(cve)(bundle)(E_top)(E_bot), inner ysep=0pt] (bandE) {};
\end{scope}
\node[bandlabel] at (bandE.north) {Evidence Acquisition};

\node[engD]  (anchor) at ({0.50\liveW}, 0 |- cve.north) [anchor=north] {\archIcon{cArchFailBorder}{target}~Anchor: failure $F$\\and final error $Cn_1$};
\node[eng]   (q1)     at ([yshift=-\gapInterNodeArr]anchor.south) [anchor=north] {\archIcon{cArchIndigoBorder}{ask}~Q1: which \textsc{operation}?\\the class menu $T_B[W_n]$ is scored};
\node[eng]   (q2)     at ([yshift=-\gapInterNodeArr]q1.south) [anchor=north] {\archIcon{cArchAmberBorder}{ask}~Q2: which \textsc{operand} facet\\is wrong? (name, data, type)};
\node[eng]   (q3)     at ([yshift=-\gapInterNodeArr]q2.south) [anchor=north] {\archIcon{sSky800}{branch}~Q3: \textsc{bug} or \textsc{fault}?\\(stopping test, $Cs_n \in T_C$)};
\node[eng]   (q4)     at ([yshift=-\gapInterNodeArrLbl]q3.south) [anchor=north] {\archIcon{cArchCorpusText}{loop}~Q4: the fault becomes the\\consequence one link back};
\node[stopn] (root)   at ([yshift=-\gapInterNodeNoArr]q4.south) [anchor=north] {\archIconRev{sDeepAbyss}{bug}~Bug: root found,\\chain complete};

\coordinate (D_top) at ([yshift=\gapDblLineBandFirstNode]anchor.north);
\coordinate (D_bot) at ([yshift=-\gapLastNodeBandBorder]root.south);
\begin{scope}[on background layer]
\node[band, fit=(anchor)(root)(D_top)(D_bot), inner ysep=0pt] (bandD) {};
\end{scope}
\node[bandlabel] at (bandD.north) {Backward Derivation\\Engine (per-link loop)};

\draw[arr] (anchor) -- (q1);
\draw[arr] (q1) -- (q2);
\draw[arr] (q2) -- (q3);
\draw[arr] (q3) -- node[pos=0.5, right=3pt, font=\small \itshape, text=cArchSlateText] {fault} (q4);

\draw[arrR] (q3.east) -- ++(0.5,0) coordinate (exitR) |- (root.east);
\node[lbl, rotate=-90] at ($(exitR)!0.60!(exitR|-root.east)$) {\textit{bug}};

\draw[arr] (q4.west) -- ++(-0.5,0) coordinate (fbL) |- (q1.west);
\node[lbl, rotate=90, align=center, font=\itshape] at ($(fbL)!0.5!(fbL|-q1.west)$) {climb one link};

\draw[arr] ($(bundle.north east)!0.3!(bundle.south east)$) --
node[lbl, pos=0.50] {$E$} (bandD.west |- {$(bundle.north east)!0.3!(bundle.south east)$});

\node[store] (tabs) at ({0.86\liveW}, 0 |- cve.north) [anchor=north] {\archIcon{sDeeperPeach}{table}~Closed tables\\$T_A..T_H$};
\node[store] (onto) at ([yshift=-\gapInterNodeNoArr]tabs.south) [anchor=north] {\archIcon{sDarkMutedTeal}{chain}~OWL orthogonality};
\node[store] (llm)  at ([yshift=-\gapInterNodeNoArr]onto.south) [anchor=north] {\archIcon{cArchCorpusBorder}{robot}~Constrained oracle $\Omega$};

\coordinate (S_top) at ([yshift=\gapDblLineBandFirstNode]tabs.north);
\coordinate (S_bot) at ([yshift=-\gapLastNodeBandBorder]llm.south);
\begin{scope}[on background layer]
\node[band, fit=(tabs)(llm)(S_top)(S_bot), inner ysep=0pt] (bandS) {};
\end{scope}
\node[bandlabel] at (bandS.north) {Symbolic Store\\and Oracle};

\draw[arr] (bandS.west|-tabs) -- node[lbl, pos=0.5, above=1.5pt, align=center, font=\itshape] {exact\\lookups} (bandD.east|-tabs);
\draw[arr] (bandD.east|-llm) -- node[lbl, pos=0.5, above=1.5pt, align=center, rotate=90, anchor=west, yshift=0pt, font=\itshape] {menus,\\evidence} (bandS.west|-llm);

\node[ver]  (voc) at ({0.86\liveW}, 0 |- bandS.south) [anchor=north, yshift=-\gapInterBand] {\archIcon{cArchGoodBorder}{checkcircle}~Vocabulary check $P_{\mathrm{voc}}$};
\node[ver]  (str) at ([yshift=-\gapInterNodeArr]voc.south) [anchor=north] {\archIcon{cArchIndigoBorder}{tree}~SHACL structure $P_{\mathrm{str}}$};
\node[ver]  (ort) at ([yshift=-\gapInterNodeArr]str.south) [anchor=north] {\archIcon{cArchAmberBorder}{checkdouble}~Orthogonality $P_{\mathrm{ort}}$};
\node[verD] (out) at ([yshift=-\gapInterNodeArr]ort.south) [anchor=north] {\archIcon{sSky900}{rosette}~Conformant chain $\mathcal{C}$};

\coordinate (V_top) at ([yshift=\gapDblLineBandFirstNode]voc.north);
\coordinate (V_bot) at ([yshift=-\gapLastNodeBandBorder]out.south);
\begin{scope}[on background layer]
\node[band, fit=(voc)(out)(V_top)(V_bot), inner ysep=0pt] (bandV) {};
\end{scope}
\node[bandlabel] at (bandV.north) {Neuro-Symbolic\\Verification};

\draw[arrG] (voc) -- (str);
\draw[arrG] (str) -- (ort);
\draw[arrG] (ort) -- (out);

\draw[{Stealth[scale=1.2]}-{Stealth[scale=1.2]}, draw=cArchCorpusBorder, thick, dashed]
  ($(bandS.south)+(-1.55,0)$) -- ($(bandV.north)+(-1.55,0)$);
\draw[{Stealth[scale=1.2]}-{Stealth[scale=1.2]}, draw=cArchCorpusBorder, thick, dashed]
  ($(bandS.south)+(1.55,0)$) -- ($(bandV.north)+(1.55,0)$);
\node[font=\small \itshape, text=cArchSlateText, align=center, fill=white, inner sep=1pt]
  at ($(bandS.south)!0.42!(bandV.north)$)
  {shared vocabulary,\\shapes, restrictions};

\draw[arr] (bandD.east|-str) -- (bandV.west|-str);
\node[lbl, align=center, rotate=90, anchor=west, font=\itshape]
  at ($(exitR|-str)!0.5!(bandV.west|-str)+(0,3pt)$) {candidate\\chain};

\coordinate (vrp) at ($(bandV.south)+(0,-0.6)$);
\draw[arrR] (bandV.south) -- (vrp) -- (bandD.south|-vrp) -- (bandD.south);
\node[lbl, font=\itshape] at ($(vrp)!0.5!(bandD.south|-vrp)$) {violation report, regeneration};

\end{tikzpicture}%
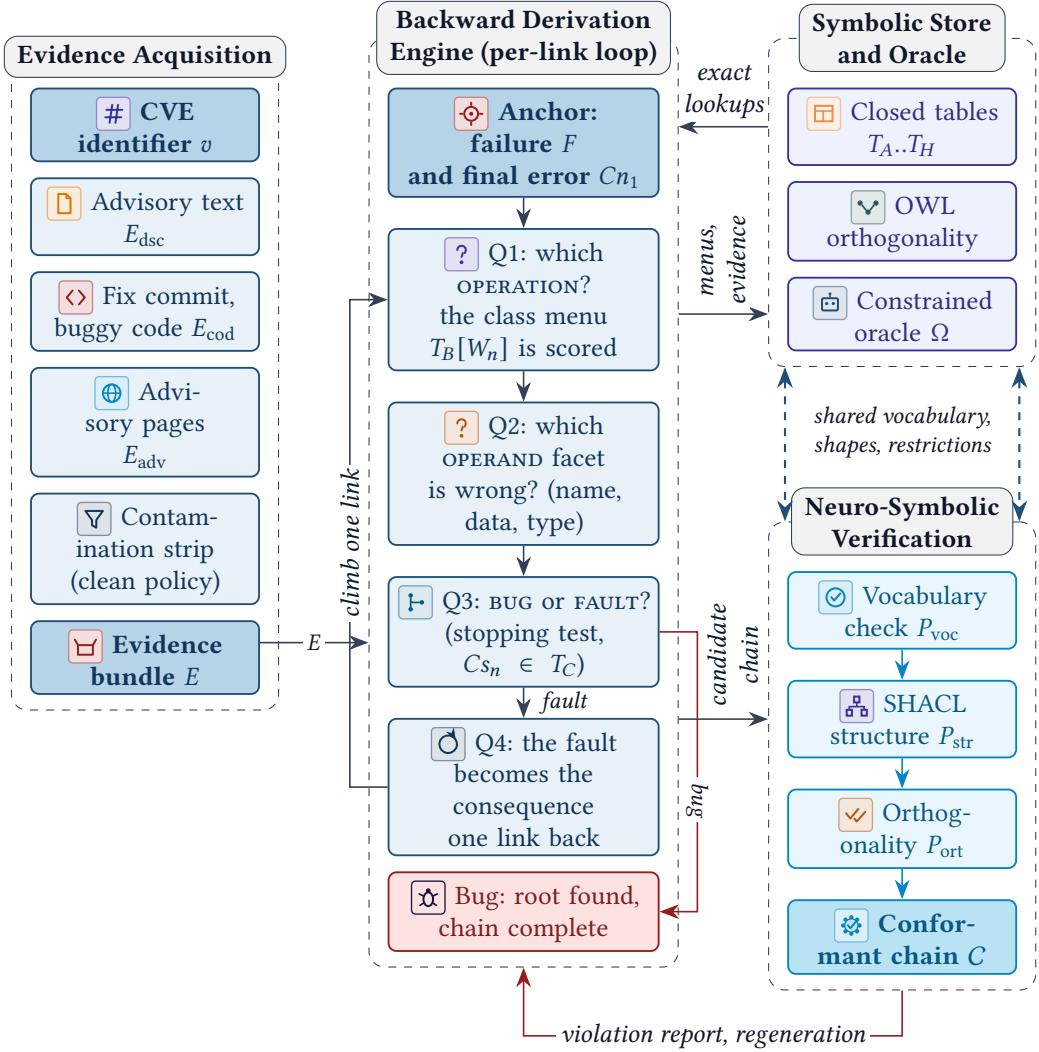
\floatnote{\textit{Algorithm~\ref{alg:bf_derivation}, notation of Table~\ref{tab:notation}. The engine derives one weakness per link; exact lookups hit the closed tables while inference routes to the oracle $\Omega$, and the verifier gates every chain before release.}}
\caption{Architectural process view of the backward derivation pipeline.}
\label{fig:bf_derivation_architecture}
\Description{Three dashed columns. Left, Evidence Acquisition: five source boxes (CVE identifier, advisory text, fix commit, advisory pages, contamination strip) feed one evidence bundle. Center, Backward Derivation Engine: a per-link loop that anchors on the failure, then asks four numbered questions (which operation, which fault, bug or fault, and the fault becoming the next consequence) with a loop-back arrow, ending in a red box marking the bug found and the chain complete. Right, a Symbolic Store and Oracle above a Neuro-Symbolic Verification block: closed tables and OWL orthogonality constrain the oracle, and four predicates (vocabulary, SHACL structure, orthogonality, linking) gate the candidate chain into a conformant-chain box. A red violation-report arrow returns from the verifier to the engine.}
\end{inlinefloat}
F2 at the per-axis decision level. The reduction is real but partial, and this comparison claims no more than that. At the combination level, the joint space of valid cause, operation, consequence, and attribute values remains large and long-tailed, so a system that predicts a full chain still confronts a skewed target; faceting relocates the intractability to smaller per-axis decisions; it does not remove it outright~\cite{he2009imbalanced,zhang2023longtailed,10.1145/3815425,li2025out}. Consistent with this reading, the inter-rater study (bottom panel of Figure~\ref{fig:bf_fix_panels}) records almost perfect agreement on the small cause and operation axes ($\kappa=\wtKappaCause$ and $\kappa=\wtKappaOperation$) and only fair agreement on the attribute axis ($\kappa=\wtKappaAttribute$), which carries the largest and most ambiguous per-axis value set. F2 is therefore mitigated but not eliminated: BF lowers the per-axis target to a finite, small set, while the residual skew at the combination level remains a learnability problem that a CVE-to-BF system is intended to measure.

        \subsubsection{Explicit cause and operation axes address sink-only labeling (F3)}
            Sink-only labeling (F3) carries the operational test given in Section~\mbox{\ref{subsec:definitions}}. BF expresses cause and operation as first-class axes that are distinct from the consequence~\cite{bojanova2024bf}. The specification records cause and consequence as separate elements of every weakness: ``Bugs and faults are causes of security weaknesses, and errors and final errors are their consequences''~\cite{bojanova2024bf}. The specification names consequence of sink-only labeling explicitly: ``Some [CVEs] list the final error at the sink as the root cause instead of the bug or hardware-induced fault that starts the chain. Focusing on the final error helps identify mitigation techniques, but the actual root cause must be known and fixed to resolve the vulnerability''~\cite{bojanova2024bf}. Sink-only labeling assigns the category of the last weakness alone; BF additionally records the first, so that two CVEs which share a sink but differ in their upstream cause receive different descriptions, as the cause and operation axes each carry an independent value. The information that sink-only labeling discards is thus retained by construction, on axes that are distinct from consequence.

        \subsubsection{The cause-operation-consequence chain addresses the absence of a causal chain (F4)}

            The absence of a causal chain (F4) is fixed in Section~\mbox{\ref{subsec:definitions}}. CWE supplies curated exceptions through its Named Chains and Composites (CWE-709~\cite{cwe709}, CWE-678~\cite{cwe678}), which enumerate specific pre-named chains in place of a general by-construction rule under which every vulnerability is a cause-to-consequence chain~\cite{cwe_rcm_guidance}. BF records a vulnerability as a cause-to-consequence chain of weaknesses~\cite{bojanova2024bf}. BF gives the chain a syntactic form: each weakness is one cause-operation-consequence instance, and the consequence of one weakness propagates as the cause of the next, so that ``The BF formalism supports a deeper understanding of vulnerabilities as chains of weaknesses that adhere to strict causation, propagation, and composition rules''~\cite{bojanova2024bf}. The cause-to-consequence link is the mechanism that carries the chain forward: a vulnerability ``starts with a bug or hardware-induced fault, propagates through errors that become faults, and ends with a final error that introduces an exploit vector toward a failure''~\cite{bojanova2024bf}. The specification represents Heartbleed in exactly this form, as a sequence of BF weakness states and a generated weakness chain~\cite{bojanova2024bf}, which shows that the multi-stage vulnerability named in Section~\ref{subsec:definitions} and depicted in Figure~\ref{fig:heartbleed_chain} is losslessly representable as a chain, not reduced to a single CWE. The same representation is provided for the BadAlloc pattern that includes CVE-2021-21834~\cite{bojanova2024bf}, with the explicit five-stage chain DVR $\curvearrowright$ TCM $\curvearrowright$ MMN $\curvearrowright$ MAD $\curvearrowright$ MUS. A chain representation eliminates the absence of a causal chain by design; reconstructing the full chain for a given CVE from its prose description remains a practical inference task, which a CVE-to-BF system would be intended to test. 


\definecolor{cNotationVioletBorder}{HTML}{D3BDF4}
\definecolor{cNotationVioletFillD}{HTML}{E5E2FE}
\definecolor{cNotationVioletFillL}{HTML}{F5F3FF}
\definecolor{cNotationVioletText}{HTML}{210D42}
\definecolor{cBfFixBorder}{HTML}{B9D97A}
\definecolor{cBfFixAccent}{HTML}{76B900}
\definecolor{cBfFixGreenBorder}{HTML}{166534}
\definecolor{cBfFixFillD}{HTML}{E4F1C8}
\definecolor{cBfFixGreenFillD}{HTML}{86EFAC}
\definecolor{cBfFixFillL}{HTML}{F4F9EA}
\definecolor{cBfFixGreenFillL}{HTML}{DCFCE7}
\definecolor{cBfFixText}{HTML}{223A06}
\definecolor{cBfFixGreenText}{HTML}{14532D}
\definecolor{cBfFixGrayBorder}{HTML}{808080}
\definecolor{cBfFixGrayFill}{HTML}{F4F4F4}
\definecolor{cBfFixGrayText}{HTML}{404040}
\definecolor{cKappaMistBorder}{HTML}{808080} %
\definecolor{cKappaMistFillD}{HTML}{F4F4F4} %
\definecolor{cKappaMistText}{HTML}{404040} %
\definecolor{cKappaIndigoBorder}{HTML}{3730A3} %
\definecolor{cKappaIndigoFillD}{HTML}{C7D2FE} %
\definecolor{cKappaIndigoText}{HTML}{312E81} %
\begin{inlinefloat}
\pgfmathsetlengthmacro{\PW}{0.49\liveW}\pgfmathsetlengthmacro{\PX}{0.51\liveW}
\begin{tikzpicture}[font=\normalsize,
  frame/.style={draw=cBfFixBorder, rounded corners=2.5pt, line width=0.9pt},
  badge/.style={draw=cBfFixAccent, fill=cBfFixFillD, text=cBfFixText,
                rounded corners=1.2pt, line width=0.6pt, font=\normalsize\bfseries,
                inner xsep=3.5pt, inner ysep=1.8pt},
  plab/.style={text=cBfFixText, font=\normalsize\bfseries, align=center},
  mini/.style={draw=cBfFixBorder, fill=cBfFixFillL, text=cBfFixText,
               rounded corners=1.2pt, line width=0.7pt, align=center,
               font=\normalsize, inner xsep=3pt, inner ysep=2pt, minimum height=0.55cm},
  minid/.style={mini, fill=cBfFixFillD, draw=cBfFixAccent, line width=1.0pt},
  minig/.style={draw=cBfFixGrayBorder, fill=cBfFixGrayFill, text=cBfFixGrayText,
                rounded corners=1.2pt, line width=0.7pt, align=center, font=\normalsize,
                inner xsep=3pt, inner ysep=2pt, minimum height=0.55cm},
  note/.style={text=cBfFixGrayText, font=\small, align=center},
  arr/.style={-{Stealth[scale=1.30]}, line width=0.9pt, draw=cBfFixAccent},
  darr/.style={-{Stealth[scale=1.30]}, line width=0.6pt, dashed, draw=cBfFixAccent}]
\draw[frame] (0,0) rectangle (\PW,-3.9);
\node[badge, anchor=north west] at (0.12,-0.12) {F1};
\node[minig] (x1op) at ({0.25*\PW},-1.60) {operation:\\\textbf{Read}};
\node[minid] (x1cl) at ({0.72*\PW},-1.60) {class: \textbf{MUS}\\(exactly one)};
\draw[arr] (x1op.east) -- (x1cl.west);
\node[note] at ({0.5*\PW},-2.80) {disjoint operation sets:\\one class per operation};
\node[plab] at ({0.5*\PW},-3.50) {orthogonal weakness classes};
\draw[frame] (\PX,0) rectangle (\liveW,-3.9);
\node[badge, anchor=north west] at ({\PX+0.12cm},-0.12) {F2};
\node[mini, font=\small] (x2a) at ({\PX+0.15*\PW},-1.05) {DVL DVR};
\node[mini, font=\small] (x2b) at ({\PX+0.50*\PW},-1.05) {MAD MMN MUS};
\node[mini, font=\small] (x2c) at ({\PX+0.85*\PW},-1.05) {TCV TCM $\ldots$};
\node[minid] (x2m) at ({\PX+0.5*\PW},-2.15) {(cause, operation) $\rightarrow$ consequence};
\draw[arr] (x2a.south) -- (x2m.155);
\draw[arr] (x2b.south) -- (x2m.north);
\draw[arr] (x2c.south) -- (x2m.25);
\node[note] at ({\PX+0.5*\PW},-2.95) {small closed class set; values\\composed from per-axis menus};
\node[plab] at ({\PX+0.5*\PW},-3.50) {structured composition};
\draw[frame] (0,-4.3) rectangle (\PW,-8.2);
\node[badge, anchor=north west] at (0.12,-4.42) {F3};
\node[minid] (x3a) at ({0.18*\PW},-5.85) {cause\\(the bug)};
\node[mini] (x3b) at ({0.5*\PW},-5.85) {operation};
\node[minid] (x3c) at ({0.82*\PW},-5.85) {consequence\\(the sink)};
\draw[arr] (x3a.east) -- (x3b.west);
\draw[arr] (x3b.east) -- (x3c.west);
\node[note] at ({0.5*\PW},-7.00) {both ends recorded:\\first-class cause and operation};
\node[plab] at ({0.5*\PW},-7.80) {cause is never collapsed into sink};
\draw[frame] (\PX,-4.3) rectangle (\liveW,-8.2);
\node[badge, anchor=north west] at ({\PX+0.12cm},-4.42) {F4};
\node[mini] (x4a) at ({\PX+0.17*\PW},-5.85) {$W_1$};
\node[mini] (x4b) at ({\PX+0.39*\PW},-5.85) {$W_2$};
\node[mini] (x4c) at ({\PX+0.61*\PW},-5.85) {$W_3$};
\node[minid] (x4f) at ({\PX+0.84*\PW},-5.85) {IEX};
\draw[darr] (x4a.east) -- (x4b.west);
\draw[darr] (x4b.east) -- (x4c.west);
\draw[arr] (x4c.east) -- (x4f.west);
\node[note] at ({\PX+0.5*\PW},-7.00) {consequence $\curvearrowright$ cause propagation:\\the order is part of the record};
\node[plab] at ({\PX+0.5*\PW},-7.80) {per-instance causal chain};
\end{tikzpicture}

\vspace{4pt}
\definecolor{cKappaSteelBorder}{HTML}{ADBCD3}
\definecolor{cKappaSteelHdr}{HTML}{DCE4EE}
\definecolor{cKappaSteelText}{HTML}{0D162D}
\colorlet{tblHeaderBg}{cKappaSteelHdr}
\colorlet{tblHeaderFg}{cKappaSteelText}
\colorlet{tblHeaderRule}{cKappaSteelBorder}
\colorlet{tblSubHeaderBg}{cKappaSteelHdr}
\begin{lrbox}{\tbltmpbox}%
\begin{tikzpicture}
\matrix (m) [tblmat,
  row sep=0.3pt,
  nodes={inner ysep=2.5pt, minimum height=0.5cm},
  column 1/.style={nodes={text width=0.40\liveW, align=left}},
  column 2/.style={nodes={text width=0.22\liveW, align=center}},
  column 3/.style={nodes={text width=0.22\liveW, align=center}},
]{
  |[tblhdr]| & |[tblhdr]| & |[tblhdr]| \\
  |[tblhdr]| & |[tblhdr]| Agreement & |[tblhdr]| Cohen's $\kappa$ \\
  \hspace*{18pt}Cause       & \wtAgreePctCause\%       & |[\wtKappaStyleCause]|       \wtKappaCause \\
  \hspace*{18pt}Operation   & \wtAgreePctOperation\%   & |[\wtKappaStyleOperation]|   \wtKappaOperation \\
  \hspace*{18pt}Consequence & \wtAgreePctConsequence\% & |[\wtKappaStyleConsequence]| \wtKappaConsequence \\
  \hspace*{18pt}Attribute   & \wtAgreePctAttribute\%   & |[\wtKappaStyleAttribute]|   \wtKappaAttribute \\
  \hspace*{18pt}Pooled (all \wtIrrCells\ cells) & \wtAgreePctPooled\% & |[\wtKappaStylePooled]| \wtKappaPooled \\
};
\draw[tblHeaderBg, line width=1.5pt] (m.west|-m-2-1.north) -- (m-1-2.north west|-m-2-1.north);
\node[anchor=center, font=\bfseries, text=tblHeaderFg, align=center]
  at ($(m-1-1.north)!0.5!(m-2-1.south)$) {BF Taxonomic Axis};
\node[tblhdr, anchor=center] at ($(m-1-2.center)!0.5!(m-1-3.center)$) {Inter-rater Reliability Metrics};
\draw[Rmid] ($(m-1-2.north west|-m-2-1.north)+(2pt,0)$) -- ($(m.east|-m-2-1.north)+(-2pt,0)$);
\draw[Rtop] (m.north west) -- (m.north east);
\draw[Rmid] (m.west|-m-3-1.north) -- (m.east|-m-3-1.north);
\draw[Rmid] (m.west|-m-7-1.north) -- (m.east|-m-7-1.north);
\draw[Rbot] (m.south west) -- (m.south east);
\begin{scope}[on background layer]
  \fill[tblSubHeaderBg] (m.west|-m-2-1.north) rectangle (m.east|-m-3-1.north);
\end{scope}
\end{tikzpicture}%
\end{lrbox}
\usebox{\tbltmpbox}
\floatnote[c]{\lkbandlegend}
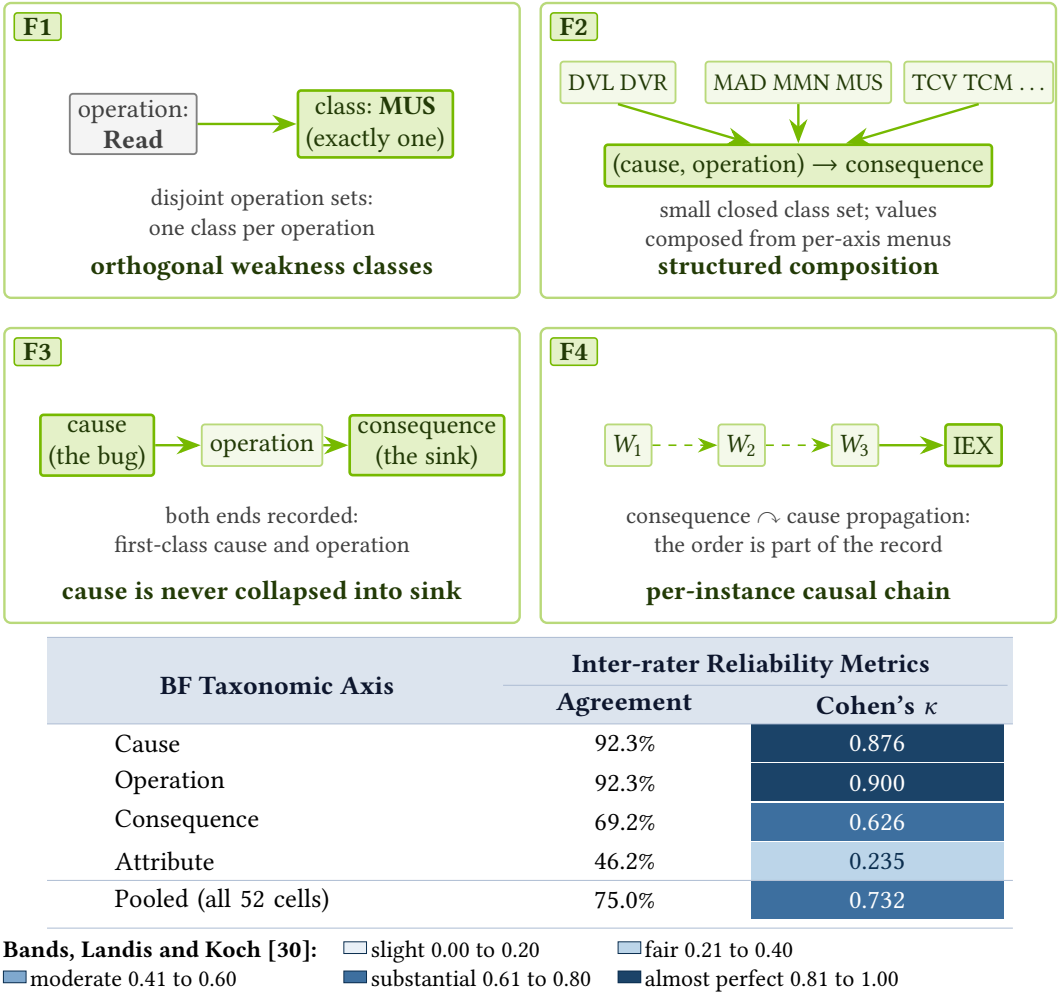
\caption{Top: the BF design property addressing each failure, replicating
Figure~\ref{fig:cwe_failure_panels} panel for panel; the grounding
specification language per cell is in its bottom matrix.
Bottom: inter-rater agreement on the blind BF mappings, over \wtIrrCVEs\
CVEs and \wtIrrCells\ axis cells.}
\label{fig:bf_fix_panels}
\label{tab:bf_kappa}
\Description{Top: a two-by-two grid of green-bordered panels F1 to F4 mirroring the failure panels. F1 shows one operation resolving to exactly one class; F2 shows three class boxes composing into a single cause, operation, consequence triple; F3 shows a chain in which the cause, operation, and consequence boxes are all filled; F4 shows weakness boxes W1, W2, and W3 joined by propagation arrows into the failure. Bottom: a table with five rows (cause, operation, consequence, attribute, pooled) and two metric columns, percent agreement and Cohen's kappa, with the kappa cells shaded from light to dark blue by Landis and Koch band and a band legend beneath.}
\end{inlinefloat}

    Across the four failures, each BF property addresses one failure directly, and the correspondence is one-to-one by construction. The bottom panel of Figure~\ref{fig:cwe_failure_panels} records this matrix, with the primary correspondence on the diagonal. The inter-rater study (bottom panel of Figure~\ref{fig:bf_fix_panels}) offers a first, partial confirmation of the orthogonality-to-determinism link. The mapping shows that BF addresses each failure by design; whether each is reduced in measured practice is the question that a CVE-to-BF pipeline would answer, and the inter-rater result indicates that the reduction holds first on the cause and operation axes.
\FloatBarrier
\section{Case Studies}\label{sec:worked_examples}

The four case studies below are stratified, one per BF class type, and selected from the \wtPoolSlots-CVE candidate pool in Table~\mbox{\ref{tab:candidate_pool}} by failure density, defined as the number of structural CWE failures a candidate exposes. All four share a pattern: the CWE label moved or was contested, whereas the BF classification did not change under that reassignment, because the disjoint operation sets of the BF classes admit one class and one value per axis (Section~\ref{subsec:bf_addresses_failures}).

\definecolor{cWorkedMapVioletBorder}{HTML}{F0E8FC} %
\definecolor{cWorkedMapVioletFillD}{HTML}{E5E2FE} %
\definecolor{cWorkedMapVioletFillL}{HTML}{F6F4FF} %
\definecolor{cWorkedMapVioletText}{HTML}{210D42} %
\begin{inlinefloat}

\colorlet{tblHeaderBg}{cWorkedMapVioletFillD}
\colorlet{tblHeaderFg}{cWorkedMapVioletText}
\colorlet{tblHeaderRule}{cWorkedMapVioletBorder}
\colorlet{tblSubHeaderBg}{cWorkedMapVioletFillL}

\begin{lrbox}{\tbltmpbox}%
\begin{tikzpicture}
\matrix (m) [tblmat,
  column 1/.style={nodes={text width=0.17\liveW, align=left}},
  column 2/.style={nodes={text width=0.22\liveW, align=left}},
  column 3/.style={nodes={text width=0.24\liveW, align=left}},
  column 4/.style={nodes={text width=\dimexpr0.37\liveW-32pt\relax, align=left}},
]{
  |[tblhdr]| \textbf{CVE} & |[tblhdr]| \textbf{CWE issue} & |[tblhdr]| \textbf{BF chain anchor} & |[tblhdr]| \textbf{Illustrated resolution} \\
  CVE-2021-3156 (Sudo Baron Samedit) & Overlapping boundary and overflow labels; label moved across hierarchy levels & DVL $\rightarrow$ MAD $\rightarrow$ MUS $\rightarrow$ ACE & Preserves the input-validation root and the memory chain leading to code execution. \\
  CVE-2015-0235 (GHOST) & Boundary label hides wrong-size allocation upstream & MMN $\rightarrow$ MUS & Records the allocation-stage cause before the out-of-bounds memory use. \\
  CVE-2021-21834 (GPAC) & Integer-overflow and allocation labels name different stages of one chain & DVR $\rightarrow$ TCM $\rightarrow$ MMN $\rightarrow$ MAD $\rightarrow$ MUS & Makes the verify-to-compute-to-allocate-to-position-to-write sequence explicit (BadAlloc pattern). \\
  CVE-2014-0160 (Heartbleed) & Over-read sink label hides missing verification & DVR $\rightarrow$ MAD $\rightarrow$ MUS $\rightarrow$ IEX (with converging MUS Clear) & Restores the missing-verification cause before the information exposure. \\
};

\draw[Rtop] (m.north west) -- (m.north east);
\draw[Rmid] (m.west|-m-2-1.north) -- (m.east|-m-2-1.north);
\draw[Rmid] (m.west|-m-3-1.north) -- (m.east|-m-3-1.north);
\draw[Rmid] (m.west|-m-4-1.north) -- (m.east|-m-4-1.north);
\draw[Rmid] (m.west|-m-5-1.north) -- (m.east|-m-5-1.north);
\draw[Rbot] (m.south west) -- (m.south east);

\begin{scope}[on background layer]
  \fill[cWorkedMapVioletFillL] (m.west|-m-3-1.north) rectangle (m.east|-m-4-1.north);
  \fill[cWorkedMapVioletFillL] (m.west|-m-5-1.north) rectangle (m.south east);
\end{scope}

\end{tikzpicture}%
\end{lrbox}
\usebox{\tbltmpbox}
\captionof{table}{Case study used to contrast CWE labeling with BF chain
specification; full chain prose is in
Section~\ref{sec:worked_examples}.}
\label{tab:worked_example_mappings}
\end{inlinefloat}

\subsection{CVE-2021-3156 (Sudo Baron Samedit), BF\_INP $\rightarrow$ BF\_MEM}\label{ex:sudo}

    \paragraph{CVE summary}
        A heap-based buffer overflow in Sudo arises when an argument ending in a single backslash is processed in shell mode, so the parser miscomputes the length and writes past the allocated buffer.
    
    \paragraph{Original CWE assignment}
        Original record characterizes the defect as an off-by-one error (CWE-193)~\cite{nvd_cve20213156}.\footnote{CWE-193:\url{https://cwe.mitre.org/data/definitions/193.html}.}
    
    \paragraph{Current CWE assignment}
        Primarily assigned to CWE-122~\cite{nvd_cve20213156}, with CWE-193 retained as a secondary label.\footnote{CWE-122:\url{https://cwe.mitre.org/data/definitions/122.html}.}

\definecolor{cSudoAmberText}{HTML}{78350F} %
\definecolor{cSudoBfText}{HTML}{12365B} %
\definecolor{cSudoCauseBorder}{HTML}{B45309} %
\definecolor{cSudoCauseColor}{HTML}{FEF3C7} %
\definecolor{cSudoConseqBorder}{HTML}{1F4E79} %
\definecolor{cSudoConseqColor}{HTML}{E8F1F8} %
\definecolor{cSudoFailText}{HTML}{7F1D1D} %
\definecolor{cSudoFailureBorder}{HTML}{334155} %
\definecolor{cSudoFailureColor}{HTML}{F1F5F9} %
\definecolor{cSudoFinalErrBorder}{HTML}{991B1B} %
\definecolor{cSudoFinalErrColor}{HTML}{FEE2E2} %
\definecolor{cSudoIndigoText}{HTML}{312E81} %
\definecolor{cSudoOpBorder}{HTML}{3730A3} %
\definecolor{cSudoOpColor}{HTML}{EEF2FF} %
\definecolor{cSudoSlateText}{HTML}{0F172A} %
\begin{inlinefloat}
\begin{tikzpicture}[
    font=\normalsize,
    wbox/.style={rectangle, rounded corners=2pt, text width=0.255\liveW, inner xsep=3pt, inner ysep=3pt,
                 align=center, draw, line width=0.8pt},
    cause/.style  ={fill=cSudoCauseColor,    draw=cSudoCauseBorder,    text=cSudoAmberText},
    op/.style     ={fill=cSudoOpColor,       draw=cSudoOpBorder,       text=cSudoIndigoText},
    conseq/.style ={fill=cSudoConseqColor,   draw=cSudoConseqBorder,   text=cSudoBfText},
    final/.style  ={fill=cSudoFinalErrColor, draw=cSudoFinalErrBorder, text=cSudoFailText},
    failure/.style={fill=cSudoFailureColor,  draw=cSudoFailureBorder,  text=cSudoSlateText},
    clabel/.style ={text=black!62, font=\small, anchor=south west, inner ysep=1pt},
    plabel/.style ={text=black!62, font=\small, fill=white, inner sep=1.2pt},
    arr/.style    ={-{Stealth[scale=1.3]}, semithick, gray},
    farr/.style   ={-{Stealth[scale=1.3]}, semithick, gray},
    darr/.style   ={-{Stealth[scale=0.9]}, dashed, gray, thin}
]
\node[text=black!62, font=\normalsize\bfseries] at ({0.15\liveW}, 0) {CAUSE};
\node[text=black!62, font=\normalsize\bfseries] at ({0.50\liveW}, 0) {OPERATION};
\node[text=black!62, font=\normalsize\bfseries] at ({0.85\liveW}, 0) {CONSEQUENCE};
\draw[gray!35, line width=0.5pt] (0,-0.25) -- (\liveW,-0.25);

\node[wbox, cause]  (c1) at ({0.15\liveW}, -1.20) {\textbf{Erroneous Code} \\ Bug (code defect)};
\node[clabel, yshift=2pt] at (c1.north west) {BF class: DVL};
\node[wbox, op]     (o1) at ({0.50\liveW}, -1.20) {\textbf{Validate} \\ Check escape characters};
\node[wbox, conseq] (e1) at ({0.85\liveW}, -1.20) {\textbf{Invalid Data} \\ Error - propagates};
\draw[arr] (c1.east) -- (o1.west);
\draw[arr] (o1.east) -- (e1.west);

\node[wbox, cause]  (c2) at ({0.15\liveW}, -3.20) {\textbf{Wrong Size} \\ Fault (from W1 error)};
\node[clabel, xshift=-6pt, yshift=2pt] at (c2.north west) {BF class: MAD};
\node[wbox, op]     (o2) at ({0.50\liveW}, -3.20) {\textbf{Reposition} \\ Move pointer by size};
\node[wbox, conseq] (e2) at ({0.85\liveW}, -3.20) {\textbf{Overbound Pointer} \\ Error - propagates};
\draw[arr] (c2.east) -- (o2.west);
\draw[arr] (o2.east) -- (e2.west);

\node[wbox, cause]  (c3) at ({0.15\liveW}, -5.20) {\textbf{Overbound Pointer} \\ Fault (from W2 error)};
\node[clabel, xshift=-6pt, yshift=2pt] at (c3.north west) {BF class: MUS};
\node[wbox, op]     (o3) at ({0.50\liveW}, -5.20) {\textbf{Write} \\ Write via pointer};
\node[wbox, final] (e3) at ({0.85\liveW}, -5.20) {\textbf{Buffer Overflow} \\ Final Error};
\draw[arr] (c3.east) -- (o3.west);
\draw[arr] (o3.east) -- (e3.west);

\draw[darr] (e1.south) -- ++(0,-0.45) -| ([xshift=1.15cm]c2.north);
\node[plabel] at ({0.50\liveW}, -2.20) {Invalid Data $\rightarrow$ Wrong Size};

\draw[darr] (e2.south) -- ++(0,-0.45) -| ([xshift=1.15cm]c3.north);
\node[plabel] at ({0.50\liveW}, -4.20) {Overbound Pointer $\rightarrow$ Overbound Pointer};

\node[wbox, failure, text width=0.42\liveW] (fail) at ({0.50\liveW}, -6.75) {\textbf{Failure: ACE} \\ Arbitrary Code Execution};
\draw[farr] (e3.south) -- ++(0,-0.25) -| (fail.north);

\draw[draw=gray!35, rounded corners=2pt, line width=0.8pt] (0, -7.50) rectangle (\liveW, -9.05);
\node[text=black!62, font=\small\bfseries, anchor=south west] at (0, -7.50) {LEGEND};
\node[rectangle, rounded corners=1pt, fill=cSudoCauseColor, draw=cSudoCauseBorder, minimum width=0.6cm, minimum height=0.30cm, anchor=west] at (0.15, -7.92) {};
\node[anchor=west, font=\small] at (0.85, -7.92) {Cause (Bug or Fault)};
\node[rectangle, rounded corners=1pt, fill=cSudoOpColor, draw=cSudoOpBorder, minimum width=0.6cm, minimum height=0.30cm, anchor=west] at ({0.5\liveW+0.15cm}, -7.92) {};
\node[anchor=west, font=\small] at ({0.5\liveW+0.85cm}, -7.92) {Operation (BF class identifier)};
\node[rectangle, rounded corners=1pt, fill=cSudoConseqColor, draw=cSudoConseqBorder, minimum width=0.6cm, minimum height=0.30cm, anchor=west] at (0.15, -8.32) {};
\node[anchor=west, font=\small] at (0.85, -8.32) {Consequence (Error, propagates)};
\node[rectangle, rounded corners=1pt, fill=cSudoFinalErrColor, draw=cSudoFinalErrBorder, minimum width=0.6cm, minimum height=0.30cm, anchor=west] at ({0.5\liveW+0.15cm}, -8.32) {};
\node[anchor=west, font=\small] at ({0.5\liveW+0.85cm}, -8.32) {Final Error (exploit entry point)};
\draw[darr] (0.15, -8.72) -- ++(0.6, 0);
\node[anchor=west, font=\small] at (0.85, -8.72) {Propagation between weaknesses};
\end{tikzpicture}%

\floatnote[c]{\itshape The annotators' four-axis label (cause Erroneous Code, operation Write, consequence Buffer Overflow, attribute Heap) reads the chain at its sink weakness; both assigned it identically (Section~\ref{sec:scope}).}
\caption{BF specification of the Sudo Baron Samedit vulnerability (CVE-2021-3156).}
\label{fig:sudo_chain}
\end{inlinefloat}

    \paragraph{Analysis of the reassignment}
        The label moved from the root-cause off-by-one to the heap sink as analysis deepened. Because CWE itself chains a root-cause weakness to its sink~\cite{cwe709}, the reassignment is best read as a choice of which end to surface. Hence, both ends remain accurate descriptions of distinct stages of one chain~\cite{nvd_cve20213156}.
    
    \paragraph{CWE failures illustrated}
        It illustrates non-orthogonal hierarchy (F1) as five CWE entries plausibly apply: CWE-119 at class level, CWE-787 and CWE-120 at base level, CWE-122 as a variant of CWE-787, and CWE-193 as root-cause off-by-one. It illustrates sink-only labeling (F3) because the primary assignment, CWE-122, describes the final heap corruption, while upstream parser bug is not expressible by CWE-122 alone. It exemplifies the absence of a causal chain (F4) because the single per-CVE CWE field collapses the chain to one label and offers no device to express that the off-by-one (CWE-193) causes out-of-bounds write (CWE-787) that manifests as heap overflow (CWE-122).
    
    \paragraph{BF classification}
        Two annotators classified this CVE as cause \emph{Erroneous Code}, operation \emph{Write}, consequence \emph{Buffer Overflow}, and attribute \emph{Heap}~\cite{bojanova2024bf}. The upstream stages of the chain map to DVL and MAD before MUS Write sink, and the failure class is ACE, since the heap write yields local privilege escalation to root~\cite{qualys2021baronsamedit,bojanova2024bf} (see Figure~\mbox{\ref{fig:sudo_chain}}).
    
    \paragraph{BF properties that address each illustrated failure}
        For F1, the orthogonal weakness classes of Section~\ref{subsec:bf_addresses_failures} address it. For F3, the cause axis records the erroneous code that the sink-only CWE-122 omits. For F4, the defect is recorded as a chain whose consequence feeds the next cause, which the single CWE cannot encode.

\subsection{CVE-2015-0235 (GHOST), BF\_MEM}\label{ex:ghost}

    \paragraph{CVE summary}
        GHOST is a heap-based buffer overflow in the GNU C library's \texttt{gethostbyname} and \texttt{gethostbyname2} functions~\cite{qualys2015ghost}. A specially crafted hostname triggers an undersized allocation in \texttt{\_\_nss\_hostname\_digits\_dots}: the size computation totals three buffer entities, whereas the subsequent pointer layout prepares four, so the width of one alias pointer is omitted from the allocated size, and the \texttt{strcpy} of the hostname then writes past the buffer~\cite{qualys2015ghost}.

\definecolor{cGhostAmberText}{HTML}{78350F} %
\definecolor{cGhostBfText}{HTML}{12365B} %
\definecolor{cGhostCauseBorder}{HTML}{B45309} %
\definecolor{cGhostCauseColor}{HTML}{FEF3C7} %
\definecolor{cGhostConseqBorder}{HTML}{1F4E79} %
\definecolor{cGhostConseqColor}{HTML}{E8F1F8} %
\definecolor{cGhostFailText}{HTML}{7F1D1D} %
\definecolor{cGhostFailureBorder}{HTML}{334155} %
\definecolor{cGhostFailureColor}{HTML}{F1F5F9} %
\definecolor{cGhostFinalErrBorder}{HTML}{991B1B} %
\definecolor{cGhostFinalErrColor}{HTML}{FEE2E2} %
\definecolor{cGhostIndigoText}{HTML}{312E81} %
\definecolor{cGhostOpBorder}{HTML}{3730A3} %
\definecolor{cGhostOpColor}{HTML}{EEF2FF} %
\definecolor{cGhostSlateText}{HTML}{0F172A} %
\begin{inlinefloat}
\begin{tikzpicture}[
    font=\normalsize,
    wbox/.style={rectangle, rounded corners=2pt, text width=0.255\liveW, inner xsep=3pt, inner ysep=3pt,
                 align=center, draw, line width=0.8pt},
    cause/.style  ={fill=cGhostCauseColor,    draw=cGhostCauseBorder,    text=cGhostAmberText},
    op/.style     ={fill=cGhostOpColor,       draw=cGhostOpBorder,       text=cGhostIndigoText},
    conseq/.style ={fill=cGhostConseqColor,   draw=cGhostConseqBorder,   text=cGhostBfText},
    final/.style  ={fill=cGhostFinalErrColor, draw=cGhostFinalErrBorder, text=cGhostFailText},
    failure/.style={fill=cGhostFailureColor,  draw=cGhostFailureBorder,  text=cGhostSlateText},
    clabel/.style ={text=black!62, font=\small, anchor=south west, inner ysep=1pt},
    plabel/.style ={text=black!62, font=\small, fill=white, inner sep=1.2pt},
    arr/.style    ={-{Stealth[scale=1.3]}, semithick, gray},
    farr/.style   ={-{Stealth[scale=1.3]}, semithick, gray},
    darr/.style   ={-{Stealth[scale=0.9]}, dashed, gray, thin}
]
\node[text=black!62, font=\normalsize\bfseries] at ({0.15\liveW}, 0) {CAUSE};
\node[text=black!62, font=\normalsize\bfseries] at ({0.50\liveW}, 0) {OPERATION};
\node[text=black!62, font=\normalsize\bfseries] at ({0.85\liveW}, 0) {CONSEQUENCE};
\draw[gray!35, line width=0.5pt] (0,-0.25) -- (\liveW,-0.25);

\node[wbox, cause]  (c1) at ({0.15\liveW}, -1.20) {\textbf{Wrong Size} \\ Fault (computed size)};
\node[clabel] at (c1.north west) {BF class: MMN};
\node[wbox, op]     (o1) at ({0.50\liveW}, -1.20) {\textbf{Allocate} \\ Allocate with wrong size};
\node[wbox, conseq] (e1) at ({0.85\liveW}, -1.20) {\textbf{Insufficient Size} \\ Error - propagates};
\draw[arr] (c1.east) -- (o1.west);
\draw[arr] (o1.east) -- (e1.west);

\node[wbox, cause]  (c2) at ({0.15\liveW}, -3.20) {\textbf{Insufficient Size} \\ Fault (from W1 error)};
\node[clabel] at (c2.north west) {BF class: MUS};
\node[wbox, op]     (o2) at ({0.50\liveW}, -3.20) {\textbf{Write} \\ Write past end};
\node[wbox, final] (e2) at ({0.85\liveW}, -3.20) {\textbf{Buffer Overflow} \\ Final Error};
\draw[arr] (c2.east) -- (o2.west);
\draw[arr] (o2.east) -- (e2.west);

\draw[darr] (e1.south) -- ++(0,-0.45) -| ([xshift=1.15cm]c2.north);
\node[plabel] at ({0.50\liveW}, -2.20) {Insufficient Size $\rightarrow$ Insufficient Size};

\node[wbox, failure, text width=0.42\liveW] (fail) at ({0.50\liveW}, -4.75) {\textbf{Failure: ACE} \\ arbitrary code execution via the heap overflow};
\draw[farr] (e2.south) -- ++(0,-0.25) -| (fail.north);

\draw[draw=gray!35, rounded corners=2pt, line width=0.8pt] (0, -5.50) rectangle (\liveW, -7.05);
\node[text=black!62, font=\small\bfseries, anchor=south west] at (0, -5.50) {LEGEND};
\node[rectangle, rounded corners=1pt, fill=cGhostCauseColor, draw=cGhostCauseBorder, minimum width=0.6cm, minimum height=0.30cm, anchor=west] at (0.15, -5.92) {};
\node[anchor=west, font=\small] at (0.85, -5.92) {Cause (Bug or Fault)};
\node[rectangle, rounded corners=1pt, fill=cGhostOpColor, draw=cGhostOpBorder, minimum width=0.6cm, minimum height=0.30cm, anchor=west] at ({0.5\liveW+0.15cm}, -5.92) {};
\node[anchor=west, font=\small] at ({0.5\liveW+0.85cm}, -5.92) {Operation (BF class identifier)};
\node[rectangle, rounded corners=1pt, fill=cGhostConseqColor, draw=cGhostConseqBorder, minimum width=0.6cm, minimum height=0.30cm, anchor=west] at (0.15, -6.32) {};
\node[anchor=west, font=\small] at (0.85, -6.32) {Consequence (Error, propagates)};
\node[rectangle, rounded corners=1pt, fill=cGhostFinalErrColor, draw=cGhostFinalErrBorder, minimum width=0.6cm, minimum height=0.30cm, anchor=west] at ({0.5\liveW+0.15cm}, -6.32) {};
\node[anchor=west, font=\small] at ({0.5\liveW+0.85cm}, -6.32) {Final Error (exploit entry point)};
\draw[darr] (0.15, -6.72) -- ++(0.6, 0);
\node[anchor=west, font=\small] at (0.85, -6.72) {Propagation between weaknesses};
\end{tikzpicture}%

\floatnote[c]{\itshape From a wrong-size allocation to a write past the allocated buffer; the failure class is ACE. Reference stratum: axis values follow the BF specification.}
\caption{BF specification of the GHOST glibc vulnerability (CVE-2015-0235) as a chain of weaknesses.}
\label{fig:ghost_chain}
\end{inlinefloat}

    \paragraph{Original CWE assignment}
        The NVD record originally labeled CVE-2015-0235 with the Class-level entry CWE-119\mbox{~\cite{cwe119}}, Improper Restriction of Operations within the Bounds of a Memory Buffer.\footnote{CWE-119: \url{https://cwe.mitre.org/data/definitions/119.html}.}
    
    \paragraph{Current CWE assignment}
        The label was later reassigned to CWE-787, Out-of-bounds Write\mbox{~\cite{nvd_cve20150235}},\footnote{CWE-787: \url{https://cwe.mitre.org/data/definitions/787.html}.} with CWE-122 as the variant-level descendant; CWE-787 is a Base and CWE-119 a Class~\cite{cwe122,cwe787,cwe119}.
    
    \paragraph{Analysis of the reassignment}
        The available labels record the sink, not the upstream cause. The root cause of the defect is that the allocator allocates an insufficient buffer size; the out-of-bounds write is the manifestation, not the bug.
    
    \paragraph{CWE failures illustrated}
        This CVE shows non-orthogonal hierarchy (F1) through the co-applicability of CWE-119, CWE-787, and CWE-122 to the same defect; it exemplifies sink-only labeling (F3) because the allocation-stage cause, the undersized buffer, is not expressed by any of the CWEs actually assigned to this CVE. It illustrates the absence of a causal chain (F4) because the allocation cause and the use sink are two distinct stages of a chain that the CWE assignment collapses into one label.
    
    \paragraph{BF classification}
        The BF specification represents this case through the memory bugs model~\cite{bojanova_mem_bugs}. The first BF state is in the MMN class as a $\langle$\emph{Wrong Size}, \emph{Allocate}$\rangle$ $\rightarrow$ \emph{Insufficient Size} weakness. The second BF state is in the MUS class as a $\langle$\emph{Insufficient Size}, \emph{Write}$\rangle$ $\rightarrow$ \emph{Buffer Overflow} weakness, with the attribute on the Heap address state. Figure~\mbox{\ref{fig:ghost_chain}} depicts the chain, with the allocation defect upstream of the write.
    
    \paragraph{BF properties that address each illustrated failure}
        For F1, orthogonal weakness classes address the failure as before. For F3, the MMN Allocate operation and its Wrong Size cause record the allocation-stage bug. For F4, the cause-operation-consequence chain represents the allocate-then-write sequence as two linked weaknesses rather than a single sink label.

\subsection{CVE-2021-21834 (GPAC), BF\_DAT $\rightarrow$ BF\_MEM (BadAlloc pattern)}\label{ex:gpac}

    \paragraph{CVE summary}
        CVE-2021-21834 is an integer overflow that leads to a heap-based buffer overflow in the GPAC multimedia framework. A crafted input causes a size computation to wrap around and produce a value much smaller than the intended buffer size; the write overruns the allocation~\cite{talos2021gpac}. The vulnerability follows the BadAlloc pattern~\cite{bojanova2024bf}.
    
    \paragraph{Original CWE assignment}
        The defect has been associated with CWE-190, and with CWE-787~\cite{cwe190,cwe787},\footnote{CWE-190:
        \url{https://cwe.mitre.org/data/definitions/190.html}; CWE-787:
        \url{https://cwe.mitre.org/data/definitions/787.html}.} and with the
        broader CWE-119.
    
    \paragraph{Current CWE assignment}
        CWE-190 names the integer overflow stage; CWE-787 names the out-of-bounds write stage; CWE-119 sits at the class level above both.

    \paragraph{Analysis of the reassignment}
        NIST SP 800-231~\cite{bojanova2024bf} makes the same point that each label reads one stage of one chain, for the structurally analogous CVE-2018-5907, where the entire chain is CWE-20 $\rightarrow$ CWE-190 $\rightarrow$ CWE-119.
\definecolor{cGpacAmberText}{HTML}{78350F} %
\definecolor{cGpacBfText}{HTML}{12365B} %
\definecolor{cGpacCauseBorder}{HTML}{B45309} %
\definecolor{cGpacCauseColor}{HTML}{FEF3C7} %
\definecolor{cGpacConseqBorder}{HTML}{1F4E79} %
\definecolor{cGpacConseqColor}{HTML}{E8F1F8} %
\definecolor{cGpacFailText}{HTML}{7F1D1D} %
\definecolor{cGpacFailureBorder}{HTML}{334155} %
\definecolor{cGpacFailureColor}{HTML}{F1F5F9} %
\definecolor{cGpacFinalErrBorder}{HTML}{991B1B} %
\definecolor{cGpacFinalErrColor}{HTML}{FEE2E2} %
\definecolor{cGpacIndigoText}{HTML}{312E81} %
\definecolor{cGpacOpBorder}{HTML}{3730A3} %
\definecolor{cGpacOpColor}{HTML}{EEF2FF} %
\definecolor{cGpacSlateText}{HTML}{0F172A} %
\begin{inlinefloat}
\begin{tikzpicture}[
    font=\normalsize,
    wbox/.style={rectangle, rounded corners=2pt, text width=0.255\liveW, inner xsep=3pt, inner ysep=3pt,
                 align=center, draw, line width=0.8pt},
    cause/.style  ={fill=cGpacCauseColor,    draw=cGpacCauseBorder,    text=cGpacAmberText},
    op/.style     ={fill=cGpacOpColor,       draw=cGpacOpBorder,       text=cGpacIndigoText},
    conseq/.style ={fill=cGpacConseqColor,   draw=cGpacConseqBorder,   text=cGpacBfText},
    final/.style  ={fill=cGpacFinalErrColor, draw=cGpacFinalErrBorder, text=cGpacFailText},
    failure/.style={fill=cGpacFailureColor,  draw=cGpacFailureBorder,  text=cGpacSlateText},
    clabel/.style ={text=black!62, font=\small, anchor=south west, inner ysep=1pt},
    plabel/.style ={text=black!62, font=\small, fill=white, inner sep=1.2pt},
    arr/.style    ={-{Stealth[scale=1.3]}, semithick, gray},
    farr/.style   ={-{Stealth[scale=1.3]}, semithick, gray},
    darr/.style   ={-{Stealth[scale=0.9]}, dashed, gray, thin}
]
\node[text=black!62, font=\normalsize\bfseries] at ({0.15\liveW}, 0) {CAUSE};
\node[text=black!62, font=\normalsize\bfseries] at ({0.50\liveW}, 0) {OPERATION};
\node[text=black!62, font=\normalsize\bfseries] at ({0.85\liveW}, 0) {CONSEQUENCE};
\draw[gray!35, line width=0.5pt] (0,-0.25) -- (\liveW,-0.25);

\node[wbox, cause]  (c1) at ({0.15\liveW}, -1.20) {\textbf{Missing/Erroneous Code} \\ Bug (absent size check)};
\node[clabel] at (c1.north west) {BF class: DVR};
\node[wbox, op]     (o1) at ({0.50\liveW}, -1.20) {\textbf{Verify} \\ size input check};
\node[wbox, conseq] (e1) at ({0.85\liveW}, -1.20) {\textbf{Inconsistent Value} \\ Error - propagates};
\draw[arr] (c1.east) -- (o1.west);
\draw[arr] (o1.east) -- (e1.west);

\node[wbox, cause]  (c2) at ({0.15\liveW}, -3.20) {\textbf{Wrong Argument} \\ Fault (from W1 error)};
\node[clabel] at (c2.north west) {BF class: TCM};
\node[wbox, op]     (o2) at ({0.50\liveW}, -3.20) {\textbf{Calculate} \\ co64 atom size calc};
\node[wbox, conseq] (e2) at ({0.85\liveW}, -3.20) {\textbf{Wrap Around} \\ Error - propagates};
\draw[arr] (c2.east) -- (o2.west);
\draw[arr] (o2.east) -- (e2.west);

\node[wbox, cause]  (c3) at ({0.15\liveW}, -5.20) {\textbf{Wrong Size} \\ Fault (from W2 error)};
\node[clabel] at (c3.north west) {BF class: MMN};
\node[wbox, op]     (o3) at ({0.50\liveW}, -5.20) {\textbf{Allocate} \\ under-sized buffer};
\node[wbox, conseq] (e3) at ({0.85\liveW}, -5.20) {\textbf{Insufficient Size} \\ Error - propagates};
\draw[arr] (c3.east) -- (o3.west);
\draw[arr] (o3.east) -- (e3.west);

\node[wbox, cause]  (c4) at ({0.15\liveW}, -7.20) {\textbf{Insufficient Size} \\ Fault (from W3 error)};
\node[clabel] at (c4.north west) {BF class: MAD};
\node[wbox, op]     (o4) at ({0.50\liveW}, -7.20) {\textbf{Reposition} \\ pointer past bound};
\node[wbox, conseq] (e4) at ({0.85\liveW}, -7.20) {\textbf{Overbound Pointer} \\ Error - propagates};
\draw[arr] (c4.east) -- (o4.west);
\draw[arr] (o4.east) -- (e4.west);

\node[wbox, cause]  (c5) at ({0.15\liveW}, -9.20) {\textbf{Overbound Pointer} \\ Fault (from W4 error)};
\node[clabel] at (c5.north west) {BF class: MUS};
\node[wbox, op]     (o5) at ({0.50\liveW}, -9.20) {\textbf{Write} \\ write past end};
\node[wbox, final] (e5) at ({0.85\liveW}, -9.20) {\textbf{Buffer Overflow} \\ Final Error};
\draw[arr] (c5.east) -- (o5.west);
\draw[arr] (o5.east) -- (e5.west);

\draw[darr] (e1.south) -- ++(0,-0.45) -| ([xshift=1.15cm]c2.north);
\node[plabel] at ({0.50\liveW}, -2.20) {Inconsistent Value $\rightarrow$ Wrong Argument};

\draw[darr] (e2.south) -- ++(0,-0.45) -| ([xshift=1.15cm]c3.north);
\node[plabel] at ({0.50\liveW}, -4.20) {Wrap Around $\rightarrow$ Wrong Size};

\draw[darr] (e3.south) -- ++(0,-0.45) -| ([xshift=1.15cm]c4.north);
\node[plabel] at ({0.50\liveW}, -6.20) {Insufficient Size $\rightarrow$ Insufficient Size};

\draw[darr] (e4.south) -- ++(0,-0.45) -| ([xshift=1.15cm]c5.north);
\node[plabel] at ({0.50\liveW}, -8.20) {Overbound Pointer $\rightarrow$ Overbound Pointer};

\node[wbox, failure, text width=0.42\liveW] (fail) at ({0.50\liveW}, -10.75) {\textbf{Failure: ACE} \\ remote code execution (RCE)};
\draw[farr] (e5.south) -- ++(0,-0.25) -| (fail.north);

\draw[draw=gray!35, rounded corners=2pt, line width=0.8pt] (0, -11.50) rectangle (\liveW, -13.05);
\node[text=black!62, font=\small\bfseries, anchor=south west] at (0, -11.50) {LEGEND};
\node[rectangle, rounded corners=1pt, fill=cGpacCauseColor, draw=cGpacCauseBorder, minimum width=0.6cm, minimum height=0.30cm, anchor=west] at (0.15, -11.92) {};
\node[anchor=west, font=\small] at (0.85, -11.92) {Cause (Bug or Fault)};
\node[rectangle, rounded corners=1pt, fill=cGpacOpColor, draw=cGpacOpBorder, minimum width=0.6cm, minimum height=0.30cm, anchor=west] at ({0.5\liveW+0.15cm}, -11.92) {};
\node[anchor=west, font=\small] at ({0.5\liveW+0.85cm}, -11.92) {Operation (BF class identifier)};
\node[rectangle, rounded corners=1pt, fill=cGpacConseqColor, draw=cGpacConseqBorder, minimum width=0.6cm, minimum height=0.30cm, anchor=west] at (0.15, -12.32) {};
\node[anchor=west, font=\small] at (0.85, -12.32) {Consequence (Error, propagates)};
\node[rectangle, rounded corners=1pt, fill=cGpacFinalErrColor, draw=cGpacFinalErrBorder, minimum width=0.6cm, minimum height=0.30cm, anchor=west] at ({0.5\liveW+0.15cm}, -12.32) {};
\node[anchor=west, font=\small] at ({0.5\liveW+0.85cm}, -12.32) {Final Error (exploit entry point)};
\draw[darr] (0.15, -12.72) -- ++(0.6, 0);
\node[anchor=west, font=\small] at (0.85, -12.72) {Propagation between weaknesses};
\end{tikzpicture}%

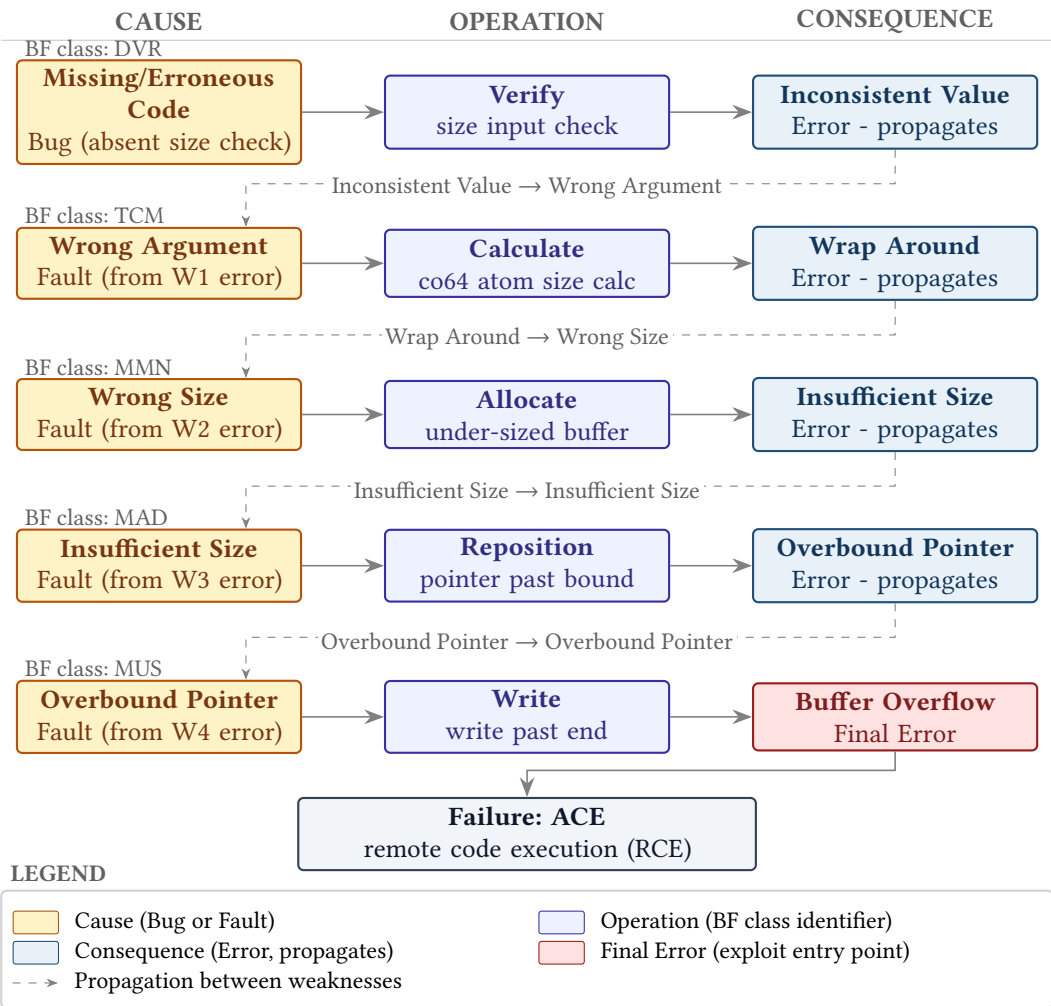
\floatnote[c]{\itshape The five-stage BadAlloc pattern: an unverified size input wraps in the \texttt{co64} size arithmetic, under-allocates the buffer, repositions the pointer past its bound, and the write overflows; the failure class is ACE. Reference stratum: axis values follow the BF specification.}
\caption{BF specification of the GPAC MPEG-4 vulnerability (CVE-2021-21834) as a BadAlloc-pattern chain.}
\label{fig:gpac_chain}
\end{inlinefloat}
      
    \paragraph{CWE failures illustrated}
        This CVE exemplifies the absence of a causal chain (F4): the chain spans five BF weakness stages (DVR, TCM, MMN, MAD, MUS) and a failure stage (ACE / RCE), and no single CWE entry can compose them. It also shows the non-orthogonal hierarchy (F1) through the co-applicability of CWE-119, CWE-190, and CWE-787, as well as sink-only labeling (F3).
    
    \paragraph{BF classification}
        NIST SP 800-231~\cite{bojanova2024bf} specifies the BadAlloc pattern chain explicitly as DVR $\curvearrowright$ TCM $\curvearrowright$ MMN $\curvearrowright$ MAD $\curvearrowright$ MUS. Reading CVE-2021-21834 against this template, the first weakness is a $\langle$\emph{Missing/Erroneous Code}, \emph{Verify}$\rangle$ $\rightarrow$ \emph{Inconsistent Value} weakness in DVR; the second is a $\langle$\emph{Wrong Argument}, \emph{Calculate}$\rangle$ $\rightarrow$ \emph{Wrap Around} weakness in TCM; the third is a $\langle$\emph{Wrong Size}, \emph{Allocate}$\rangle$ $\rightarrow$ \emph{Insufficient Size} weakness in MMN; the fourth is a $\langle$\emph{Insufficient Size}, \emph{Reposition}$\rangle$ $\rightarrow$ \emph{Overbound Pointer} weakness in MAD; and the fifth is a $\langle$\emph{Overbound Pointer}, \emph{Write}$\rangle$ $\rightarrow$ \emph{Buffer Overflow} weakness in MUS. Figure~\mbox{\ref{fig:gpac_chain}} depicts the five stages.

    \paragraph{BF properties that address each illustrated failure}
        For F1, orthogonal weakness classes map each stage to a single class. For F3, the explicit cause axis records the \emph{Missing/Erroneous Code} bug at the head of the chain. For F4, the cause-consequence chain composes the five weaknesses into one representation: \emph{Inconsistent Value} $\curvearrowright$ \emph{Wrong Argument}; \emph{Wrap Around} $\curvearrowright$ \emph{Wrong Size}; \emph{Insufficient Size} $\curvearrowright$ \emph{Reposition fault}; \emph{Overbound Pointer} $\curvearrowright$ \emph{Write fault}.

\subsection{CVE-2014-0160 (Heartbleed), BF\_INP $\rightarrow$ BF\_MEM, with convergence}\label{ex:heartbleed}

    \paragraph{CVE summary}
        Heartbleed (CVE-2014-0160) was a high-severity information-disclosure vulnerability in the TLS and DTLS heartbeat extension of the OpenSSL cryptographic library, with a CVSS v3.1 base score of 7.5~\cite{nvd_cve20140160}. In OpenSSL 1.0.1 through 1.0.1f, the heartbeat handler omits a bounds check before a \texttt{memcpy()} whose length is taken from an attacker-controlled payload-length field~\cite{owasp_heartbleed,durumeric2014matter}. With each request, the responder discloses up to 64KB of memory through buffer over-reads~\cite{nvd_cve20140160,cisa2014heartbleed,bojanova2024bf}.
    
    \paragraph{Original CWE assignment}
        The NVD record originally labeled CVE-2014-0160 with the Class-level entry CWE-119\mbox{~\cite{cwe119}}, Improper Restriction of Operations within the Bounds of a Memory Buffer.\footnote{CWE-119: \url{https://cwe.mitre.org/data/definitions/119.html}.}

\definecolor{cHeartbleedAmberText}{HTML}{78350F} %
\definecolor{cHeartbleedBfText}{HTML}{12365B} %
\definecolor{cHeartbleedCauseBorder}{HTML}{B45309} %
\definecolor{cHeartbleedCauseColor}{HTML}{FEF3C7} %
\definecolor{cHeartbleedConseqBorder}{HTML}{1F4E79} %
\definecolor{cHeartbleedConseqColor}{HTML}{E8F1F8} %
\definecolor{cHeartbleedFailText}{HTML}{7F1D1D} %
\definecolor{cHeartbleedFailureBorder}{HTML}{334155} %
\definecolor{cHeartbleedFailureColor}{HTML}{F1F5F9} %
\definecolor{cHeartbleedFinalErrBorder}{HTML}{991B1B} %
\definecolor{cHeartbleedFinalErrColor}{HTML}{FEE2E2} %
\definecolor{cHeartbleedIndigoText}{HTML}{312E81} %
\definecolor{cHeartbleedOpBorder}{HTML}{3730A3} %
\definecolor{cHeartbleedOpColor}{HTML}{EEF2FF} %
\definecolor{cHeartbleedSlateText}{HTML}{0F172A} %
\begin{inlinefloat}
\begin{tikzpicture}[
    font=\normalsize,
    wbox/.style={rectangle, rounded corners=2pt, text width=0.255\liveW, inner xsep=3pt, inner ysep=3pt,
                 align=center, draw, line width=0.8pt},
    cause/.style  ={fill=cHeartbleedCauseColor,    draw=cHeartbleedCauseBorder,    text=cHeartbleedAmberText},
    op/.style     ={fill=cHeartbleedOpColor,       draw=cHeartbleedOpBorder,       text=cHeartbleedIndigoText},
    conseq/.style ={fill=cHeartbleedConseqColor,   draw=cHeartbleedConseqBorder,   text=cHeartbleedBfText},
    final/.style  ={fill=cHeartbleedFinalErrColor, draw=cHeartbleedFinalErrBorder, text=cHeartbleedFailText},
    failure/.style={fill=cHeartbleedFailureColor,  draw=cHeartbleedFailureBorder,  text=cHeartbleedSlateText},
    clabel/.style ={text=gray, font=\small, anchor=south west, inner ysep=1pt},
    plabel/.style ={text=gray, font=\small, fill=white, inner sep=1.2pt},
    arr/.style    ={-{Stealth[scale=1.3]}, semithick, gray},
    farr/.style   ={-{Stealth[scale=1.3]}, semithick, gray},
    darr/.style   ={-{Stealth[scale=0.9]}, dashed, gray, thin}
]
\node[text=gray, font=\normalsize\bfseries] at ({0.15\liveW}, 0) {CAUSE};
\node[text=gray, font=\normalsize\bfseries] at ({0.50\liveW}, 0) {OPERATION};
\node[text=gray, font=\normalsize\bfseries] at ({0.85\liveW}, 0) {CONSEQUENCE};
\draw[gray!35, line width=0.5pt] (0,-0.25) -- (\liveW,-0.25);

\node[wbox, cause]  (c1) at ({0.15\liveW}, -1.20) {\textbf{Missing Code} \\ Bug (code defect)};
\node[clabel] at (c1.north west) {BF class: DVR};
\node[wbox, op]     (o1) at ({0.50\liveW}, -1.20) {\textbf{Verify} \\ Check payload length};
\node[wbox, conseq] (e1) at ({0.85\liveW}, -1.20) {\textbf{Inconsistent Value} \\ Error - propagates};
\draw[arr] (c1.east) -- (o1.west);
\draw[arr] (o1.east) -- (e1.west);

\node[wbox, cause]  (c2) at ({0.15\liveW}, -3.20) {\textbf{Wrong Size} \\ Fault (from W1 error)};
\node[clabel] at (c2.north west) {BF class: MAD};
\node[wbox, op]     (o2) at ({0.50\liveW}, -3.20) {\textbf{Reposition} \\ Move pointer by size};
\node[wbox, conseq] (e2) at ({0.85\liveW}, -3.20) {\textbf{Overbound Pointer} \\ Error - propagates};
\draw[arr] (c2.east) -- (o2.west);
\draw[arr] (o2.east) -- (e2.west);

\node[wbox, cause]  (c3) at ({0.15\liveW}, -5.20) {\textbf{Overbound Pointer} \\ Fault (from W2 error)};
\node[clabel] at (c3.north west) {BF class: MUS};
\node[wbox, op]     (o3) at ({0.50\liveW}, -5.20) {\textbf{Read} \\ Read via pointer};
\node[wbox, final] (e3) at ({0.85\liveW}, -5.20) {\textbf{Buffer Over-Read} \\ Final Error};
\draw[arr] (c3.east) -- (o3.west);
\draw[arr] (o3.east) -- (e3.west);

\draw[darr] (e1.south) -- ++(0,-0.45) -| ([xshift=1.15cm]c2.north);
\node[plabel] at ({0.50\liveW}, -2.20) {Inconsistent Value $\rightarrow$ Wrong Size};

\draw[darr] (e2.south) -- ++(0,-0.45) -| ([xshift=1.15cm]c3.north);
\node[plabel] at ({0.50\liveW}, -4.20) {Overbound Pointer $\rightarrow$ Overbound Pointer};

\node[wbox, failure, text width=0.42\liveW] (fail) at ({0.50\liveW}, -6.75) {\textbf{Failure: IEX} \\ Information Exposure};
\draw[farr] (e3.south) -- ++(0,-0.25) -| (fail.north);

\draw[draw=gray!35, rounded corners=2pt, line width=0.8pt] (0, -7.50) rectangle (\liveW, -9.05);
\node[text=gray, font=\small\bfseries, anchor=south west] at (0, -7.50) {LEGEND};
\node[rectangle, rounded corners=1pt, fill=cHeartbleedCauseColor, draw=cHeartbleedCauseBorder, minimum width=0.6cm, minimum height=0.30cm, anchor=west] at (0.15, -7.92) {};
\node[anchor=west, font=\small] at (0.85, -7.92) {Cause (Bug or Fault)};
\node[rectangle, rounded corners=1pt, fill=cHeartbleedOpColor, draw=cHeartbleedOpBorder, minimum width=0.6cm, minimum height=0.30cm, anchor=west] at ({0.5\liveW+0.15cm}, -7.92) {};
\node[anchor=west, font=\small] at ({0.5\liveW+0.85cm}, -7.92) {Operation (BF class identifier)};
\node[rectangle, rounded corners=1pt, fill=cHeartbleedConseqColor, draw=cHeartbleedConseqBorder, minimum width=0.6cm, minimum height=0.30cm, anchor=west] at (0.15, -8.32) {};
\node[anchor=west, font=\small] at (0.85, -8.32) {Consequence (Error, propagates)};
\node[rectangle, rounded corners=1pt, fill=cHeartbleedFinalErrColor, draw=cHeartbleedFinalErrBorder, minimum width=0.6cm, minimum height=0.30cm, anchor=west] at ({0.5\liveW+0.15cm}, -8.32) {};
\node[anchor=west, font=\small] at ({0.5\liveW+0.85cm}, -8.32) {Final Error (exploit entry point)};
\draw[darr] (0.15, -8.72) -- ++(0.6, 0);
\node[anchor=west, font=\small] at (0.85, -8.72) {Propagation between weaknesses};
\end{tikzpicture}%

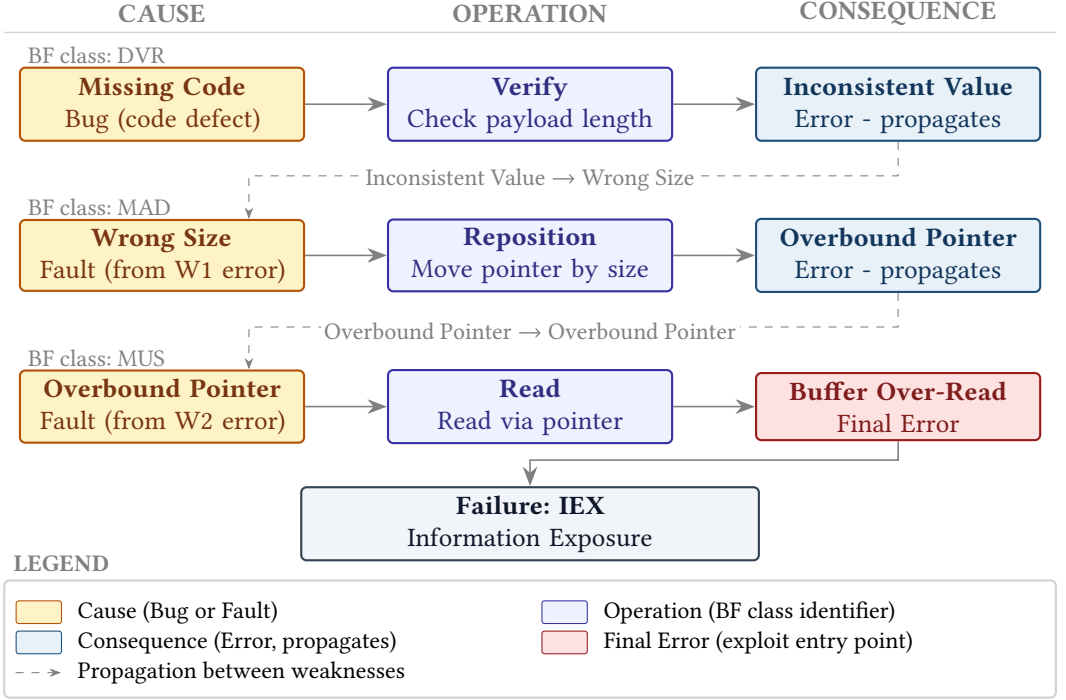
\caption{BF formal specification of the Heartbleed vulnerability (CVE-2014-0160) as a chain of weaknesses.}
\label{fig:heartbleed_chain}
\end{inlinefloat}

    \paragraph{Current CWE assignment}
        The label was later narrowed to CWE-125, Out-of-bounds Read\mbox{~\cite{cwe125}}, with\footnote{CWE-125: \url{https://cwe.mitre.org/data/definitions/125.html}.} related labels CWE-126 (Buffer Over-read) and CWE-20 (Improper Input Validation) sometimes referenced as upstream causes.
    
    \paragraph{Analysis of the reassignment}
        NIST SP 800-231 records the issue plainly: ``CVE-2014-0160 Heartbleed lists the final error at the sink, buffer over-read, as the root cause, while it is missing input verification that leads to pointer reposition over the upper bound and then to buffer over-read''~\cite{bojanova2024bf}.
    
    \paragraph{CWE failures illustrated}
        This CVE illustrates sink-only labeling (F3): the assigned CWE describes the over-read manifestation, not the missing input verification that produced it. It illustrates the absence of a causal chain (F4) because the defect spans three weaknesses in the main chain and a second converging chain (MUS Clear) that the single CWE-125 cannot represent. It exemplifies non-orthogonal hierarchy (F1) through the co-applicability of CWE-125, CWE-126, and CWE-20 to different stages of the same defect.
    
    \paragraph{BF classification}
        NIST SP 800-231~\cite{bojanova2024bf} specifies the BF chain for Heartbleed. The main chain comprises three weaknesses:
        
        \begin{itemize}[leftmargin=*,itemsep=0pt,topsep=2pt]
          \item $\langle$\emph{Missing Code}, \emph{Verify}$\rangle$ $\rightarrow$
                \emph{Inconsistent Value} in DVR;
          \item $\langle$\emph{Wrong Size}, \emph{Reposition}$\rangle$ $\rightarrow$
                \emph{Overbound Pointer} in MAD;
          \item $\langle$\emph{Overbound Pointer}, \emph{Read}$\rangle$ $\rightarrow$
                \emph{Buffer Over-Read} in MUS.
        \end{itemize}
        A converging chain in MUS records the parallel weakness $\langle$\emph{Missing Code}, \emph{Clear}$\rangle$ $\rightarrow$ \emph{Not Cleared Object}. The two chains converge at the \emph{IEX} failure: ``Either the missing Verify bug or the missing Clear bug has to be fixed to avoid this security failure''~\cite{bojanova2024bf}.
    
    \paragraph{BF properties that address each illustrated failure}
        For F3, the explicit cause axis records the \emph{Missing Code} bug at the Verify operation. For F4, the cause-consequence chain combines the three main-chain weaknesses and the converging MUS Clear weakness into a single specification, without collapsing them into CWE-125. For F1, the disjoint operation sets across DVR, MAD, MUS, and the IEX failure class prevent freedom of ancestor-or-sibling labels.
\FloatBarrier
\section{Empirical Evaluation}\label{sec:evaluation}

Study~1 measures the reproducibility of manual CVE-to-BF mapping. Study~2 measures the determinism of automated CVE-to-BF mapping. Both studies report reproducibility, not human-validated accuracy.

    \subsection{Study 1: Inter-Rater Reliability of Manual Mapping}\label{subsec:study1}

    \paragraph{Design}
        Two annotators independently and anonymously mapped \wtIrrCVEs\ CVEs onto the four BF axes of cause, operation, consequence, and attribute. Agreement is quantified per axis with Cohen's $\kappa$~\cite{cohen1960kappa} and interpreted against the Landis-Koch bands~\cite{landis1977observer}. The statistic measures chance-corrected reproducibility of the mapping procedure.
    
    \paragraph{Results}
        Cause and operation are reproducible at almost-perfect agreement, with $\kappa=\wtKappaCause$ and $\kappa=\wtKappaOperation$ respectively; consequence reaches substantial agreement at $\kappa=\wtKappaConsequence$; attribute reaches fair agreement at $\kappa=\wtKappaAttribute$; and the pooled statistic over all \wtIrrCells\ cells is $\kappa=\wtKappaPooled$. The pattern supports the orthogonality argument on the axes where the specification constrains the choice most tightly, and it locates the principal limitation on the attribute axis.
    
    \subsection{Study 2: Automated Determinism of CVE-to-BF Mapping}\label{subsec:study2}
    
        The study is implemented using an open-source, reproducible FastAPI application for real-time agentic inference~\cite{mitra2026agentic}. Because a determinism figure measured on one deployment cannot separate what the task contributes from what the model and its serving stack contribute, the study executes the same generation design as \wtArmCount\ \emph{combinations}: each combination is one model deployment at its own prompt budget.
    
    \paragraph{Experiment Setup and Configuration}
        Two open-weight deployments participate: \gemmaModel{} (the primary deployment) and \gptossModel{}, each invoked through the vLLM OpenAI-compatible endpoint. Every model call uses temperature 0, and a single prompt-budget regime completes the design. Under the \emph{per-model} budget (combinations \texttt{individual-\allowbreak max-\allowbreak capacity}), the budget is derived from each deployment's own context window, so each deployment reads as much evidence as it can hold and neither is held to the other's ceiling. Evidence follows a clean-evidence policy: the model receives only the NVD advisory text and the fix-commit patch material.
        
\subsection{The Formal Derivation Procedure}\label{subsec:derivation_algorithm}

    The evidence-based generation of Study~2 instantiates one underlying procedure, which Algorithm~\ref{alg:bf_derivation} states in the notation of Table~\ref{tab:notation} and Figure~\ref{fig:bf_derivation_architecture} renders as an architectural process. The procedure operationalizes backward bug identification as defined in SP 800-231: the walk starts at the observable failure, fills one weakness triple per link, and terminates when a cause resolves to a bug rather than a fault~\cite{bojanova2024bf}. The closed tables $T_A$ to $T_H$ are transcriptions of the vocabulary tables of SP 800-231, that is the class types and member classes, the operations per class, the bug values, the fault and error values with their type dimension, the final-error values with the failure classes, and the operation and operand attribute values; the transcription adds nothing and removes nothing, so every value the algorithm can generate already exists in the standard. A fixed sequence of five lookups then derives each link. The class assignment is the deterministic map $\kappa$, because the closed value sets assign each error value to exactly one producing class; the operation, cause, and attribute assignments are semantic-inference selections; and the stopping test at line 12 is a set membership test against the bug table $T_C$, because a fault is a good operation over a bad operand and so always points one link further back. In contrast, a bug is the improper operation itself and therefore ends the chain. When the promoted fault belongs to a different class type than the current link, the algorithm records a class-type crossing, which SP 800-231 permits as a meaningful value propagation between class types~\cite{bojanova2024bf}.

\definecolor{cAlgHeaderBg}{HTML}{B7D3E8} %
\definecolor{cAlgHeaderFg}{HTML}{12365B} %
\definecolor{cAlgHeaderRule}{HTML}{1F4E79} %
\definecolor{cAlgSubHeaderBg}{HTML}{E8F1F8} %
\begin{algorithm}[!htb]
\colorlet{tblHeaderFg}{cAlgHeaderFg}
\caption{\thd{: Backward BF Chain Derivation with Neuro-Symbolic Verification}}
\label{alg:bf_derivation}
\small
\begin{algorithmic}[1]
\REQUIRE CVE $v$; evidence $E=(E_{\mathrm{dsc}},E_{\mathrm{cod}},E_{\mathrm{adv}})$;
\hspace*{\algorithmicindent} closed tables $T_A..T_H$; layer bound $N_{\max}$
\ENSURE chain $\mathcal{C}=\langle W_{N}\,{\bfprop}\cdots{\bfprop}\,W_{1}\rangle\!\rightarrow\!F$; audit $\mathcal{R}$
\STATE \COMMENT{\textbf{Derivation phase} (backward, sink to root)}
\STATE $F \leftarrow \Omega^{*}\!\big(q_{\mathrm{flr}},\, T_A[\_\mathrm{FLR}] \mid E_{\mathrm{dsc}}\big)$ \COMMENT{failure classes}
\STATE $Cn_1 \leftarrow \Omega\big(q_{\mathrm{err}},\, \mathrm{values}(T_E) \mid E_{\mathrm{dsc}}\big)$ \COMMENT{final error}
\STATE $n \leftarrow 1$
\REPEAT
  \STATE $W_n \leftarrow \kappa(Cn_n)$ \COMMENT{deterministic: the consequence fixes the class}
  \STATE $Op_n \leftarrow \Omega\big(q_{\mathrm{op}},\, T_B[W_n] \mid E_{\mathrm{dsc}},E_{\mathrm{cod}}\big)$
  \STATE \COMMENT{the full operation menu of $W_n$ is scored; one is kept}
  \STATE $Cs_n \leftarrow \Omega\big(q_{\mathrm{cs}},\, T_C \cup T_D[W_n] \mid E_{\mathrm{cod}},\mathrm{Obj}(Op_n)\big)$
  \STATE $A_n \leftarrow \Omega\text{-filled attributes over } T_G[W_n] \text{ and } T_H$
  \STATE $\mathcal{R} \leftarrow \mathcal{R} \cup
     \{\mathrm{record}(n,\text{query},\text{result},\text{anchor})\}$
  \IF{$Cs_n \in T_C$}
     \STATE \textbf{break} \COMMENT{a bug is an improper operation: root found}
  \ENDIF
  \STATE $Cn_{n+1} \leftarrow Cs_n$ \COMMENT{the error out of $W_{n+1}$ equals the fault into $W_n$}
  \IF{$\tau(\kappa(Cn_{n+1})) \neq \tau(W_n)$}
     \STATE $\mathcal{R} \leftarrow \mathcal{R}\cup\{\mathrm{crossing}(n)\}$
     \STATE \COMMENT{meaningful value propagation across class types}
  \ENDIF
  \STATE $n \leftarrow n+1$
\UNTIL{$n > N_{\max}$}
\STATE \COMMENT{\textbf{Verification phase} (symbolic, deterministic)}
\STATE $P_{\mathrm{voc}} \leftarrow \forall x \in \mathcal{C}: x \in T_A..T_H$ \COMMENT{closed vocabulary}
\STATE $P_{\mathrm{str}} \leftarrow \mathcal{C} \models \Sigma$
\COMMENT{$\Sigma$: SHACL shapes; one cause, operation, consequence per $W$; one sink}
\STATE $P_{\mathrm{ort}} \leftarrow \forall n: \mathrm{class}(Op_n)=W_n$
\COMMENT{OWL restrictions: the operation fixes the class}
\STATE $P_{\mathrm{lnk}} \leftarrow \forall n>1: Cs_n = Cn_{n+1}$
\COMMENT{linking rule}
\IF{$\neg(P_{\mathrm{voc}} \wedge P_{\mathrm{str}} \wedge P_{\mathrm{ort}} \wedge P_{\mathrm{lnk}})$}
  \STATE \textbf{return} $(\bot,\ \mathcal{R}\,\cup\,\{\text{violation report}\})$
\ENDIF
\RETURN $(\mathcal{C},\ \mathcal{R})$
\end{algorithmic}
\end{algorithm}

    The split between the two lookup modes defines the neuro-symbolic boundary of the architecture. A deterministic lookup, such as $\kappa$ or the stopping test, is a set or function evaluation over the closed tables and never involves a model. A semantic-inference selection is the oracle call $\Omega(q, M \mid E)$ of Table~\ref{tab:notation}, and the apparatus constrains it three ways: the prompt presents the complete menu $M$ of legal values, decoding is constrained by a JSON schema whose value domain is exactly $M$, and the temperature is zero. The oracle therefore selects among vocabulary entries and can never introduce a value from outside the closed sets; when the model transport fails, or the constrained decoder rejects the schema, the step raises a typed error, and the run halts rather than degrading to unconstrained generation. The verification phase then re-establishes every property symbolically, independently of the oracle. The vocabulary predicate $P_{\mathrm{voc}}$ confirms closed-set membership. The structural predicate $P_{\mathrm{str}}$ validates the chain against a set of Shapes Constraint Language (SHACL) shapes~\cite{w3c_shacl_2017}, which require exactly one cause, one operation, and one consequence per weakness and exactly one sink weakness per chain. The orthogonality predicate $P_{\mathrm{ort}}$ confirms that each selected operation belongs to its link's class, which the accompanying Web Ontology Language (OWL) ontology encodes as a class restriction per operation~\cite{w3c_owl2_2012}. The linking predicate $P_{\mathrm{lnk}}$ confirms that the cause of each link equals the consequence of the link behind it. A violation returns to the derivation engine as a structured report and the affected steps regenerate, so an accepted chain always satisfies all four predicates; the vocabulary, the ontology, the shapes, and every conformant chain are released with the codebase as machine-readable artifacts, and each accepted chain carries the full audit trail $\mathcal{R}$ of its per-step queries, results, and evidence anchors.
        
\definecolor{cSamplingVivOrangeBorder}{HTML}{D9B58F} %
\definecolor{cSamplingVivOrangeFillD}{HTML}{F6E3D3} %
\definecolor{cSamplingVivOrangeFillL}{HTML}{FFFCFD} %
\definecolor{cSamplingVivOrangeText}{HTML}{1B0903} %
\begin{inlinefloat}

\colorlet{tblHeaderBg}{cSamplingVivOrangeFillD}
\colorlet{tblHeaderFg}{cSamplingVivOrangeText}
\colorlet{tblHeaderRule}{cSamplingVivOrangeBorder}
\colorlet{tblSubHeaderBg}{cSamplingVivOrangeFillL}

\def\sampRowH{0.45}
\def\sampRowSep{0.3}
\definecolor{cSamplingZebra}{HTML}{FBF4EE}
\begin{lrbox}{\tbltmpbox}%
\begin{tikzpicture}
\matrix (m) [tblmat,
  row sep=\sampRowSep pt,
  nodes={minimum height=\sampRowH cm},
  column 1/.style={nodes={text width=\dimexpr0.62\liveW-24pt\relax, align=left}},
  column 2/.style={nodes={align=center, text width=0.19\liveW}},
  column 3/.style={nodes={align=center, text width=0.19\liveW}},
]{
  |[tblhdr]| Frame construction & |[tblhdr]| Removed & |[tblhdr]| Remaining \\
  Baseline corpus (CVE JSON 5.0 records) & & \wtFrameParsed \\
  E1: record not in PUBLISHED state & \wtFrameRmEone & \wtFrameRemEone \\
  E2: description under \wtFrameMinWords\ words & \wtFrameRmEtwo & \wtFrameRemEtwo \\
  E3: no reference link & \wtFrameRmEthree & \wtFrameRemEthree \\
  E4: NIST-authored BF specification & \wtFrameRmEfour & \wtFrameRemEfour \\
  |[tblhdr]| Stratum & |[tblhdr]| Frame size & |[tblhdr]| Allocated \\
  S1: \_INP family & \wtStratumNSone & \wtStratumAllocSone \\
  S2: \_MEM family & \wtStratumNStwo & \wtStratumAllocStwo \\
  S3: \_DAT family & \wtStratumNSthree & \wtStratumAllocSthree \\
  S4a: CWE outside current BF scope & \wtStratumNSfoura & \wtStratumAllocSfoura \\
  S4b: no usable CWE in the record & \wtStratumNSfourb & \wtStratumAllocSfourb \\
  \textbf{Total} & \textbf{\wtStrataTotal} & \textbf{\wtDrawN} \\
};

\begin{scope}[on background layer]
  \foreach \r in {3,5,9,11}{
    \fill[cSamplingZebra] (m.west|-m-\r-1.north) rectangle (m.east|-m-\r-1.south);}
\end{scope}
\draw[Rtop] (m.north west) -- (m.north east);
\draw[Rmid] (m.west|-m-2-1.north) -- (m.east|-m-2-1.north);
\draw[Rmid] (m.west|-m-7-1.north) -- (m.east|-m-7-1.north);
\draw[Rmid] (m.west|-m-8-1.north) -- (m.east|-m-8-1.north);
\draw[Rmid] (m.west|-m-13-1.north) -- (m.east|-m-13-1.north);
\draw[Rbot] (m.south west) -- (m.south east);

\begin{scope}[on background layer]
  \fill[tblSubHeaderBg] (m.west|-m-7-1.north) rectangle (m.east|-m-8-1.north);
\end{scope}

\end{tikzpicture}%
\end{lrbox}
\usebox{\tbltmpbox}
\captionof{table}{Sampling frame and stratified allocation of the evaluation
draw (seed \wtDrawSeed); allocation is proportional with a fixed
largest-remainder rule.}
\label{tab:sampling}
\end{inlinefloat}

    \paragraph{The generation design}
    
        The pipeline is evidence-based and stepwise, where for each CVE, the model derives a BF weakness chain from the CVE's evidence bundle alone, with one constrained oracle call per derivation step, over repeated independent rounds. Each generated chain passes through the symbolic verifier of Section~\ref{subsec:derivation_algorithm} and every stored chain carries its deployment identifier and budget mode, so a combination is identified from the data rather than folder.
    
    \paragraph{CVE selection}
        The sampling frame is derived from a seeded, stratified probability sample of 1,000 CVEs. The population is every published record of the curated CVE corpus that does not appear in a frozen 26-CVE exclusion set. Stratification uses a two-layer CWE-to-BF-family crosswalk, where an unassigned record moves between strata, thereby costing subgroup precision. One deviation is recorded: the NVD enrichment API was unreachable, so the primary CWE was resolved, which inflates the no-usable-CWE stratum S4b. Each deployment's evaluation corpus is then resolved from this frame: the \wtVerifiedN\ CVEs whose NIST-verified BF chains are published are explicitly named, and the frame's leading records are graded by the confidence of their retrievable evidence until the high- and medium-confidence quotas are filled. Each deployment's resolved corpus is therefore the verified class plus its own graded selections, and the per-combination totals that carry \mbox{\wtWindowRounds} completed rounds are reported with the results rather than fixed in advance. The verified \wtVerifiedN\ add a reading that the drawn records are excluded by design: for these CVEs, a derived chain can be compared against a published answer chain, and that comparison is reported case by case, never as a pooled accuracy score. Every combination targets \wtWindowRounds\ independent rounds per CVE. Cross-combination comparisons are computed on a balanced window of CVEs that carry \wtWindowRounds\ completed rounds on \emph{every} combination; within-combination readings are computed per combination over that combination's own CVEs with \wtWindowRounds\ completed rounds, capped at \wtWindowRounds.
    
    \paragraph{Metrics}
        Chain-level determinism compares every pair among a group's chains (a group is one CVE's completed rounds within one combination) after canonicalization to the ordered weakness tuples of cause, operation, consequence, and finality plus the failure multiset: exact determinism $D$ is the fraction of identical pairs, structural similarity $S$ scores partial agreement on a 0 to 1 scale, a per-axis decomposition over eight independent axes (chain length, failure set, and the root and sink values of cause, operation, and consequence) locates where variation lives, and each group's $D$ carries a Wilson 95\% interval~\cite{wilson1927probable}. The weakness class is not a coordinate for comparison.

\definecolor{cNotationIceBorder}{HTML}{A9BFCF} %
\definecolor{cNotationIceFillD}{HTML}{D9E7F0} %
\definecolor{cNotationIceFillL}{HTML}{F6FCFF} %
\definecolor{cNotationIceText}{HTML}{0A232D} %
\begin{inlinefloat}
\colorlet{tblHeaderBg}{cNotationIceFillD}
\colorlet{tblHeaderFg}{cNotationIceText}
\colorlet{tblHeaderRule}{cNotationIceBorder}
\colorlet{tblSubHeaderBg}{cNotationIceFillL}

\def\notRowH{0.45}
\def\notRowSep{0.3}
\definecolor{cNotationZebra}{HTML}{F3F8FB}
\begin{lrbox}{\tbltmpbox}%
\begin{tikzpicture}
\matrix (m) [tblmat,
  row sep=\notRowSep pt,
  nodes={minimum height=\notRowH cm},
  column 1/.style={nodes={text width=0.11\liveW, align=left}},
  column 2/.style={nodes={text width=\dimexpr0.89\liveW-16pt\relax, align=left}},
]{
  |[tblhdr]| & |[tblhdr]| \\
  |[tblhdr]| Symbol & |[tblhdr]| Description \\
  $v$ & CVE identifier under analysis \\
  $E$ & clean evidence bundle $(E_{\mathrm{dsc}},E_{\mathrm{cod}},E_{\mathrm{adv}})$: advisory text, buggy source with its fix commit, and referenced advisory pages \\
  $T_A$ & BF class types and their member classes \\
  $T_B$ & BF operations per class (one operation name belongs to exactly one class) \\
  $T_C$ & bug values (code bugs and specification bugs) \\
  $T_D$ & fault and error values with their type dimension \\
  $T_E$ & final-error values and security-failure classes \\
  $T_G$ & operation-attribute values (Mechanism, Source Code, Execution Space) \\
  $T_H$ & operand-attribute values over the eight operand facets \\
  $W_n$ & weakness $n$, counted backward from the sink: $(Cs_n, Op_n, Cn_n, A_n)$, that is cause, operation, consequence, attributes \\
  $F$ & the failure classes of the chain, $F \subseteq T_A[\_\mathrm{FLR}]$ \\
  $\kappa$ & class map: $\kappa(e)$ is the unique class whose error set contains the value $e$, per the orthogonality of the closed value sets \\
  $\tau$ & class-type map: $\tau(c)$ is the class type of class $c$ in $T_A$ \\
  $\Omega$ & constrained semantic-inference oracle: $\Omega(q, M \mid E) \in M$ returns exactly one element of the closed menu $M$ for question $q$ under evidence $E$; $\Omega^{*}$ is its multi-select form, $\Omega^{*}(q, M \mid E) \subseteq M$ \\
  $P_{\mathrm{voc}}, P_{\mathrm{str}},$ & verification predicates: closed-vocabulary membership, structural \\
  $P_{\mathrm{ort}}, P_{\mathrm{lnk}}$ & shape conformance, operation-class orthogonality, and link consistency \\
  $\mathcal{C}$, $\mathcal{R}$ & the derived chain and its audit trail \\
  $N_{\max}$ & layer bound (nine, the number of BF weakness classes) \\
};

\node[tblhdr, anchor=center, yshift=1pt ] at ($(m-1-1.center)!0.61!(m-1-2.center)$) {Notation Directory};

\begin{scope}[on background layer]
  \foreach \r in {4,6,8,10,12,14,16,18,20}{
    \fill[cNotationZebra] (m.west|-m-\r-1.north) rectangle (m.east|-m-\r-1.south);}
\end{scope}
\draw[Rtop, line width=0.04pt] (m.north west) -- (m.north east);
\draw[Rmid, line width=0.03pt] (m.west|-m-2-1.north) -- (m.east|-m-2-1.north); %
\draw[Rmid, line width=0.03pt] (m.west|-m-3-1.north) -- (m.east|-m-3-1.north); %
\draw[Rbot, line width=0.04pt] (m.south west) -- (m.south east);

\begin{scope}[on background layer]
  \fill[tblSubHeaderBg] (m.west|-m-2-1.north) rectangle (m.east|-m-3-1.north);
\end{scope}
\end{tikzpicture}%
\end{lrbox}
\usebox{\tbltmpbox}
\captionof{table}{Notation for Algorithm~\ref{alg:bf_derivation}; the closed
vocabulary tables $T_A$ to $T_H$ are transcribed verbatim from NIST SP
800-231~\cite{bojanova2024bf}.}
\label{tab:notation}
\end{inlinefloat}
    
    \paragraph{Justification of the metric suite}
        The structural similarity is a single normalized agreement count under optimal alignment: for a chain pair $(a,b)$ the weakness sequences are aligned by dynamic programming with per-field partial credit, and \begin{equation} S(a,b)=\frac{m(a,b)+\lvert F_a\cap F_b\rvert} {4\cdot\max(\lvert W_a\rvert,\lvert W_b\rvert)+\max(\lvert F_a\rvert,\lvert F_b\rvert)}, \label{eq:structural_similarity} \end{equation} where $m(a,b)$ counts matched weakness fields under the optimal alignment, $F$ is the failure multiset, $W$ the weakness sequence, and four is the number of independently compared fields per weakness (cause, operation, consequence, finality). $S$ equals one exactly on identical canonical chains. Two readings of a chain are reported below. The \mbox{\emph{whole chain}} is every compared field of every weakness; the \mbox{\emph{root-to-sink class-chain}} is the sequence of BF class types alone, read from the root weakness to the sink, so two chains share a class-chain when they pass through the same classes whatever they assign beneath them. The round count per CVE follows from the clustered pair structure: pairs grow as $\binom{k}{2}$, while groups remain independent units, so many moderate-$k$ groups dominate a few large-$k$ groups. At \wtWindowRounds\ rounds, $D$ is therefore a per-CVE indicator on the admissible grid \wtRoundGridTeX{} rather than a continuous rate.
    
    \subsection{Study 2 Results}\label{subsec:study2_results}
\begin{inlinefloat}
\begin{lrbox}{\tbltmpbox}%
\begin{tikzpicture}
\matrix (m) [tblmat,
  row sep=0pt,
  nodes={inner xsep=2.5pt, inner ysep=3pt, minimum height=0.55cm, font=\small},
  column 1/.style={nodes={text width=0.19\liveW, align=left}},
  column 2/.style={nodes={text width=0.11\liveW, align=center}},
  column 3/.style={nodes={text width=0.10\liveW, align=center}},
  column 4/.style={nodes={text width=0.11\liveW, align=center}},
  column 5/.style={nodes={text width=0.11\liveW, align=center}},
  column 6/.style={nodes={text width=0.11\liveW, align=center}},
  column 7/.style={nodes={text width=0.16\liveW, align=center}},
]{
  |[tblhdr, font=\bfseries\small]| Deployment & |[tblhdr, font=\bfseries\small]| Prompt budget & |[tblhdr, font=\bfseries\small]| Groups $n$ & |[tblhdr, font=\bfseries\small]| $D$ mean & |[tblhdr, font=\bfseries\small]| $D$ median & |[tblhdr, font=\bfseries\small]| $S$ mean & |[tblhdr, font=\bfseries\small]| Fully deterministic \\
  Gemma 4 31B & Own & 85 & |[lkAlmost]| 0.890 & |[lkAlmost]| 1.000 & |[lkAlmost]| 0.968 & 72 of 85 \\
  GPT-OSS 120B & Own & 85 & |[lkSlight]| 0.094 & |[lkSlight]| 0.000 & |[lkModerate]| 0.577 & 3 of 85 \\
};
\draw[Rtop] (m.north west) -- (m.north east);
\draw[Rmid] (m.west|-m-2-1.north) -- (m.east|-m-2-1.north);
\draw[Rbot] (m.south west) -- (m.south east);
\end{tikzpicture}%
\end{lrbox}
\usebox{\tbltmpbox}
\floatnote[c]{\lkbandlegend}
\providecommand{\wtArmsCaption}{Determinism of the evidence-based derivation for each of the 2 model-budget combinations, over one balanced window of 85 CVEs at 3 rounds.}
\caption{\wtArmsCaption}
\label{tab:walkthrough_arms}
\providecommand{\wtArmsDescription}{A table with one row per model-budget combination and columns for deployment, prompt budget, number of CVE groups, mean and median determinism, mean structural similarity, and the count of fully deterministic groups. The three unit-interval columns are shaded by Landis and Koch band from light to dark blue; the count columns are unshaded. A band legend sits beneath the grid.}
\Description{\wtArmsDescription}
\end{inlinefloat}

    \paragraph{The balanced cross-combination window}
        Figure~\ref{tab:walkthrough_arms} and Figure~\ref{tab:multimodel_summary} report determinism per combination over the balanced window of CVEs carrying \wtWindowRounds\ completed rounds, which admits a window of size \wtWindowCVEs\ at \wtWindowRounds\ rounds per combination. On that window \gemmaModel{} reproduces identical chains for most CVEs ($D=\wtDmeanGemmaOwnCap$, $S=\wtSmeanGemmaOwnCap$), while \gptossModel{} reproduces almost none ($D=\wtDmeanGptOssOwnCap$ with $S=\wtSmeanGptOssOwnCap$). The \gptossModel{} similarity score sits well above zero, and the per-axis block of Figure~\ref{tab:multimodel_summary} places its disagreement on the whole-chain, sink, and failure axes rather than on the root.
    
\definecolor{cSummSteelBorder}{HTML}{ADBCD3}
\definecolor{cSummSteelFillD}{HTML}{E4EBF2}
\definecolor{cSummSteelFillL}{HTML}{F3F6F6}
\definecolor{cSummSteelText}{HTML}{0D162D}

\definecolor{cSummSteelBorder}{HTML}{ADBCD3}
\definecolor{cSummSteelFillD}{HTML}{E4EBF2}
\definecolor{cSummSteelFillL}{HTML}{F3F6F6}
\definecolor{cSummSteelText}{HTML}{0D162D}

\begin{inlinefloat}

\colorlet{tblHeaderBg}{cSummSteelFillD}
\colorlet{tblHeaderFg}{cSummSteelText}
\colorlet{tblHeaderRule}{cSummSteelBorder}
\colorlet{tblSubHeaderBg}{cSummSteelFillL}

\begin{lrbox}{\tbltmpbox}%
\begin{tikzpicture}
\matrix (m) [
  tblmat,
  inner sep=0pt, 
  column sep=0pt,
  row sep=0pt,
  nodes={inner xsep=4pt, inner ysep=2.6pt, minimum height=0.52cm,
         font=\normalsize},
  column 1/.style={nodes={text width=0.44\liveW, align=left}},
  column 2/.style={nodes={text width=0.22\liveW, align=center}},
  column 3/.style={nodes={text width=0.22\liveW, align=center}}
]{
  |[tblhdr]| Deployment & |[tblhdr]| Gemma 4 31B & |[tblhdr]| GPT-OSS 120B \\
  \textbf{Prompt budget} & Own & Own \\
  \textbf{CVE groups}, $n$ & 85 & 85 \\
  \textbf{Rounds}, $R$ & 3 & 3 \\
  Exact determinism $D$, mean & |[lkAlmost]| 0.890 & |[lkSlight]| 0.094 \\
  $D$, median & |[lkAlmost]| 1.000 & |[lkSlight]| 0.000 \\
  $D$, minimum & |[lkSlight]| 0.000 & |[lkSlight]| 0.000 \\
  Fully deterministic share & |[lkAlmost]| 0.847 & |[lkSlight]| 0.035 \\
  Structural similarity $S$, mean & |[lkAlmost]| 0.968 & |[lkModerate]| 0.577 \\
  \emph{Per-axis agreement} &  &  \\
  \hspace{2mm}chain length & |[lkAlmost]| 0.976 & |[lkModerate]| 0.502 \\
  \hspace{2mm}root cause & |[lkAlmost]| 0.992 & |[lkSubstantial]| 0.678 \\
  \hspace{2mm}root operation & |[lkAlmost]| 0.918 & |[lkFair]| 0.333 \\
  \hspace{2mm}root consequence & |[lkAlmost]| 0.953 & |[lkFair]| 0.345 \\
  \hspace{2mm}sink cause & |[lkAlmost]| 0.969 & |[lkModerate]| 0.514 \\
  \hspace{2mm}sink operation & |[lkAlmost]| 0.992 & |[lkSubstantial]| 0.729 \\
  \hspace{2mm}sink consequence & |[lkAlmost]| 0.953 & |[lkSubstantial]| 0.784 \\
  \hspace{2mm}failure set & |[lkAlmost]| 0.984 & |[lkModerate]| 0.573 \\
};

\draw[Rtop] (m.north west) -- (m.north east);
\draw[Rmid] (m.west|-m-2-1.north) -- (m.east|-m-2-1.north); %
\draw[Rmid] (m.west|-m-5-1.north) -- (m.east|-m-5-1.north); %
\draw[Rmid] (m.west|-m-10-1.north) -- (m.east|-m-10-1.north); %
\draw[Rmid] (m.west|-m-11-1.north) -- (m.east|-m-11-1.north); %
\draw[Rbot] (m.south west) -- (m.south east);

\begin{scope}[on background layer]
  \fill[tblSubHeaderBg] (m.west|-m-2-1.north) rectangle (m.east|-m-5-1.north);
  \fill[tblSubHeaderBg] (m.west|-m-10-1.north) rectangle (m.east|-m-11-1.north);
\end{scope}

\end{tikzpicture}%
\end{lrbox}
\usebox{\tbltmpbox}
\floatnote{\lkbandlegend\\[1pt]\textit{Every combination runs at its deployment's own context ceiling (\emph{Own}), so a difference across a row is attributable to the deployment. Configuration rows are not shaded. Every value is measured.}}
\providecommand{\wtSummaryCaption}{Determinism of the evidence-based derivation by model-budget combination, over one balanced window of 85 CVEs at 3 rounds.}
\caption{\wtSummaryCaption}
\label{tab:multimodel_summary}
\providecommand{\wtSummaryDescription}{A two-column comparison with one column per model-budget combination. The configuration rows (prompt budget, CVE groups, rounds) are unshaded. The determinism rows (mean, median, minimum, fully deterministic share, structural similarity) and the per-axis agreement rows (chain length through failure set) are shaded from light to dark blue by Landis and Koch band. A band legend and a reading note sit beneath the grid.}
\Description{\wtSummaryDescription}
\end{inlinefloat}
    
    \paragraph{Within-combination determinism}
        Figure~\ref{fig:det_axes_panel} extends the per-axis reading to each combination's full coverage: \wtOwnNGemmaOwnCap\ CVEs on \gemmaModel{} and \wtOwnNGptOssOwnCap\ on \gptossModel{}. \gemmaModel{} holds every axis high; on \gptossModel{} the profile sits far lower on every axis. Figure~\ref{fig:spine_vs_whole} states the two-number rule: the count of CVEs whose class-chain reproduces against the count whose whole chain does. On \gemmaModel{} the class-chain reproduces identically for more CVEs than the whole chain does (\wtOwnSpineKGemmaOwnCap\ against \wtOwnWholeKGemmaOwnCap\ of \wtOwnNGemmaOwnCap), so the residual variation lives below the class-chain in attribute-level and value-level fields. The gap is widest where determinism is lowest: on \gptossModel{} the class-chain reproduces for \wtOwnSpineKGptOssOwnCap\ CVEs against \wtOwnWholeKGptOssOwnCap\ whole chains.
    
\definecolor{cSvwSlateDark}{HTML}{384557}  %
\definecolor{cSvwSlateMid}{HTML}{7489A1}   %
\definecolor{cSvwOrangeDark}{HTML}{7C2F06} %
\definecolor{cSvwGrid}{HTML}{E9EEF5}       %
\definecolor{cSvwText}{HTML}{293445}       %
\def\svwArms{%
  1.20/\wtOwnWholeShareGemmaOwnCap/\wtOwnSpineShareGemmaOwnCap/cSvwSlateDark,
  0.60/\wtOwnWholeShareGptOssOwnCap/\wtOwnSpineShareGptOssOwnCap/cSvwOrangeDark}
\begin{inlinefloat}
\pgfmathsetlengthmacro{\PL}{0.60\liveW}
\begin{tikzpicture}[font=\normalsize]
\foreach \gv/\gf in {0/0, 0.25/0.25, 0.50/0.5, 0.75/0.75, 1.0/1}{
  \draw[cSvwGrid, line width=0.4pt] ({\gf*\PL},-0.05) -- ({\gf*\PL},1.80);
  \node[anchor=north, text=cSvwText, inner sep=1.5pt] at ({\gf*\PL},-0.08) {\gv};}
\draw[cSvwText, line width=0.5pt] (0,-0.05) -- (0,1.80);
\foreach \y/\w/\s/\hue in \svwArms {%
  \draw[\hue!55, line width=1.0pt] ({\w*\PL},\y) -- ({\s*\PL},\y);
  \fill[\hue] ({\w*\PL},\y) circle (2pt);
  \draw[\hue, line width=0.9pt, fill=white] ({\s*\PL},\y) circle (2pt);
}
\node[anchor=east, text=cSvwText, inner sep=2pt] at (-0.15,1.20)
  {\gemmaModel{} ($n=\wtOwnNGemmaOwnCap$)};
\node[anchor=east, text=cSvwText, inner sep=2pt] at (-0.15,0.60)
  {\gptossModel{} ($n=\wtOwnNGptOssOwnCap$)};
\fill[cSvwSlateDark] (0.10,2.30) circle (2pt);
\node[anchor=west, text=cSvwText] at (0.24,2.30) {whole chain identical};
\draw[cSvwSlateDark, line width=0.9pt, fill=white] ({0.55*\PL},2.30) circle (2pt);
\node[anchor=west, text=cSvwText] at ({0.55*\PL+0.14cm},2.30) {class-chain identical};
\node[anchor=north, text=cSvwText] at ({0.5*\PL},-0.48)
  {share of CVEs with all rounds identical};
\end{tikzpicture}%
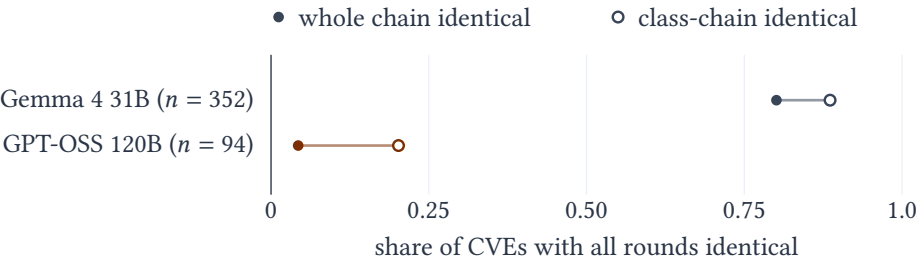
\floatnote{\itshape Share of CVEs whose repeated rounds produce an identical whole chain (filled) against an identical
class-chain (open). Both deployments read to their own context ceiling. Computed per combination over that
combination's own CVEs with \wtWindowRounds{} completed rounds (per-combination $n$ at left). }
\caption{The two determinism numbers of the evidence-based derivation, one row per combination.}
\label{fig:spine_vs_whole}
\Description{A dot plot with one row per model-budget combination and a horizontal axis from 0 to 1 giving the share of CVEs whose repeated rounds are identical. Each row carries two markers joined by a thin line: a filled diamond for the whole chain and an open diamond for the class-chain, so the gap between the two markers shows how much of the variation lies below the class-chain.}
\end{inlinefloat}
    
    \paragraph{Determinism by evidence-confidence stratum}
        Figure~\ref{fig:det_by_stratum} compares the fully deterministic share across the evidence-confidence strata. On \gemmaModel{} the high-confidence stratum is fully deterministic for \wtOwnHighFullyGemmaOwnCap\ of \wtOwnHighNGemmaOwnCap\ CVEs (mean $D$ \wtOwnHighDGemmaOwnCap) while the medium-confidence stratum is fully deterministic for \wtOwnMedFullyGemmaOwnCap\ of \wtOwnMedNGemmaOwnCap\ (mean $D$ \wtOwnMedDGemmaOwnCap). On \gptossModel{} both strata are fully deterministic for a small minority (\wtOwnHighFullyGptOssOwnCap\ of \wtOwnHighNGptOssOwnCap\ against \wtOwnMedFullyGptOssOwnCap\ of \wtOwnMedNGptOssOwnCap), on counts small enough that the comparison is descriptive only. The strata are observed rather than assigned, so these are differences across groups, not effects of the group, and Appendix~\ref{sec:threats_a} records the corresponding conditions.
\definecolor{cGagSlateDark}{HTML}{384557}  %
\definecolor{cGagOrangeDark}{HTML}{7C2F06} %
\definecolor{cGagGrid}{HTML}{E9EEF5}       %
\definecolor{cGagText}{HTML}{293445}       %
\begin{inlinefloat}
\pgfmathsetlengthmacro{\PL}{0.33\liveW}
\begin{tikzpicture}[font=\small]
\pgfmathsetmacro\gagTop{(\wtGoldRowCount+1)*0.42}
\foreach \gl/\gf in \wtRoundGridFracs {%
  \draw[cGagGrid, line width=0.4pt] ({\gf*\PL},0.10) -- ({\gf*\PL},\gagTop);
  \node[anchor=north, text=cGagText, inner sep=1.5pt] at ({\gf*\PL},0.02) {$\gl$};}
\draw[cGagText, line width=0.5pt] (0,0.10) -- (0,\gagTop);
\foreach \sr in \wtGoldSeps {%
  \pgfmathsetmacro\sy{(\wtGoldRowCount-\sr+0.5)*0.42}
  \draw[gray!45, dashed, line width=0.3pt] ({-0.19\liveW},\sy) -- ({0.78\liveW},\sy);}
\foreach \r/\lbl/\sp in \wtGoldRowLabels {%
  \pgfmathsetmacro\py{(\wtGoldRowCount-\r+1)*0.42}
  \node[anchor=east, text=cGagText, inner sep=1pt] at (-0.24,\py) {\lbl};
  \node[anchor=west, text=cGagText, inner sep=1pt] at ({\PL+0.2cm},\py) {\sp};
}
\foreach \r/\v in \wtGoldMarksGemmaOwnCap {%
  \pgfmathsetmacro\py{(\wtGoldRowCount-\r+1)*0.42+0.07}
  \fill[cGagSlateDark] ({\v*\PL},\py) circle (1.7pt);}
\foreach \r/\v in \wtGoldMarksGptOssOwnCap {%
  \pgfmathsetmacro\py{(\wtGoldRowCount-\r+1)*0.42-0.07}
  \fill[cGagOrangeDark] ({\v*\PL-1.7pt},{\py cm-1.7pt}) rectangle ({\v*\PL+1.7pt},{\py cm+1.7pt});}
\pgfmathsetmacro\gagHdr{\gagTop+0.18}
\node[anchor=west, text=cGagText] at ({\PL+0.2cm},\gagHdr) {verified class-chain (measured)};
\pgfmathsetmacro\gagLegA{\gagTop+0.62}
\fill[cGagSlateDark] ({-0.18\liveW},\gagLegA) circle (1.7pt);
\node[anchor=west, text=cGagText] at ({-0.175\liveW},\gagLegA) {\gemmaModel{}};
\fill[cGagOrangeDark] ({0.06\liveW-1.7pt},{\gagLegA cm-1.7pt}) rectangle ({0.06\liveW+1.7pt},{\gagLegA cm+1.7pt});
\node[anchor=west, text=cGagText] at ({0.065\liveW},\gagLegA) {\gptossModel{}};
\node[anchor=west, text=cGagText] at ({0.32\liveW},\gagLegA) {both at their own context ceiling};
\node[anchor=north, text=cGagText, align=center] at ({0.5*\PL},-0.40)
  {share of \wtWindowRounds{} rounds whose class-chain equals the verified class-chain};
\end{tikzpicture}%
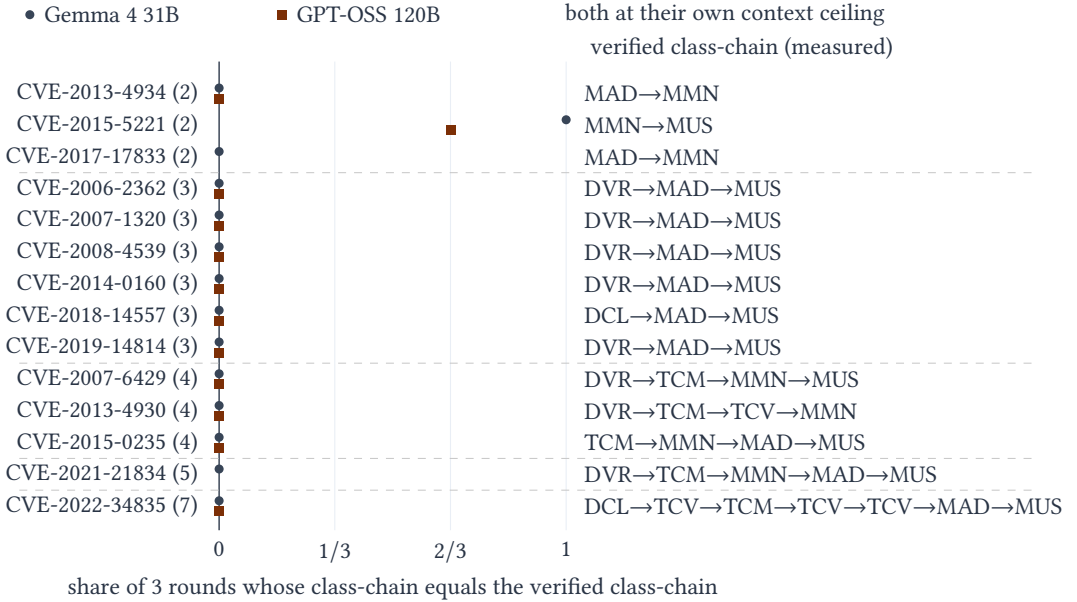
\floatnote{\itshape Over the NIST-verified CVEs with a published verified class-chain (\wtGoldRowCount\ out of \wtVerifiedN), chain length in
parentheses, up to one mark per combination and CVE on the round grid \wtRoundGridTeX; a combination without
\wtWindowRounds{} completed rounds for a CVE carries no mark on that row. The right column prints the
published verified class-chain. }
\caption{Measured per-CVE verified-class-chain agreement over the NIST-verified CVEs, sorted by verified chain length.}
\label{fig:gold_agreement}
\Description{A dot plot with one row per NIST-verified CVE, sorted by the length of its published chain, with the chain length in parentheses after the CVE identifier. The horizontal axis from 0 to 1 gives the share of rounds whose class-chain equals the published chain, with tick marks at 0, one third, two thirds, and 1. Each row carries up to one filled marker per combination (triangle for one deployment, diamond for the other); dashed horizontal rules separate the chain-length groups, and the right column prints the published chain as a sequence of class abbreviations.}
\end{inlinefloat}

\definecolor{cDvgSlateDark}{HTML}{384557}  %
\definecolor{cDvgOrangeDark}{HTML}{7C2F06} %
\definecolor{cDvgGrid}{HTML}{E9EEF5}       %
\definecolor{cDvgText}{HTML}{293445}       %
\begin{inlinefloat}
\pgfmathsetlengthmacro{\PL}{0.62\liveW}\def\PH{5.2}
\begin{tikzpicture}[font=\small]
\foreach \gl/\gf in \wtRoundGridFracs {%
  \draw[cDvgGrid, line width=0.4pt] ({\gf*\PL},0) -- ({\gf*\PL},\PH);
  \node[anchor=north, text=cDvgText, inner sep=1.5pt] at ({\gf*\PL},-0.30) {$\gl$};}
\foreach \gl/\gf in \wtRoundGridFracs {%
  \pgfmathsetmacro\gy{\gf*\PH}
  \draw[cDvgGrid, line width=0.4pt] (0,\gy) -- (\PL,\gy);
  \node[anchor=east, text=cDvgText, inner sep=1.5pt] at (-0.30,\gy) {$\gl$};}
\draw[cDvgText, line width=0.5pt] (0,0) -- (\PL,0);
\draw[cDvgText, line width=0.5pt] (0,0) -- (0,\PH);
\draw[gray!60, dashed, line width=0.5pt] ({0.8*\PL},0) -- ({0.8*\PL},\PH);
\node[anchor=south, text=gray, inner sep=1.5pt] at ({0.8*\PL},{\PH+0.04}) {$D=0.8$};
\draw[gray!60, dashed, line width=0.5pt] (0,2.60) -- (\PL,2.60);
\node[anchor=west, text=gray, inner sep=1.5pt, align=left] at ({\PL+0.06cm},2.60)
  {agreement\\$=0.5$};
\foreach \x/\y in \wtGoldPtsGemmaOwnCap {%
  \fill[cDvgSlateDark] ({\x*\PL+4.5pt},{\y*\PH cm+4.5pt}) circle (1.8pt);}
\foreach \x/\y in \wtGoldPtsGptOssOwnCap {%
  \fill[cDvgOrangeDark] ({\x*\PL+2.7pt},{\y*\PH cm-6.3pt}) rectangle ({\x*\PL+6.3pt},{\y*\PH cm-2.7pt});}
\fill[cDvgSlateDark] (0.10,{\PH+1.15}) circle (1.8pt);
\node[anchor=west, text=cDvgText] at (0.22,{\PH+1.15})
  {\gemmaModel{} ($n=\wtGoldNGemmaOwnCap$)};
\fill[cDvgOrangeDark] ({0.50*\PL-1.8pt},{\PH cm+1.15cm-1.8pt}) rectangle ({0.50*\PL+1.8pt},{\PH cm+1.15cm+1.8pt});
\node[anchor=west, text=cDvgText] at ({0.50*\PL+0.12cm},{\PH+1.15})
  {\gptossModel{} ($n=\wtGoldNGptOssOwnCap$)};
\node[anchor=west, text=cDvgText] at (0.10,{\PH+0.70})
  {both at their own context ceiling};
\node[anchor=north, text=cDvgText] at ({0.5*\PL},-0.72)
  {whole-chain exact determinism $D$ (\wtRoundGridTeX)};
\node[anchor=south, rotate=90, text=cDvgText] at (-1.15,2.60)
  {verified-class-chain agreement ratio};
\end{tikzpicture}%
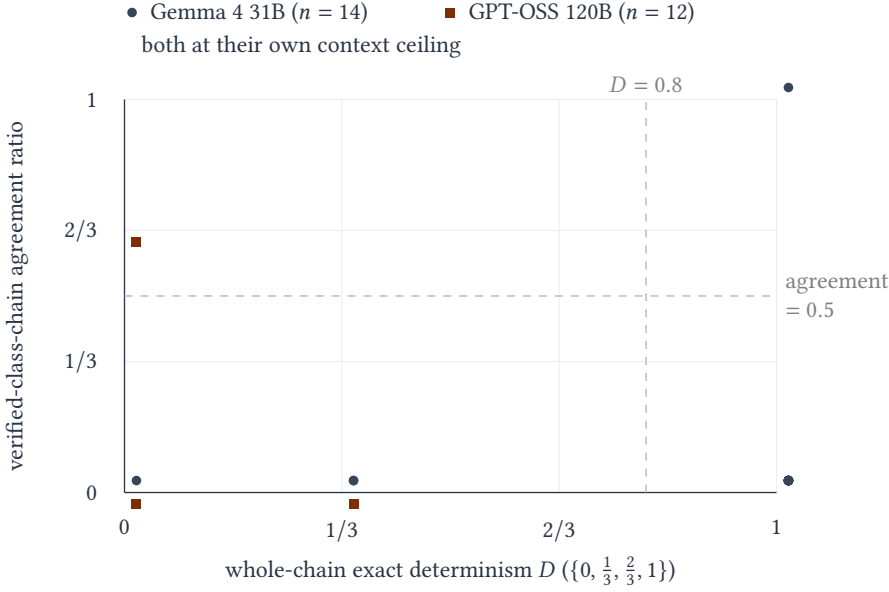
\floatnote{\itshape Whole-chain exact determinism $D$ against the per-CVE share of rounds whose class-chain equals the
published verified class-chain, over the verified CVEs with \wtWindowRounds{} completed rounds in that combination
and a published verified class-chain (per-combination $n$ in the legend). Both measures live on the round grid
\wtRoundGridTeX{} and marks carry a small fixed dodge per combination. The $y$ series is a per-CVE indicator, not a pooled accuracy
score.}
\caption{Measured determinism against verified-class-chain agreement for the NIST-verified CVEs, one mark per CVE and combination.}
\label{fig:det_vs_gold}
\Description{A scatter plot with whole-chain exact determinism on the horizontal axis and verified-class-chain agreement share on the vertical axis, both on the grid 0, one third, two thirds, 1, with dashed grid lines at the intermediate values. Each mark is one CVE and combination, a triangle for one deployment and a diamond for the other, with a small fixed horizontal offset per combination. A labeled dashed horizontal line gives the mean agreement, and a legend across the top names the two deployments.}
\end{inlinefloat}

    \paragraph{The verified-corpus reading}
        Figure~\ref{fig:bug_sequence_paths} shows which class-chains each deployment builds against the class-chains the verified corpus contains; Figure~\ref{fig:gold_agreement} reports the share of rounds whose class-chain equals the published class-chain; and Figure~\ref{fig:det_vs_gold} plots that per-CVE indicator against whole-chain determinism. The three readings agree: the derived class-chains concentrate on short two- and three-class paths where the verified corpus concentrates on \mbox{DVR $\to$ MAD $\to$ MUS} and its relatives, and a reliably reproduced chain most often reproduces a class-chain that differs from the published one. The mark cluster in the lower right region of Figure~\ref{fig:det_vs_gold}, high $D$ at zero agreement, is therefore the modal outcome: the pipeline is stable on a rendering of the vulnerability that is legal BF and is not the published rendering, a behavior this paper reads against the framework properties recorded in Appendix~\ref{sec:threats_a}.

    \subsection{Interpretation and Scope}\label{subsec:study2_interpretation}
    
    First, determinism under a baseline configuration is a property of the deployment, not of the task, so the task admits deterministic automation but does not guarantee it. Second, where determinism is high, the residual variation is localized below the class-chain, in attribute-level fields, which extends the Study~1 pattern: the axes the specification constrains most tightly reproduce best in both studies. Third, stability and fidelity are distinct properties: the determinism measured here is a claim about repeatability under the baseline configurations and the evidence-based input representation, never a claim of accuracy relative to expert-adjudicated chains. Appendix~\ref{sec:threats_a} records the corresponding threats.

\definecolor{cBspRampDark}{HTML}{3D6F9F}   %
\definecolor{cBspRampMid}{HTML}{A8C9E4}    %
\definecolor{cBspRampLight}{HTML}{DCEAF6}  %
\definecolor{cBspSlateDark}{HTML}{1B4368}  %
\definecolor{cBspOrangeDark}{HTML}{6B7686} %
\definecolor{cBspText}{HTML}{22303D}       %
\def\bspVerified{%
  1/{DVR,MAD,MUS}/5/cBspRampMid/cBspText,
  2/{MAD,MMN}/2/cBspRampLight/cBspText,
  3/{MMN,MUS}/1/cBspRampLight/cBspText,
  4/{DCL,MAD,MUS}/1/cBspRampLight/cBspText,
  5/{DVR,TCM,MMN,MUS}/1/cBspRampLight/cBspText,
  6/{DVR,TCM,TCV,MMN}/1/cBspRampLight/cBspText,
  7/{TCM,MMN,MAD,MUS}/1/cBspRampLight/cBspText,
  8/{DVR,TCM,MMN,MAD,MUS}/1/cBspRampLight/cBspText,
  9/{DCL,TCV,TCM,TCV,TCV,MAD,MUS}/1/cBspRampLight/cBspText}

\FloatBarrier
\section{Scope and Boundary Conditions}\label{sec:scope}

    Consistent with reporting-guideline practice for systematic reviews~\cite{page2021prisma}, the paper's claims apply within stated boundaries, recorded here so the comparison is judged against what was established, not against a wider construction.

\paragraph{BF coverage scope}
    The BF taxonomy as of NIST SP 800-231, July 2024, defines a finite set of weakness and failure class types: Input/Output Check (INP); Memory (MEM); Data Type (DAT); and Failure (FLR)~\cite{bojanova2024bf}. BF covers a defined subset of the vulnerability classes that CWE covers, and CWE entries outside this set are out of scope for the comparison.

\paragraph{Restriction of comparison}
    The paper's comparison applies within BF's current coverage. CVEs whose root cause lies outside the four BF class types named above are addressed in future work, as BF extends across the forthcoming NIST class-type specifications~\cite{bojanova2024bf}, not by the comparison reported here.

\paragraph{Study scope}
    The paper presents four case studies, one per BF class type, and an inter-rater study computed over $N=\wtIrrCVEs$ CVEs that form the clean unmapped stratum: CVEs that carry neither a published BF mapping nor prior 

\definecolor{cDbsSlateDark}{HTML}{384557}   %
\definecolor{cDbsSlateFill}{HTML}{BACDE2}   %
\definecolor{cDbsOrangeDark}{HTML}{7C2F06}  %
\definecolor{cDbsOrangeFill}{HTML}{F6BC9C}  %
\definecolor{cDbsGrid}{HTML}{E9EEF5}        %
\definecolor{cDbsText}{HTML}{293445}        %
\definecolor{cDapBlueDark}{HTML}{1B4368}   %
\definecolor{cDapGrid}{HTML}{EAF2FB}       %
\definecolor{cDapText}{HTML}{0B2E4F}       %
\begin{inlinefloat}
\begin{minipage}[c]{0.44\liveW}
\centering
\figboxscript{%
\begin{tikzpicture}[font=\small]
\def\dbsLaneW{0.50}     %
\def\dbsLanes{2}        %
\def\dbsStride{1.95}    %
\pgfmathsetmacro{\dbsClusterW}{\dbsLanes*\dbsLaneW}
\pgfmathsetmacro{\dbsHalf}{\dbsClusterW/2}
\newcommand{\dbsMark}[7]{%
  \ifnum#7=0
    \draw[gray!70, line width=0.7pt] (#3+0.06,0.05) -- (#3+0.34,0.05);
    \node[anchor=north, text=cDbsText, inner sep=0.5pt] at (#3+0.20,-0.05) {$0$};
  \else
    \pgfmathsetmacro\yt{#4*2.4}
    \pgfmathsetmacro\ylo{#5*2.4}
    \pgfmathsetmacro\yhi{#6*2.4}
    \ifnum#7<5
      \fill[#2] (#3+0.20,\yt) circle (1.6pt);
    \else
      \fill[#1] (#3,0) rectangle (#3+0.40,\yt);
      \draw[#2, line width=0.4pt] (#3,0) rectangle (#3+0.40,\yt);
    \fi
    \draw[#2, line width=0.5pt] (#3+0.20,\ylo) -- (#3+0.20,\yhi);
    \draw[#2, line width=0.5pt] (#3+0.13,\ylo) -- (#3+0.27,\ylo);
    \draw[#2, line width=0.5pt] (#3+0.13,\yhi) -- (#3+0.27,\yhi);
    \node[anchor=north, text=cDbsText, inner sep=0.5pt] at (#3+0.20,-0.05) {#7};
  \fi}
\newcommand{\dbsFrame}[1]{%
  \pgfmathsetmacro\pw{0.10+(#1-1)*\dbsStride+\dbsClusterW+0.10}
  \foreach \gy in {0.0,0.6,1.2,1.8,2.4}{
    \draw[cDbsGrid, line width=0.4pt] (0,\gy) -- (\pw,\gy);
  }
  \draw[cDbsText, line width=0.5pt] (0,0) -- (\pw,0);
}
\begin{scope}
\dbsFrame{3}
\foreach \gy/\gl in {0.0/0, 0.6/0.25, 1.2/0.50, 1.8/0.75, 2.4/1.0}{
  \node[anchor=east, text=cDbsText, inner sep=1.5pt] at (-0.06,\gy) {\small\gl};}
\dbsMark{cDbsSlateFill}{cDbsSlateDark}{0.10}{\wtOwnVerShareGemmaOwnCap}{\wtOwnVerLoGemmaOwnCap}{\wtOwnVerHiGemmaOwnCap}{\wtOwnVerNGemmaOwnCap}
\dbsMark{cDbsOrangeFill}{cDbsOrangeDark}{0.60}{\wtOwnVerShareGptOssOwnCap}{\wtOwnVerLoGptOssOwnCap}{\wtOwnVerHiGptOssOwnCap}{\wtOwnVerNGptOssOwnCap}
\node[anchor=north, text=cDbsText, inner sep=1pt] at (0.60,-0.30) {verified};
\dbsMark{cDbsSlateFill}{cDbsSlateDark}{2.05}{\wtOwnHighShareGemmaOwnCap}{\wtOwnHighLoGemmaOwnCap}{\wtOwnHighHiGemmaOwnCap}{\wtOwnHighNGemmaOwnCap}
\dbsMark{cDbsOrangeFill}{cDbsOrangeDark}{2.55}{\wtOwnHighShareGptOssOwnCap}{\wtOwnHighLoGptOssOwnCap}{\wtOwnHighHiGptOssOwnCap}{\wtOwnHighNGptOssOwnCap}
\node[anchor=north, text=cDbsText, inner sep=1pt, align=center] at (2.55,-0.30) {high\\confidence};
\dbsMark{cDbsSlateFill}{cDbsSlateDark}{4.00}{\wtOwnMedShareGemmaOwnCap}{\wtOwnMedLoGemmaOwnCap}{\wtOwnMedHiGemmaOwnCap}{\wtOwnMedNGemmaOwnCap}
\dbsMark{cDbsOrangeFill}{cDbsOrangeDark}{4.50}{\wtOwnMedShareGptOssOwnCap}{\wtOwnMedLoGptOssOwnCap}{\wtOwnMedHiGptOssOwnCap}{\wtOwnMedNGptOssOwnCap}
\node[anchor=north, text=cDbsText, inner sep=1pt, align=center] at (4.50,-0.30) {medium\\confidence};
\node[anchor=north, text=cDbsText, align=center, text width=5.1cm] at (2.55,-1.20)
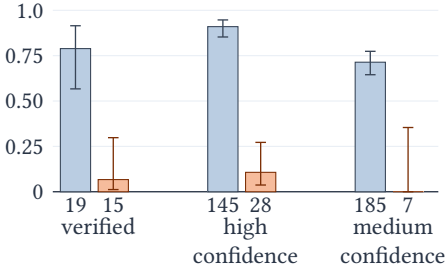
  {(a) fully deterministic share by evidence-confidence stratum};
\end{scope}
\begin{scope}[yshift=-5.20cm]
\dbsFrame{2}
\foreach \gy/\gl in {0.0/0, 0.6/0.25, 1.2/0.50, 1.8/0.75, 2.4/1.0}{
  \node[anchor=east, text=cDbsText, inner sep=1.5pt] at (-0.06,\gy) {\small\gl};}
\dbsMark{cDbsSlateFill}{cDbsSlateDark}{0.10}{\wtOwnVerShareGemmaOwnCap}{\wtOwnVerLoGemmaOwnCap}{\wtOwnVerHiGemmaOwnCap}{\wtOwnVerNGemmaOwnCap}
\dbsMark{cDbsOrangeFill}{cDbsOrangeDark}{0.60}{\wtOwnVerShareGptOssOwnCap}{\wtOwnVerLoGptOssOwnCap}{\wtOwnVerHiGptOssOwnCap}{\wtOwnVerNGptOssOwnCap}
\node[anchor=north, text=cDbsText, inner sep=1pt] at (0.60,-0.30) {verified};
\dbsMark{cDbsSlateFill}{cDbsSlateDark}{2.05}{\wtOwnDrawnShareGemmaOwnCap}{\wtOwnDrawnLoGemmaOwnCap}{\wtOwnDrawnHiGemmaOwnCap}{\wtOwnDrawnNGemmaOwnCap}
\dbsMark{cDbsOrangeFill}{cDbsOrangeDark}{2.55}{\wtOwnDrawnShareGptOssOwnCap}{\wtOwnDrawnLoGptOssOwnCap}{\wtOwnDrawnHiGptOssOwnCap}{\wtOwnDrawnNGptOssOwnCap}
\node[anchor=north, text=cDbsText, inner sep=1pt, align=center] at (2.55,-0.30) {drawn\\complement};
\node[anchor=north, text=cDbsText, align=center, text width=5.1cm] at (2.55,-1.20)
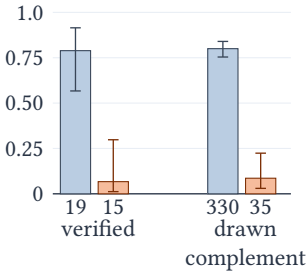
  {(b) fully deterministic share by corpus source};
\end{scope}
\begin{scope}[yshift=-7.60cm]
\fill[cDbsSlateFill] (0.00,-0.09) rectangle (0.30,0.09);
\draw[cDbsSlateDark, line width=0.4pt] (0.00,-0.09) rectangle (0.30,0.09);
\node[anchor=west, text=cDbsText] at (0.38,0) {\gemmaModel{}};
\fill[cDbsOrangeFill] (2.55,-0.09) rectangle (2.85,0.09);
\draw[cDbsOrangeDark, line width=0.4pt] (2.55,-0.09) rectangle (2.85,0.09);
\node[anchor=west, text=cDbsText] at (2.93,0) {\gptossModel{}};
\node[anchor=north west, text=cDbsText, align=left, text width=5.3cm, inner xsep=0pt] at (0.00,-0.22)
  {both at their own context ceiling; whiskers: Wilson 95\%;
   $n$ under each mark; point when $n<5$; dash when $n=0$};
\end{scope}
\end{tikzpicture}%
}
\end{minipage}%
\begin{minipage}[c]{0.56\liveW}
\centering
\figboxscript{%
\begin{tikzpicture}[font=\small]
\def\dapCellH{0.46}     %
\def\dapRowSeam{0.02}   %
\def\dapCellW{1.90}     %
\def\dapColSeam{0.04}   %
\def\dapSepLW{0.5}      %
\pgfmathsetmacro{\dapPitchY}{\dapCellH+\dapRowSeam}
\pgfmathsetmacro{\dapPitchX}{\dapCellW+\dapColSeam}
\newcommand{\dapCell}[3]{%
  \pgfmathsetmacro\cx{#1*\dapPitchX-\dapPitchX/2}
  \pgfmathsetmacro\cy{(9-#2)*\dapPitchY}
  \pgfmathsetmacro\pct{12+#3*72}
  \pgfmathsetmacro\dark{#3>0.70 ? 1 : 0}
  \fill[cDapBlueDark!\pct!white] (\cx-\dapCellW/2,\cy-\dapCellH/2) rectangle (\cx+\dapCellW/2,\cy+\dapCellH/2);
  \draw[white, line width=\dapSepLW pt] (\cx-\dapCellW/2,\cy-\dapCellH/2) rectangle (\cx+\dapCellW/2,\cy+\dapCellH/2);
  \ifdim\dark pt>0.5pt
    \node[text=white] at (\cx,\cy) {\pgfmathprintnumber[fixed, fixed zerofill, precision=3]{#3}};
  \else
    \node[text=cDapText] at (\cx,\cy) {\pgfmathprintnumber[fixed, fixed zerofill, precision=3]{#3}};
  \fi}
\foreach \r/\t in {1/{chain length}, 2/{root cause}, 3/{root operation},
  4/{root consequence}, 5/{sink cause}, 6/{sink operation},
  7/{sink consequence}, 8/{failure set}}{
  \pgfmathsetmacro\py{(9-\r)*\dapPitchY}
  \node[anchor=east, text=cDapText, inner sep=3pt] at (-0.10,\py) {\t};}
\foreach \c/\m/\n in {%
  1/{\gemmaModel}/\wtOwnNGemmaOwnCap,
  2/{\gptossModel}/\wtOwnNGptOssOwnCap}{
  \pgfmathsetmacro\hx{\c*\dapPitchX-\dapPitchX/2}
  \node[anchor=south, text=cDapText, align=center, inner sep=2pt, text width=1.8cm]
    at (\hx,{8*\dapPitchY+\dapCellH/2+0.10}) {\m\\($n=\n$)};}
\foreach \r/\v in \wtOwnAxesGemmaOwnCap {\dapCell{1}{\r}{\v}}
\foreach \r/\v in \wtOwnAxesGptOssOwnCap {\dapCell{2}{\r}{\v}}
\begin{scope}[yshift=-0.62cm]
\node[anchor=east, text=cDapText, inner sep=4pt, align=right] at (2.10,0.10)
  {mean pairwise agreement\\share per axis:};
\foreach \i in {0,...,9}{
  \pgfmathsetmacro\lx{2.10+\i*0.28}
  \pgfmathsetmacro\lp{12+(\i+0.5)*7.2}
  \fill[cDapBlueDark!\lp!white] (\lx,-0.02) rectangle (\lx+0.28,0.22);}
\node[anchor=west, text=cDapText, inner sep=3pt] at (2.10,0.42) {0};
\node[anchor=east, text=cDapText, inner sep=3pt] at (4.90,0.42) {1};
\node[anchor=north west, text=cDapText, inner sep=4pt, font=\small,
      text width=7.3cm, align=left] at (-2.50,-0.30)
  {\textbf{1 is the expected value; higher is better.} Both combinations read up to their
   own deployment's context ceiling. A cell of 1.000 means every pair
   of rounds agreed on that axis for every CVE in that combination; 0.000 means no pair agreed.
   Each column is computed on that combination's own CVEs with \wtWindowRounds{} completed
   rounds, capped at \wtWindowRounds{}. Figure~\ref{tab:multimodel_summary} instead uses
   the balanced window every combination covers, so its $n$ is smaller.};
\end{scope}
\end{tikzpicture}%
}
\end{minipage}
\floatnote{\itshape Left: share of fully deterministic CVEs ($D=1.0$ over \wtWindowRounds{} rounds) per group and combination,
with Wilson 95\% whiskers and per-group $n$, across (a) the evidence-confidence class of the
selected CVEs (the NIST-verified \wtVerifiedN\ beside the high- and medium-confidence selections) and
(b) corpus source (verified against the confidence-selected drawn complement). Computed per combination
over that combination's own CVEs with \wtWindowRounds{} completed rounds; a group with $n<5$ renders as a
point and is read as descriptive only.}
\caption{Left: measured determinism of the evidence-based derivation by
evidence-confidence stratum, per group and combination. Right: measured
per-axis determinism profile of the evidence-based derivation, one column per
combination; reading notes and the scale key are inside the panel.}
\label{fig:det_by_stratum}
\label{fig:det_axes_panel}
\Description{Left: two stacked bar charts on a 0-to-1 vertical scale, (a) grouped by evidence-confidence stratum and (b) grouped by corpus source; each group shows a pair of bars, blue for one deployment and orange for the other, with Wilson interval whiskers and the group size printed under each bar. Right: a table with eight axis rows (chain length, root cause, root operation, root consequence, sink cause, sink operation, sink consequence, failure set) and one column per combination, each cell shaded from light to dark by its agreement share, with a shade scale bar below.}
\end{inlinefloat}
exposure to the derivation procedure's reasoning. The unmapped stratum spans three class types, with \wtIrrNINP\ CVEs in BF INP, \wtIrrNDAT\ in BF DAT, and \wtIrrNMEM\ in BF MEM; BF FLR appears on the consequence axis of multiple CVEs as Information Exposure and Injection final errors rather than as a weakness class type.

\paragraph{Annotator setup}
    Two independent human annotators produced the anonymous mappings on an identical template, and the per-axis agreement was computed on their labels. Case studies cite the resolved labels from the inter-rater study where available, and the BF specification for case-study CVEs with published NIST chains, that is, CVE-2014-0160 (Heartbleed) and the BadAlloc pattern, which includes CVE-2021-21834. The remaining case-study mappings take their BF axis values from the inter-rater study (CVE-2021-3156) and the established memory-bugs class specification (CVE-2015-0235), as stated in each case study's BF classification paragraph, and are not implied to carry an independent agreement statistic.    

\paragraph{BF maturity}
    BF is an evolving framework~\cite{bojanova2024bf,samate_bf}: NIST has announced companion class-type specifications, and the BF Taxonomy at SAMATE is maintained as a live resource independently of the specification snapshot used here. The four-to-four mapping in Section~\ref{subsec:bf_addresses_failures} and the case-study labels in Section~\ref{sec:worked_examples} are grounded in NIST SP 800-231 as the specification of record at the date of writing; a proposed CVE-to-BF system would consume the BF taxonomy as a live resource instead of hard-coding this snapshot.

\begin{inlinefloat}
\begin{tikzpicture}[font=\small]
\def\bspChipH{0.36}    %
\def\bspRowSeam{0.08}  %
\def\bspChipW{0.82}    %
\def\bspPitchX{1.04}   %
\pgfmathsetmacro{\bspPitchY}{\bspChipH+\bspRowSeam}
\newcommand{\bspRowSet}[3]{%
  \foreach \r/\seq/\c/\fc/\tc in #1 {%
    \pgfmathsetmacro\py{#3-(\r-1)*\bspPitchY}
    \foreach [count=\k] \cls in \seq {%
      \pgfmathsetmacro\xa{(\k-1)*\bspPitchX}
      \fill[\fc] (\xa,\py-\bspChipH/2) rectangle (\xa+\bspChipW,\py+\bspChipH/2);
      \node[text=\tc, inner sep=0pt] at (\xa+\bspChipW/2,\py) {\cls};
      \ifnum\k>1
        \draw[cBspText, line width=0.5pt, ->] (\xa-0.18,\py) -- (\xa-0.03,\py);
      \fi
    }
    \node[anchor=west, text=cBspText, inner sep=1pt] at (#2,\py) {$\times$\c};
  }}
\node[anchor=west, text=cBspText, font=\small\bfseries] at (-0.02,4.25)
  {NIST-verified corpus ($n=\wtGoldRowCount$ out of \wtVerifiedN)};
\bspRowSet{\bspVerified}{7.40}{3.65}
\newcommand{\bspModelPanel}[6]{%
\begin{scope}[xshift=0.60\liveW, yshift=#1]
  \fill[#2] (0.00,4.15) rectangle (0.28,4.35);
  \node[anchor=west, text=cBspText, font=\small\bfseries] at (0.36,4.25) {#3 ($n=#6$)};
  \bspRowSet{#4}{3.75}{3.65}
  \foreach \osp/\ocv in #5 {%
    \pgfmathsetmacro\oy{3.65-4*\bspPitchY}
    \draw[gray!60, dashed, line width=0.5pt] (0,\oy-\bspChipH/2) rectangle (3.55,\oy+\bspChipH/2);
    \node[text=cBspText, inner sep=0pt] at (1.775,\oy) {other: \osp\ class-chains};
    \node[anchor=west, text=cBspText, inner sep=1pt] at (3.75,\oy) {$\times$\ocv};
  }
\end{scope}}
\bspModelPanel{0cm}{cBspSlateDark}{\gemmaModel{}}{\wtSpineRowsGemmaOwnCap}{\wtSpineOtherGemmaOwnCap}{\wtSpinePanelNGemmaOwnCap}
\bspModelPanel{-2.95cm}{cBspOrangeDark}{\gptossModel{}}{\wtSpineRowsGptOssOwnCap}{\wtSpineOtherGptOssOwnCap}{\wtSpinePanelNGptOssOwnCap}
\begin{scope}[yshift=-1.70cm]
\node[anchor=west, text=cBspText] at (0.00,0) {class-chain frequency:};
\fill[cBspRampLight] (3.60,-0.10) rectangle (3.95,0.10);
\node[anchor=west, text=cBspText] at (4.02,0) {1 to 2};
\fill[cBspRampMid] (5.30,-0.10) rectangle (5.65,0.10);
\node[anchor=west, text=cBspText] at (5.72,0) {3 to 9};
\fill[cBspRampDark] (7.00,-0.10) rectangle (7.35,0.10);
\node[anchor=west, text=cBspText] at (7.42,0) {10 or more};
\end{scope}
\end{tikzpicture}%
\floatnote{\itshape Left: the distinct verified class-chains of the NIST-verified corpus (\wtGoldRowCount\ out of \wtVerifiedN),
one row per class-chain, frequency at right and encoded by chip shade on one shared blue ramp. Right: the
measured modal derived class-chains per deployment at its own context ceiling, one panel per
deployment over its own CVEs with \wtWindowRounds{} completed rounds (per-panel $n$ in the
header).}
\caption{Class-chains, root to sink, in the multi-model evaluation.}
\label{fig:bug_sequence_paths}
\Description{Three panels side by side. Left: the distinct class-chains of the NIST-verified corpus, one per row, each drawn as a sequence of class chips joined by arrows with a frequency count at the right and chip shade on a shared blue ramp. Middle and right: the same rendering of the modal derived class-chain per CVE for each deployment, each ending with a dashed row that collects the remaining chains. A frequency legend for the chip shading sits beneath.}
\end{inlinefloat}

\paragraph{Synthesis-mode boundary}
    Section~\ref{sec:related_work} and the integrated survey tables (see Table~\ref{tab:performance_reported}) cover all \wtCorpusIncluded\ papers in the corpus. Where a methodological field is not reported in the source, the table cell uses the \textit{n.r.} marker defined in the table legend. The boundary between sourced and schema-derived placement is marked in Table~\ref{tab:methodology_overview} itself through the footnoted cells so that a reader can separate authorial synthesis from verbatim reporting at the point of use.

\FloatBarrier
\section{Conclusion}\label{sec:conclusion}
    This paper presented a literature-grounded case study of NIST SP 800-231 as a target for automated vulnerability classification. The evidence suggests that current CVE-to-CWE automation faces a structural target-space problem. At the same time, the NIST Bugs Framework offers a more structured and causally meaningful representation through its compositional design. Our empirical studies show that LLM-based BF classification can achieve substantial repeatability under stable deployments, although reproduced chains do not always match published verified chains. Overall, the findings position BF as a promising and more automation-suitable target for future vulnerability classification research, with future work focused on improving per-axis validation, annotation consistency, and alignment with verified BF chains.


\FloatBarrier
\begin{acks}
This research was carried out in the PATENT Lab within the Department of Computer Science at The University of Alabama. The opinions and conclusions expressed in this paper are those of the authors and do not necessarily represent the official policies or positions of their affiliated institutions.
\end{acks}

\FloatBarrier
\bibliographystyle{ACM-Reference-Format}
\bibliography{references}

\FloatBarrier
\appendixfloatexception
\definecolor{cAcronymsHeaderBg}{HTML}{E8EFF6} %
\definecolor{cAcronymsHeaderFg}{HTML}{092030} %
\definecolor{cAcronymsHeaderRule}{HTML}{8DA5BC} %
\definecolor{cAcronymsSubHeaderBg}{HTML}{F4F7FB} %
\definecolor{cAcronymsAmberFillL}{HTML}{FEF3C7} %
\definecolor{cAcronymsAmberText}{HTML}{78350F} %
\definecolor{cAcronymsCorpusFillD}{HTML}{B7D3E8} %
\definecolor{cAcronymsCorpusText}{HTML}{12365B} %
\definecolor{cAcronymsFailFillD}{HTML}{FCA5A5} %
\definecolor{cAcronymsFailFillL}{HTML}{FEE2E2} %
\definecolor{cAcronymsFailText}{HTML}{7F1D1D} %
\definecolor{cAcronymsSlateFillD}{HTML}{CBD5E1} %
\definecolor{cAcronymsSlateFillL}{HTML}{F1F5F9} %
\definecolor{cAcronymsSlateText}{HTML}{0F172A} %
\definecolor{cAcronymsTealFillL}{HTML}{F9FEFF} %
\definecolor{cAcronymsTealText}{HTML}{061C1A} %
\definecolor{cAcronymsVioletFillL}{HTML}{F5F3FF} %
\definecolor{cAcronymsVioletText}{HTML}{4C1D95} %
\appendices

\FloatBarrier
\section{List of Acronyms}\label{app:acronyms}

This appendix collects the acronyms and abbreviations used throughout the
paper. Table~\ref{tab:acronyms_general} lists the general technical and
standards acronyms. Table~\ref{tab:acronyms_bf} tabulates the Bugs Framework
taxonomy abbreviations, ordered to replicate the taxonomy structure of
Fig.~\ref{fig:bf_taxonomy}, from the BF class types down to their
constituent classes.

\begin{inlinefloat}
\def\acrRowH{0.40}
\def\acrRowSep{0.1}
\definecolor{cAcronymsZebra}{HTML}{F3F7FA}

\begin{minipage}[t]{0.455\liveW}
\centering
\colorlet{tblHeaderBg}{cAcronymsHeaderBg}
\colorlet{tblHeaderFg}{cAcronymsHeaderFg}
\colorlet{tblHeaderRule}{cAcronymsHeaderRule}
\colorlet{tblSubHeaderBg}{cAcronymsSubHeaderBg}
\begin{lrbox}{\tbltmpbox}%
\begin{tikzpicture}[baseline=(current bounding box.north)]
\matrix (m) [tblmat,
  row sep=\acrRowSep pt,
  column sep=0pt,
  nodes={inner xsep=3pt, outer sep=0pt, minimum height=\acrRowH cm, font=\small},
  column 1/.style={nodes={text width=0.115\liveW, align=left}},
  column 2/.style={nodes={text width=0.295\liveW, align=left}},
]{
  |[tblhdr]| \textbf{Acronym} & |[tblhdr]| \textbf{Expansion} \\
  ATT\&CK & Adversarial Tactics, Techniques, and Common Knowledge \\
  BERT & Bidirectional Encoder Representations from Transformers \\
  BF & Bugs Framework \\
  CAPEC & Common Attack Pattern Enumeration and Classification \\
  CPE & Common Platform Enumeration \\
  CTI & Cyber Threat Intelligence \\
  CVE & Common Vulnerabilities and Exposures \\
  CVSS & Common Vulnerability Scoring System \\
  CWE & Common Weakness Enumeration \\
  EPSS & Exploit Prediction Scoring System \\
  JSON & JavaScript Object Notation \\
  KG & Knowledge Graph \\
  LLM & Large Language Model \\
  ML & Machine Learning \\
  NIST & National Institute of Standards and Technology \\
  NVD & National Vulnerability Database \\
  PRISMA & Preferred Reporting Items for Systematic Reviews and
          Meta-Analyses \\
  SP & Special Publication \\
};
\begin{scope}[on background layer]
  \foreach \r in {3,5,7,9,11,13,15,17,19}{
    \fill[cAcronymsZebra] (m.west|-m-\r-2.north) rectangle (m.east|-m-\r-2.south);}
\end{scope}
\draw[Rtop] (m.north west) -- (m.north east);
\draw[Rmid] (m.west|-m-2-2.north) -- (m.east|-m-2-2.north);
\draw[Rbot] (m.south west) -- (m.south east);
\end{tikzpicture}%
\end{lrbox}
\usebox{\tbltmpbox}
\captionof{table}{General technical and standards acronyms used in the paper.}
\label{tab:acronyms_general}
\end{minipage}
\begin{minipage}[t]{0.525\liveW}
\centering
\colorlet{tblHeaderBg}{cAcronymsHeaderBg}
\colorlet{tblHeaderFg}{cAcronymsHeaderFg}
\colorlet{tblHeaderRule}{cAcronymsHeaderRule}
\colorlet{tblSubHeaderBg}{cAcronymsSubHeaderBg}
\tikzset{tblhdr/.append style={font=\bfseries\small}}
\begin{lrbox}{\tbltmpbox}%
\begin{tikzpicture}[
  baseline=(current bounding box.north),
  chip/.style={align=center, minimum width=7mm,
               inner xsep=3pt, inner ysep=1.5pt, font=\small\bfseries},
  chipCT/.style={chip, fill=cAcronymsSlateFillD, text=cAcronymsSlateText},
  chipC/.style ={chip, fill=cAcronymsSlateFillL, text=cAcronymsSlateText},
  chipW/.style ={chip, fill=cAcronymsCorpusFillD, text=cAcronymsCorpusText},
  chipF/.style ={chip, fill=cAcronymsFailFillD, text=cAcronymsFailText},
  chipINP/.style={chip, fill=cAcronymsTealFillL, text=cAcronymsTealText},
  chipMEM/.style={chip, fill=cAcronymsAmberFillL, text=cAcronymsAmberText},
  chipDAT/.style={chip, fill=cAcronymsVioletFillL, text=cAcronymsVioletText},
  chipFLR/.style={chip, fill=cAcronymsFailFillL, text=cAcronymsFailText},
]
\matrix (m) [tblmat,
  row sep=\acrRowSep pt,
  column sep=0pt,
  nodes={font=\small, inner xsep=2.5pt, inner ysep=1.5pt, outer sep=0pt, minimum height=\acrRowH cm, text height=1.6ex, text depth=0.4ex},
  row 1/.style={nodes={text height=, text depth=, inner ysep=3pt}},
  column 1/.style={nodes={text width=0.055\liveW, align=left}},
  column 2/.style={nodes={align=left}},
  column 3/.style={nodes={text width=0.095\liveW, align=center}},
  column 4/.style={nodes={text width=0.095\liveW, align=center}},
]{
  |[tblhdr]| \textbf{Abbr.} & |[tblhdr]| \textbf{Name} &
  |[tblhdr]| \textbf{Category} & |[tblhdr]| \shortstack{\textbf{Class}\\\textbf{Category}} \\
  INP & Input/Output Check & |[chipCT]| CT & |[chipW]| W \\
  MEM & Memory             & |[chipCT]| CT & |[chipW]| W \\
  DAT & Data Type          & |[chipCT]| CT & |[chipW]| W \\
  FLR & Failure            & |[chipCT]| CT & |[chipF]| F \\
  DVL & Data Validation    & |[chipC]| C & |[chipINP]| INP \\
  DVR & Data Verification  & |[chipC]| C & |[chipINP]| INP \\
  MAD & Memory Addressing  & |[chipC]| C & |[chipMEM]| MEM \\
  MMN & Memory Management  & |[chipC]| C & |[chipMEM]| MEM \\
  MUS & Memory Use         & |[chipC]| C & |[chipMEM]| MEM \\
  DCL & Declaration        & |[chipC]| C & |[chipDAT]| DAT \\
  NRS & Name Resolution    & |[chipC]| C & |[chipDAT]| DAT \\
  TCV & Type Conversion    & |[chipC]| C & |[chipDAT]| DAT \\
  TCM & Type Computation   & |[chipC]| C & |[chipDAT]| DAT \\
  IEX & Information Exposure     & |[chipC]| C & |[chipFLR]| FLR \\
  ACE & Arbitrary Code Execution & |[chipC]| C & |[chipFLR]| FLR \\
  DOS & Denial of Service        & |[chipC]| C & |[chipFLR]| FLR \\
  TPR & Data Tampering           & |[chipC]| C & |[chipFLR]| FLR \\
};
\draw[Rtop] (m.north west) -- (m.north east);
\draw[Rmid] (m.west|-m-2-2.north) -- (m.east|-m-2-2.north);
\draw[Rmid, draw=gray!30] (m.west|-m-6-2.north) -- (m.east|-m-6-2.north);
\draw[Rmid, draw=gray!30] (m.west|-m-8-2.north) -- (m.east|-m-8-2.north);
\draw[Rmid, draw=gray!30] (m.west|-m-11-2.north) -- (m.east|-m-11-2.north);
\draw[Rmid, draw=gray!30] (m.west|-m-15-2.north) -- (m.east|-m-15-2.north);
\draw[Rbot] (m.south west) -- (m.south east);

\node[anchor=north west, font=\small, inner xsep=0pt, align=left,
      text width=0.525\liveW]
  at ($(m.south west)+(0,-0.18)$) {%
  \renewcommand{\arraystretch}{1.3}%
  \begin{tabular}{@{}l@{\hspace{8pt}}l@{}}
  \textbf{Category:} & \tikz[baseline=(c.base)]\node(c)[chipCT]{CT}; BF class type\\
   & \tikz[baseline=(c.base)]\node(c)[chipC]{C}; BF class\\
  \textbf{Class Category:} & \tikz[baseline=(c.base)]\node(c)[chipW]{W}; Weakness class type\\
   & \tikz[baseline=(c.base)]\node(c)[chipF]{F}; Failure class type\\
   & \tikz[baseline=(c.base)]\node(c)[chipINP]{INP};
     \tikz[baseline=(c.base)]\node(c)[chipMEM]{MEM};
     \tikz[baseline=(c.base)]\node(c)[chipDAT]{DAT};
     \tikz[baseline=(c.base)]\node(c)[chipFLR]{FLR};\\[-3.5pt]
   & parent class type of the class\\
  \end{tabular}};
\end{tikzpicture}%
\end{lrbox}
\usebox{\tbltmpbox}
\captionof{table}{Bugs Framework taxonomy abbreviations up to the class level,
matching Fig.~\ref{fig:bf_taxonomy}; the Category and Class Category
codes are expanded in the legend below the table.}
\label{tab:acronyms_bf}
\end{minipage}
\end{inlinefloat}
\FloatBarrier
\section{Threats to Validity}\label{sec:threats_a}

Several threats condition the claims presented in this paper. The threats are organized into five categories:

\subsection{Construct Validity}
Case studies interpret the reassignment of a CVE's primary CWE as evidence that the original mapping was never uniquely determined. This is indicative rather than conclusive, since a reassignment can instead reflect a deeper root-cause analysis. In addition, four case studies, representing one per BF class, are not enough to prove that these failures are systemic across the wider CVE corpus, so this generalization remains argued rather than measured.

\subsection{Internal Validity}
The case studies were selected based on failure density and anchor status; the full candidate pool of \wtPoolSlots\ CVEs is published in Table~\ref{tab:candidate_pool} and in the artifact (\texttt{bf\_example\_candidates\_full\_pool.json}). Moreover, annotator bias is bounded by the blind two-annotator design whose per-axis agreement Section~\ref{subsec:study1} reports; a family-level encoding mismatch drives the low attribute kappa, so determinism language is scoped to the cause and operation axes.

\def\qtRowH{0.44}     
\def\qtRowSep{0.3}    
\definecolor{cQtSteelBorder}{HTML}{ADBCD3}
\definecolor{cQtSteelHdr}{HTML}{DCE4EE}
\definecolor{cQtSteelZebra}{HTML}{F5F7FA}
\definecolor{cQtSteelText}{HTML}{0D162D}

\begin{inlinefloat}
\colorlet{tblHeaderBg}{cQtSteelHdr}
\colorlet{tblHeaderFg}{cQtSteelText}
\colorlet{tblHeaderRule}{cQtSteelBorder}
\begin{lrbox}{\tbltmpbox}%
\begin{tikzpicture}
\matrix (m) [tblmat,
  row sep=\qtRowSep pt,
  nodes={minimum height=\qtRowH cm},
  column 1/.style={nodes={text width=0.26\liveW, align=left}},
  column 2/.style={nodes={text width=0.38\liveW, align=left}},
  column 3/.style={nodes={text width=0.10\liveW, align=center}},
  column 4/.style={nodes={text width=0.14\liveW, align=center}},
]{
  |[tblhdr]| Short name & |[tblhdr]| Method family & |[tblhdr]| Year & |[tblhdr]| In corpus \\
  VWC-MAP & V2W-BERT + T5 pipeline & 2022 & No \\
  ThreatCompass & Graph + ML ensemble pipeline & 2025 & No \\
  CVEDrill & Fine-tuned prioritization & 2023 & Yes \\
  Text2Weak & Embedding retrieval & 2024 & Yes \\
  Pure Self-Attention & Self-attention classification & 2021 & Yes \\
  Key Term Extraction & LLM key-term extraction & 2026 & Yes \\
};
\begin{scope}[on background layer]
  \foreach \r in {3,5,7}{
    \fill[cQtSteelZebra] (m.west|-m-\r-1.north) rectangle (m.east|-m-\r-1.south);}
\end{scope}
\draw[Rtop] (m.north west) -- (m.north east);
\draw[Rmid] (m.west|-m-2-1.north) -- (m.east|-m-2-1.north);
\draw[Rbot] (m.south west) -- (m.south east);
\end{tikzpicture}%
\end{lrbox}
\usebox{\tbltmpbox}
\captionof{table}{Anchor papers used as a known-relevant validation set for search
recall, with method family, venue year, and final-corpus membership.}
\label{tab:anchors}
\end{inlinefloat}

\subsection{Limitations of the Bugs Framework as an Automation Target}
Treating SP 800-231 as the target of automation surfaces five properties of the framework that condition any CVE-to-BF automation, recorded as specification gaps. First, the attribute layer is under-specified: the specification enumerates attribute values but gives little guidance for selecting among them, and nearly all cross-round variation in the automated study concentrates in attribute assignments while the class-chain holds (Section~\ref{subsec:study2_results}). Second, the same vulnerability admits multiple legal renderings: chain length is unbounded, the causation rule's second mode is enumerated for only four propagations, and on the 42-chain verified corpus the strict value-identity reading of causation holds for 21 of 38 adjacent links, so a conforming tool must either implement an unenumerated rule or fail to reconstruct nearly half of the links. Third, a one-weakness chain whose cause resolves to a bug is legal BF while the verified chains decompose the propagation into several phases, and nothing in the specification ranks two conforming outputs of different depth. Fourth, per-class consequence sets are published for two of the nine classes and the cross-class operation flow is stated to be incomplete, so a tool cannot distinguish not-listed from not-allowed. Fifth, a chain is derivable only to the depth the available artifacts expose, so evidence richness bounds chain depth independently of model capability; the evidence ledger accompanies the artifact.

\FloatBarrier
\section{Database Search Strategies}\label{app:search}

This appendix reproduces the search protocol word-to-word that has been executed against
each of the six databases during the Identification stage, so the
candidate set can be regenerated independently. Two filters were applied
uniformly: a publication-year window of January 2018 to April 2026, and a
content-type restriction to journal and conference papers. Four databases
(arXiv, IEEE Xplore, the ACM Digital Library, and Springer Link) admit
field-tagged Boolean queries and were searched with one common keyword
concept structure adapted to each database's syntax.

\subsection{Anchor Papers: Known-Relevant Validation Set}\label{app:anchors}

The anchor papers in Table~\ref{tab:anchors} form a known-relevant
validation set which were selected before search execution to check keyword-search
recall.

An anchor is not guaranteed a place in the included corpus: VWC-MAP and
ThreatCompass are downstream pipelines in which CWE is only an
intermediate stage toward CAPEC or ATT\&CK, so under the exclusion and
safety-valve criteria of Section~\ref{sec:methodology} they are excluded
from the corpus while still functioning as recall checkpoints.

\subsection{Case-Study Candidate Pool}\label{app:candidate_pool}

Table~\ref{tab:candidate_pool} lists the \wtPoolSlots-CVE candidate pool
from which Section~\ref{sec:worked_examples} drew its four case studies,
four candidates per BF class-type family. Each candidate is scored by the
number of structural CWE failures it exposes; the selected case study
attains the maximum count in its family, with anchor status as the
tie-breaker.

\subsection{Identification Yield, Recall, and Query Protocols}\label{app:recall}

Table~\ref{tab:yield} reports the candidate records contributed per
database; pooled and deduplicated, the Identification stage yielded
\wtCorpusPooled\ distinct records. The keyword search retrieved all
\wtCorpusIncluded\ finally included papers and every anchor in
Table~\ref{tab:anchors}; no supplementary search method was required.

Table~\ref{tab:queries} reproduces the per-database queries verbatim.
Google Scholar weights early tokens heavily and degrades on long Boolean
strings, so its concept structure was decomposed into twelve short
queries run independently and merged; Semantic Scholar was searched by
keyword query with a relevance filter; arXiv was restricted to categories
cs.CR, cs.CL, and cs.LG; IEEE Xplore used its Command Search interface;
the ACM Advanced Search was applied to ``Anywhere''; and Springer Link
was searched through its field form, since its advanced search weights
title and abstract heavily.

\definecolor{cCandSteelBorder}{HTML}{ADBCD3}
\definecolor{cCandSteelHdr}{HTML}{DCE4EE}
\definecolor{cCandSteelZebra}{HTML}{F5F7FA}
\definecolor{cCandSteelGutter}{HTML}{E8EDF4}
\definecolor{cCandSteelText}{HTML}{0D162D}
\definecolor{cCandDenseLo}{HTML}{DDE7F2}
\definecolor{cCandDenseHi}{HTML}{9FBBD8}
\tikzset{
  densLo/.style={fill=cCandDenseLo, text=cCandSteelText},
  densHi/.style={fill=cCandDenseHi, text=cCandSteelText},
}
\begin{inlinebox}
    
\colorlet{tblHeaderBg}{cCandSteelHdr}
\colorlet{tblHeaderFg}{cCandSteelText}
\colorlet{tblHeaderRule}{cCandSteelBorder}
\colorlet{tblSubHeaderBg}{cCandSteelGutter}

\tikzset{tblhdr/.append style={font=\bfseries\small}}
\begin{lrbox}{\tbltmpbox}%
\begin{tikzpicture}
\matrix (m) [tblmat,
  row sep=0pt,
  nodes={font=\small, inner xsep=3pt, inner ysep=1.5pt,
         minimum height=0.32cm, align=center},
  column 1/.style={nodes={text width=0.07\liveW}},
  column 2/.style={nodes={text width=0.165\liveW}},
  column 3/.style={nodes={text width=0.15\liveW}},
  column 4/.style={nodes={text width=0.20\liveW}},
  column 5/.style={nodes={text width=0.075\liveW}},
  column 6/.style={nodes={text width=0.22\liveW}},
]{
  |[tblhdr]| Family & |[tblhdr]| CVE & |[tblhdr]| Original primary CWE & |[tblhdr]| Current primary CWE & |[tblhdr]| Failures & |[tblhdr]| Reassignment reason \\
   & CVE-2021-3156$^{\dagger\ast}$ & CWE-193 & CWE-122 (was CWE-193) & |[densHi]| 3 & Deepened root-cause analysis \\
   & CVE-2018-5907 & CWE-20 & CWE-190 & |[densHi]| 3 & Deepened root-cause analysis \\
   & CVE-2017-5638$^{\S}$ & CWE-20 & CWE-755 (was CWE-20) & |[densHi]| 3 & Deepened root-cause analysis \\
   & CVE-2014-6271 & CWE-78 & CWE-78 (unchanged) & |[densLo]| 2 & Other or unknown \\
   & CVE-2015-0235$^{\ast}$ & CWE-119 & CWE-787 & |[densHi]| 3 & Taxonomy refinement (new CWE) \\
   & CVE-2019-0708 & CWE-416 & CWE-416 (unchanged) & |[densLo]| 2 & Deepened root-cause analysis \\
   & CVE-2017-1000353 & CWE-502 & CWE-502 (unchanged) & |[densLo]| 2 & Error in initial mapping \\
   & CVE-2018-20991 & CWE-416 & CWE-416 (unchanged) & |[densLo]| 2 & Deepened root-cause analysis \\
   & CVE-2021-21834$^{\ast}$ & CWE-190 & CWE-190 (CWE-680 refined) & |[densHi]| 3 & Taxonomy refinement (new CWE) \\
   & CVE-2016-7524 & CWE-190 & CWE-190 (unchanged) & |[densLo]| 2 & Deepened root-cause analysis \\
   & CVE-2017-6500 & CWE-190 / CWE-119 & Varies$^{\P}$ & |[densLo]| 2 & Deepened root-cause analysis \\
   & CVE-2019-13297 & CWE-787 / CWE-119 & Varies$^{\P}$ & |[densLo]| 2 & Deepened root-cause analysis \\
   & CVE-2014-0160$^{\ast}$ & CWE-119 & CWE-125 & |[densHi]| 3 & Deepened root-cause analysis \\
   & CVE-2017-5638$^{\S}$ & CWE-20 & CWE-755 (was CWE-20) & |[densHi]| 3 & Deepened root-cause analysis \\
   & CVE-2021-44228 & CWE-502 & CWE-917 & |[densHi]| 3 & Deepened root-cause analysis \\
   & CVE-2014-3566 & CWE-310 & CWE-310 (unchanged) & |[densLo]| 2 & Taxonomy refinement (new CWE) \\
};

\foreach \first/\last/\name in {2/5/INP, 6/9/MEM, 10/13/DAT, 14/17/FLR}{
  \node[font=\bfseries\small, text=cCandSteelText, anchor=center]
    at ($(m-\first-1.center)!0.5!(m-\last-1.center)$) {\name};
}

\begin{scope}[on background layer]
  \fill[cCandSteelGutter] (m.west|-m-2-1.north) rectangle (m-1-2.north west|-m.south);
  \foreach \r in {3,5,7,9,11,13,15,17}{
    \fill[cCandSteelZebra] (m-1-2.north west|-m-\r-1.north)
      rectangle (m.east|-m-\r-1.south);
  }
\end{scope}

\draw[Rtop] (m.north west) -- (m.north east);
\draw[Rmid] (m.west|-m-2-1.north) -- (m.east|-m-2-1.north);
\draw[Rbot] (m.south west) -- (m.south east);
\draw[Rtop, draw=tblHeaderRule] (m-1-2.north west|-m.north) -- (m-1-2.north west|-m.south);
\draw[Rtop, draw=tblHeaderRule] (m.north west) -- (m.south west);
\draw[Rtop, draw=tblHeaderRule] (m.north east) -- (m.south east);
\foreach \r in {6,10,14}{
  \draw[Rmid] (m.west|-m-\r-1.north) -- (m.east|-m-\r-1.north);
}

\end{tikzpicture}%
\end{lrbox}
\usebox{\tbltmpbox}

\floatnote{\itshape \textbf{Reading the table.} \emph{Failures} is the number of the four
structural CWE failures the candidate exposes, shaded on a two-step scale,
\tikz[baseline=-0.30ex]{\fill[cCandDenseLo](0,0)rectangle(0.30,0.155);
\draw[cCandSteelBorder,line width=0.15pt](0,0)rectangle(0.30,0.155);}\,two and
\tikz[baseline=-0.30ex]{\fill[cCandDenseHi](0,0)rectangle(0.30,0.155);
\draw[cCandSteelBorder,line width=0.15pt](0,0)rectangle(0.30,0.155);}\,three.
Within each family the selected CVE attains the maximum count, with anchor
status as the tie-breaker. \emph{Reassignment reason} records why the current
primary CWE differs from the original, or states that it is unchanged. CWE and
reassignment fields reflect the NVD record on May 27, 2026.\\[1pt]
$^{\ast}$~Selected as the case study for its family.\quad
$^{\dagger}$~Anchor candidate already referenced in this paper; preferred on
the tie-breaker over the two other INP candidates that also reach three.\quad
$^{\S}$~CVE-2017-5638 appears under both INP and FLR: its input-check root
cause places it in INP, its failure-type consequence places it in FLR. It is
kept in both family pools and is the selected example for neither.\quad
$^{\P}$~The current primary CWE varies by source for these two records, which
share a root cause with CVE-2016-7518 and CVE-2019-13295 respectively.}
\captionof{table}{Case-study candidate pool of \wtPoolSlots\ CVEs, four per BF bug-type
family, with the four selected case studies marked.}
\label{tab:candidate_pool}
\end{inlinebox}

\begin{inlinebox}
\colorlet{tblHeaderBg}{cQtSteelHdr}
\colorlet{tblHeaderFg}{cQtSteelText}
\colorlet{tblHeaderRule}{cQtSteelBorder}
\begin{lrbox}{\tbltmpbox}%
\begin{tikzpicture}
\matrix (m) [tblmat,
  row sep=\qtRowSep pt,
  nodes={minimum height=\qtRowH cm},
  column 1/.style={nodes={text width=0.66\liveW, align=left}},
  column 2/.style={nodes={align=center, text width=0.22\liveW}},
]{
  |[tblhdr]| Database & |[tblhdr]| Records \\
  Springer Link & \wtDbSpringerLink \\
  Semantic Scholar & \wtDbSemanticScholar \\
  Google Scholar & \wtDbGoogleScholar \\
  arXiv & \wtDbArxiv \\
  IEEE Xplore & \wtDbIEEEXplore \\
  ACM Digital Library & \wtDbACMDL \\
  Pooled, deduplicated & \wtCorpusPooled \\
};
\begin{scope}[on background layer]
  \foreach \r in {3,5,7}{
    \fill[cQtSteelZebra] (m.west|-m-\r-1.north) rectangle (m.east|-m-\r-1.south);}
\end{scope}
\draw[Rtop] (m.north west) -- (m.north east);
\draw[Rmid] (m.west|-m-2-1.north) -- (m.east|-m-2-1.north);
\draw[Rmid] (m.west|-m-8-1.north) -- (m.east|-m-8-1.north);
\draw[Rbot] (m.south west) -- (m.south east);
\end{tikzpicture}%
\end{lrbox}
\usebox{\tbltmpbox}
\captionof{table}{Candidate records contributed per database at the Identification
stage, before deduplication; counts overlap across databases.}
\label{tab:yield}
\end{inlinebox}

\FloatBarrier
\par\vspace{\intextsep}
\captionof{table}{Verbatim per-database query protocols of the Identification
stage.}
\label{tab:queries}
\vspace{4pt}
\begingroup\small
\setlength{\topsep}{2pt}\setlength{\partopsep}{0pt}%
\noindent\textbf{Google Scholar}
\begin{verbatim}
Query 1:  CVE CWE classification automated mapping
Query 2:  CVE CWE prediction transformer BERT
Query 3:  CVE CWE large language model
Query 4:  NVD weakness enumeration deep learning
Query 5:  vulnerability classification CWE hierarchy
Query 6:  CVE description CWE label assignment
Query 7:  CWE-1003 classification
Query 8:  CAPEC ATT&CK CVE pipeline classification
Query 9:  vulnerability triage CWE neural
Query 10: SecureBERT vulnerability classification
Query 11: V2W-BERT CWE
Query 12: weakness type prediction CVE
\end{verbatim}
\noindent\textbf{Semantic Scholar}
\begin{verbatim}
Keyword query (2018-01 to 2026-04):
  "CVE-to-CWE classifier", "CWE label prediction",
  "automated weakness enumeration", "CWE hierarchy
  inference", "vulnerability description
  classification CWE".
\end{verbatim}
\noindent\textbf{arXiv}
\begin{verbatim}
Query 1: abs:"CVE" AND abs:"CWE" AND
         (abs:"classification" OR abs:"mapping"
          OR abs:"prediction")
Query 2: abs:"vulnerability classification"
         AND abs:"CWE"
Query 3: abs:"NVD" AND abs:"weakness" AND
         (abs:"BERT" OR abs:"LLM" OR
          abs:"transformer" OR
          abs:"large language model")
Query 4: ti:"CWE" AND
         (ti:"CVE" OR ti:"vulnerability")
Query 5: abs:"CWE-1003" OR abs:"Top-25 CWE"
         OR abs:"CWE hierarchy"
Query 6: abs:"CVE description" AND abs:"CWE"
         AND (abs:"neural" OR abs:"deep"
              OR abs:"embedding")
Query 7: abs:"SecureBERT" OR abs:"V2W-BERT"
Query 8: abs:"CTI pipeline" AND abs:"CWE"
\end{verbatim}
\noindent\textbf{IEEE Xplore}
\begin{verbatim}
("All Metadata":"CVE" AND "All Metadata":"CWE" AND
 ("All Metadata":"classification" OR
  "All Metadata":"mapping" OR
  "All Metadata":"prediction" OR
  "All Metadata":"label"))
AND
("All Metadata":"learn*" OR
 "All Metadata":"neural" OR
 "All Metadata":"BERT" OR
 "All Metadata":"transformer" OR
 "All Metadata":"large language model" OR
 "All Metadata":"LLM" OR
 "All Metadata":"embedding" OR
 "All Metadata":"retrieval")
NOT
("Document Title":"survey" AND
 NOT "Document Title":"systematic")

Second pass (title-restricted):
("Document Title":"CWE" AND "Document Title":"CVE")
OR
("Document Title":"vulnerability" AND
 "Document Title":"CWE")
OR
("Document Title":"weakness" AND "Abstract":"CVE")
\end{verbatim}
\noindent\textbf{ACM Digital Library}
\begin{verbatim}
(Abstract:(CVE AND CWE) AND
 Abstract:(classification OR mapping OR prediction
           OR assignment))
AND
(Abstract:(learning OR neural OR transformer OR BERT
           OR LLM OR "large language model" OR
           embedding OR retrieval))

Second pass (title-focused):
Title:(CWE) AND (Title:(CVE) OR
                 Title:(vulnerability) OR
                 Title:(weakness))
\end{verbatim}
\noindent\textbf{Springer Link}
\begin{verbatim}
with all of the words:
  CVE CWE

with at least one of the words:
  classification mapping prediction transformer BERT
  LLM neural embedding retrieval

without the words:
  malware-detection
  source-code-vulnerability-detection

Second pass: title contains "CWE" OR
"weakness enumeration".
\end{verbatim}
\endgroup
\par\vspace{\intextsep}

\end{document}